\documentclass[]{interact}

\usepackage{epstopdf}
\usepackage[caption=false]{subfig}
\usepackage{esint}
\usepackage{amsmath}
\usepackage[dvipsnames]{xcolor}
\usepackage[numbers,sort&compress]{natbib}
\bibpunct[, ]{[}{]}{,}{n}{,}{,}
\renewcommand\bibfont{\fontsize{10}{12}\selectfont}
\makeatletter
\def\NAT@def@citea{\def\@citea{\NAT@separator}}
\makeatother
\DeclareUnicodeCharacter{2212}{-}
\theoremstyle{plain}

\theoremstyle{definition}

\theoremstyle{remark}

\usepackage{bm}

\begin{document}

\articletype{Invited Review}

\title{Optical Vortices: Revolutionizing the field of linear and nonlinear optics}

\author{
\name{Bikash K. Das\textsuperscript{a,b,c}\thanks{Email: bikash.das@campus.technion.ac.il}, C. Granados\textsuperscript{d}\thanks{Email: cagrabu@eitech.edu.cn} and M.F. Ciappina\textsuperscript{a,b,c}\thanks{Email: marcelo.ciappina@gtiit.edu.cn}}
\affil{\textsuperscript{a}Department of Physics, Guangdong Technion-Israel Institute of Technology, 241 Daxue Rd., Shantou, 515063, Guangdong, China; \textsuperscript{b}Department of Physics, Technion-Israel Institute of Technology, Haifa, 32000, Israel; \textsuperscript{c}Guangdong Provincial Key Laboratory of Materials and Technologies for Energy Conversion, 241 Daxue Rd., Shantou, 515063, Guangdong, China; \textsuperscript{d} Eastern Institute of Technology, Ningbo 315200, China}
}

\maketitle

\begin{abstract}
Light is fundamental to mankind, as it enables us to perceive and understand the world around us. In the modern era, remarkable technological advancements have transformed light into a versatile tool for controlling a wide range of natural processes. Generally, light fields carry energy and momentum (both linear and angular). Due to the transfer of linear momentum from light to matter, the radiation pressure is exerted, whereas, the intrinsic spin angular momentum (SAM) is associated with the polarization states of light. Light fields embedded with optical orbital angular momentum (OAM)- also known as optical vortices or phase singular beams- have truly revolutionized the field of optics and extended our basic understanding of the light-matter interaction process across various scales. Optical vortices- spatially characterized by the presence of twisted phase fronts and a central intensity null- have found a myriad of applications starting from microparticle trapping and manipulation to microscopy, optical communication, and quantum information science, among others. In this review, we revisit some of the fundamental concepts on optical vortices and discuss extensively on how this new dimension of light i.e., the OAM, has been exploited in both linear and nonlinear optical regimes. We briefly discuss the different types of vortex beams, the techniques used to generate them and detect their OAM, and their propagation in free space and various material media. Particularly, we put a special emphasis on the utilization of vortex beams in nonlinear perturbative and non-perturbative regimes to explain different optical phenomena such as the second harmonic generation, sum frequency generation, parametric down-conversion, and high-order harmonic generation. The generation of vortex beams in the ultraviolet to extreme-ultraviolet spectral regimes, encoded with higher OAM values, could potentially extend their application range to areas such as high-capacity data transmission, stimulated emission depletion microscopy, phase-contrast imaging, and particle trapping in optical tweezers, among others.
\end{abstract}


\section{Introduction}
\begin{figure}[h!]
\centering 
\includegraphics[width=1\linewidth]{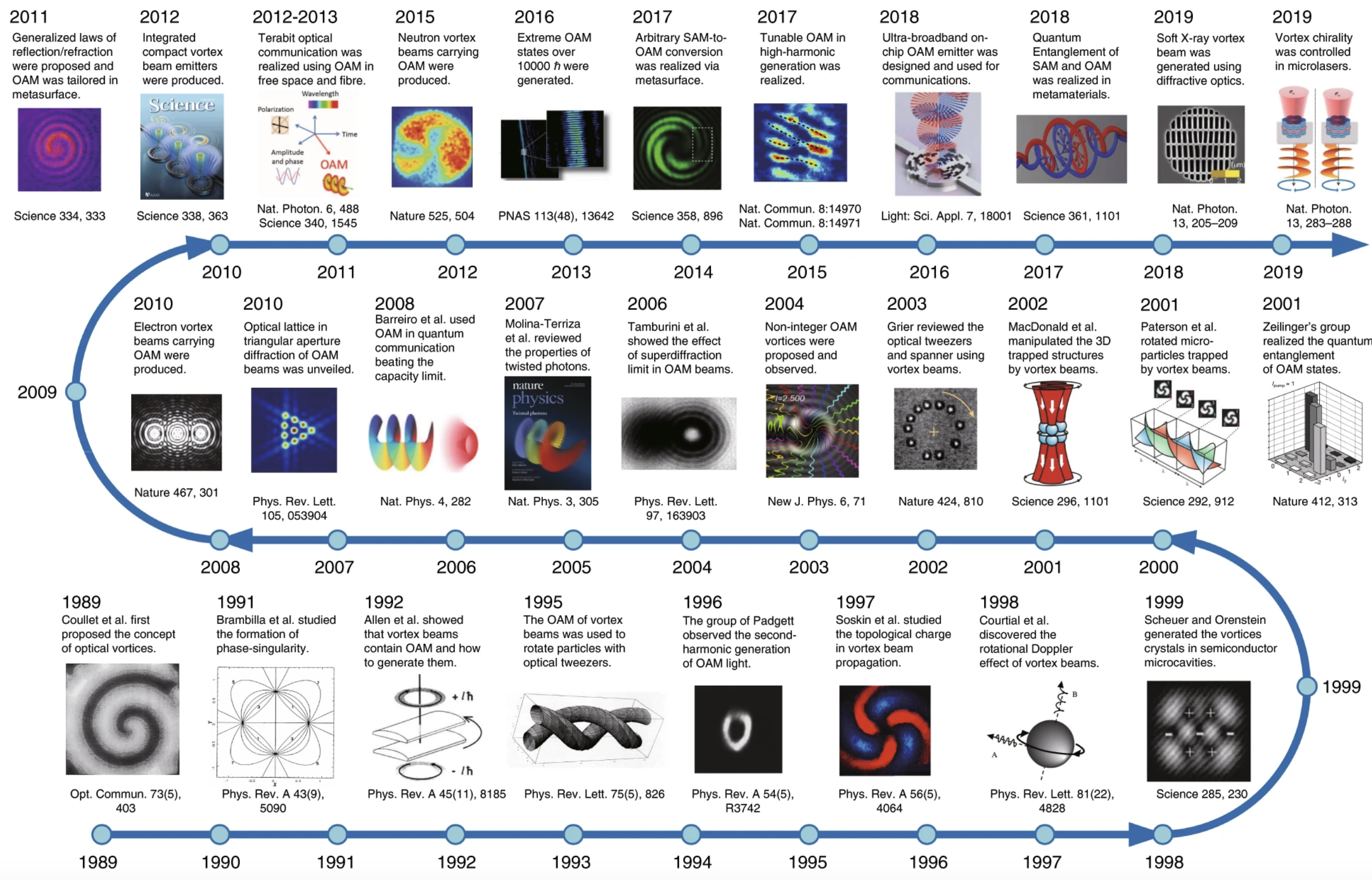}
\caption{Timeline of research developments on vortices (from 1989 to 2019). Reprinted with permission from Ref.~\cite{ShenYijie}}
\label{Fig1}
\end{figure}

\begin{figure}[h!]
\centering 
\includegraphics[width=1\linewidth]{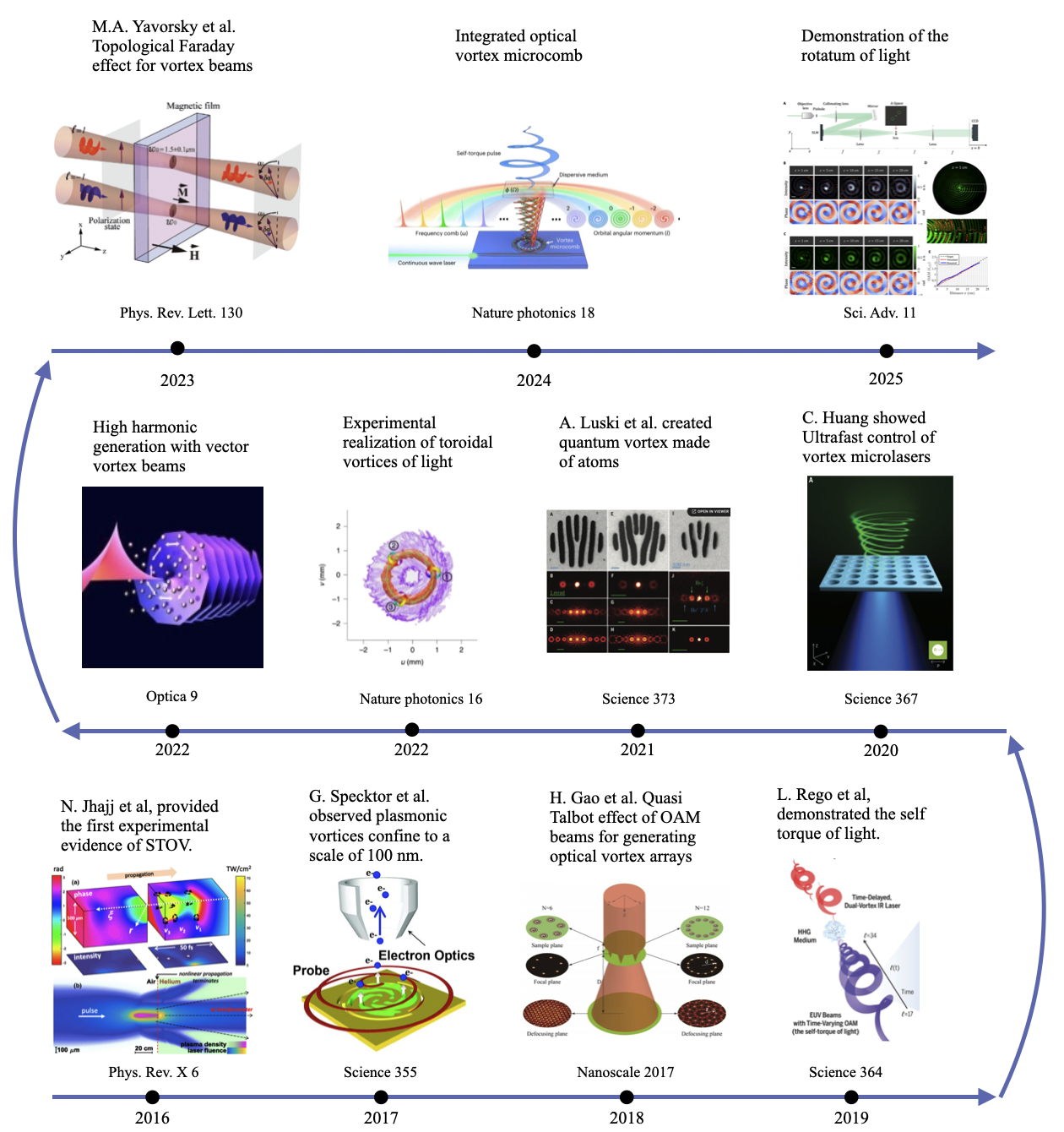}
\caption{Timeline of research developments on vortices (from 2016 to 2025). Reprinted with permission from Refs.~\cite{Jhajj,GSpektor,GaoHui,Regoself,Huang:18,Alon,WanChenhao,delasHerasVVB,Yavorsky,BoChen,AhmedHDorrah}}
\label{Fig0}
\end{figure}

Light, a transverse electromagnetic wave, carries both energy and momentum. A simple manifestation of energy transfer is sunlight warming the Earth, illustrating how electromagnetic radiation deposits energy into matter. Similarly, the linear momentum of light manifests as radiation pressure exerted on a surface, arising from the transfer of momentum from the electromagnetic field to matter during light–matter interactions~\cite{Nichols}. Electromagnetic fields with self-consistent oscillating electric and magnetic components carry spin angular momentum (SAM), which is associated with their circular polarization states (either left-handed or right-handed). In other words, this property reflects how the polarization vector or the electric field vector rotates in time as the beam propagates, forming a basis for a two-dimensional Hilbert space. In 1936, the seminal work of Beth et al. demonstrated that the SAM can be transferred from light to matter via the light-matter interaction and the effect was measured in terms of the mechanical torque arising from the photon spin~\cite{Beth}. Moreover, it was demonstrated that the SAM carries quantized values, such that the SAM of a single photon can only be $\pm\hbar$, with $\hbar$ being the reduced Planck's constant. However, until late 1980s, a little to no attention was given towards shaping the spatial part of the light beam. Drawing inspiration from their hydrodynamic counterpart, Coullet et al. figured out vortex solutions of the Maxwell-Bloch equations and introduced the concept of optical vortices (OVs) in 1989~\cite{COULLET} (see Figs.~\ref{Fig1} and~\ref{Fig0} for a timeline of research developments on vortices). Generally, OVs are known to possess a spiral or helical wavefront~\cite{Allen}. Furthermore, the spiraling of the wavefront of OVs around their propagation axis gives rise to a new quantity- the orbital angular momentum (OAM)- an additional degree of freedom which can be exploited to control and manipulate the light-matter interaction processes across various scales. In other words, light beam's phase structure spirals in the case of OAM beams, which is clearly in contrast with plane waves, where the phase structure is uniform and the wavefront is planar. Vortex beams (alternatively, called phase singular beams or spatially structured light beams or OAM beams) are characterized by the presence of a helical phase (or, azimuthal phase) term exp($il\varphi$) in their electric field distribution, where $l$ is a discrete number called topological charge (TC) (alternatively, a quantized value of OAM of $l\hbar$ per photon) and $\varphi$ is the azimuthal coordinate of the beam in the transverse plane. What really gives the photons their quantized OAM values is the azimuthal component of the Poynting vector, which spirals around the propagation axis of the beam. The topology of the wavefront is not limited to light beams; it can also be found in other types of waves, such as acoustic waves~\cite{Martynyuk-Lototska,Di-Chao}, electron waves~\cite{Uchida,Benjamin,BLIOKH20171,LloydSM}, neutron waves~\cite{Clark,Geerits}, and matter waves of neutral atoms~\cite{Lembessis,Alon}, to cite a few cases. For instance, in Ref.~\cite{Uchida}, the authors experimentally demonstrated the generation of electron vortices i.e., free electron beams with a spiraling wavefront, inside an electron microscope. Due to the undefined (non-deterministic) phase at the beam center, all vortex-beam field quantities—such as amplitude, intensity, and polarization—must vanish there, resulting in the characteristic doughnut-shaped transverse intensity distribution. Moreover, the value of the TC can be either an integer (positive or negative depending on the direction of the wavefront twist) or fractional (characterized by a dark radial opening in the transverse intensity distribution). This singularity was first introduced as a screw dislocation in wave trains, similar to crystal dislocations~\cite{Nye}. Such dislocations in light wavefronts were later found to be present universally. They have been observed not only in specially engineered laser beams, but also in laser-scattering speckle fields (a fine granular light patterns formed by the superposition of many scattered wavefronts from an irregular surface). In this case, the dark speckles correspond to optical vortices that emerge from the interference of multiple plane waves. During the propagation, they may rotate around the axis or interact with the surrounding optical vortices, repelling or attracting each other or even annihilating in a collision, or generating other type of wavefront defects. It was Allen and his collaborators who, for the first time in 1992, connected the previously unlinked dots: the presence of a phase singularity in light beams and their associated OAM~\cite{Allen}. Since then, these light beams have drawn significant attention from the optics and photonics community around the world. It is important to note that, unlike the SAM (which can be either $+\hbar$ or -$\hbar$ depending on the direction of the polarization vector's rotation), the OAM of photons can take any possible value (there is no theoretical bound to OAM), thereby, forming a basis for an infinite dimensional Hilbert space. However, in reality, the OAM value of photons is limited to tens or hundreds due to experimental constraints. Different types of vortex beams have been reported in the literature such as Laguerre-Gaussian (LG) beams~\cite{Padgett,Jie}, Bessel-Gauss (BG) beams~\cite{Xie,Zhao}, Airy beams~\cite{Hui,Junxiao,SUAREZ}, Matheiu beams~\cite{Xiaoxiao,Dongye,Loxpez}, Lorentz-Gauss (LGa) beams~\cite{Guoquan,NI,Guoquan1}, perfect optical vortex (POV) beams~\cite{Vaity,Ostrovsky,Xiujuan}, vortex Hermite-Cosh-Gaussian beams~\cite{Hricha,ELHALBA}, elegant Laguerre-Gaussian beams~\cite{Nasalski,MEI}, and Anomalous vortex beams~\cite{Yuanjie,Alahsab}, among others. Despite carrying an intrinsic OAM, these beams are classified based on the type of mathematical functions used to define their complex field amplitudes, non-diffracting propagation properties, etc. 

For instance, an LG beam is often described by the presence of an associated Laguerre polynomial with real argument and a Gaussian envelope in its field amplitude, whereas, the BG beam is explained by the presence of a Bessel function of first kind and a Gaussian envelope. Optical vortices are being rigorously generated in laboratories using many diffractive and refractive optical elements such as diffraction holograms~\cite{Janicijevic,Atencia,Carpentier}, liquid crystal spatial light modulators (LC-SLM)~\cite{Kaifei,SZATKOWSKI,Anaya}, spiral phase plates (SPP)~\cite{Cano,JANKOWSKI}, astigmatic mode converters~\cite{BEIJERSBERGEN,Ohtomo}, metasurfaces~\cite{Nanfang,Jinna,GaoHui}, photon sieves~\cite{EBRAHIMI,Nuo}, digital micromirror devices~\cite{Yue,MeiZhang}, dielectric wedges~\cite{Izdebskaya,Shvedov}, and q-plates~\cite{MELNIKOVA,Hakobyan}, to name a few. The propagation of vortex beams in different media, such as free space~\cite{ChengKe,Bareza,Bikash1,Xuyan,Zhangrong}, atmospheric turbulence~\cite{Gbur,Wuming,Yanqin,Jianhe}, oceanic turbulence~\cite{Zhenglin,Zonghua,Xinguang,Yuqi}, chiral media~\cite{Bikash2,Yuanfei,Senhua,Xiaojin}, gradient-index media~\cite{Shuai,Qusailah,Ruihuang,Hanghang,Bikash3}, strongly-nonlocal nonlinear media~\cite{Zhiping,Yiminzhou,Shijiechen,Guoquanzhou,Bikash4}, biological tissue~\cite{MeilanLuo,Meiling}, etc., have been thoroughly investigated in the past. Likewise, vortex beams are proven to be advantageous over fundamental Gaussian beams in many different applications. Apart from the phase singularity, a polarization singularity can also exist in light beams. For instance, in the case where both phase and polarization singularities co-exist, such light beams are termed as vector vortex beams i.e., vortex beams with a spatially inhomogeneous polarization profile (either radially or azimuthally)~\cite{Zhenxing,Fuyong,Zeng,Jiaquhu}. The propagation of vector vortices and their interaction with matter are extensively reported in the available literature~\cite{Wencheng,Aita,Lukin,Yizhang,Saripalli,Ishaaya,delasHerasVVB}. Vortex beams have found a myriad of applications, including microparticle trapping and manipulation~\cite{Ngjack,Alexey,Yuehantian,Mingzhou}, free-space optical communication~\cite{Weishao,Leiliu}, optical information encryption~\cite{Qingshuai,YuanYangshen}, chiral molecule resolution~\cite{Brullot}, stimulated emission depletion (STED) microscopy~\cite{BinWang,Otomo}, quantitative phase microscopy~\cite{Chaodu}, coherent diffractive imaging (CDI)~\cite{Mingli}, and quantum entanglement~\cite{Fickler,Fengyixu}, among others. All the above-mentioned discussions were concentrated around spatial or longitudinal vortex beams, where the OAM density (cross product of the position vector and the linear momentum density) is parallel to the propagation direction of vortex beams. However, the OAM density can also be directed transversal to the beam propagation direction. Such optical beams carry transverse OAM and are termed as spatiotemporal optical vortex (STOV) beams. In case of STOV beams, the spatial and temporal coordinates are coupled (spatiotemporal couplings), which is generally reflected in the azimuthal phase term. The generation, propagation, and interaction of STOV beams have been well-documented in the literature~\cite{DROR2011526,Jhajj,XinLiu,Hancock,Hyde,Junyiye,Hancock1,MartínHernández}. Furthermore, it was demonstrated that OAM modes of light can be shrunk down to sub-wavelength dimensions in the form of plasmonic vortices- on the scale of 100 nanometers- with nanometric spatial and sub-femtosecond temporal resolutions ~\cite{GSpektor}. The generation and propagation of plasmonic vortices have been the subject of considerable research interest in recent years~\cite{Tsai,Shutova_2022,YuWang2016,Hanjiaguang,Bai_2022}. In 2019, a new property of light beams- the self-torque- was introduced via extreme nonlinear light-matter interaction processes, where the OAM of the generated extreme ultraviolet radiation changes dynamically in femtosecond to sub-femtosecond temporal scales~\cite{Regoself}. The past decade has also seen the emergence of several innovative advances in OAM beams. These include the experimental demonstration of vortex microlasers with terahertz switching speeds~\cite{CanHuang}, the generation of complex toroidal vortices~\cite{WanChenhao,RenWang}, the first experimental realization of a topological Faraday effect—where OAM induces a measurable polarization rotation in light~\cite{Yavorsky}, and the demonstration of an integrated optical vortex microcomb, showcasing the potential for photonic quantum technology~\cite{BoChen}. More recently, the experimental generation of a new class of vortex beams called "optical rotatum" was demonstrated, in which the OAM experiences a quadratic chirp along the optical path, therefore, giving rise to a non-zero value of the second-order derivative of the OAM~\cite{AhmedHDorrah}.

In this review, we revisit several well-established wave equations—the homogeneous wave equation, the Helmholtz equation, and the paraxial Helmholtz equation, and analyze their solutions in different coordinate systems. We then provide an overview of (1) the main classes of vortex beams, (2) their generation and detection techniques, (3) their applications across diverse domains, (4) their propagation in free space and various material media, and (5) their role in exploring a wide range of nonlinear optical phenomena, spanning both perturbative and non-perturbative regimes. Particular emphasis is placed on nonlinear optical processes driven by vortex beams, including second-harmonic generation (SHG), sum-frequency generation (SFG), parametric down-conversion (PDC), and high-order harmonic generation (HHG), in both gas and condensed-matter targets.

\section{The Helmholtz equation and its vortex and non-vortex solutions}
\label{solutions}
In vacuum, electromagnetic waves are typically described by the homogeneous wave equation, which is a second-order (in both space and time) linear partial differential equation. Specifically, the wave equation can be written as~\cite{Maxwell}:
\begin{eqnarray}
    \nabla^{2}\bm{E}(\bm{r},t)-\frac{1}{c^{2}}\frac{\partial^2}{\partial t^2}\bm{E}(\bm{r},t)=0,
    \label{eqn1}
\end{eqnarray}
where $c=\frac{1}{\sqrt{\mu_{0}\epsilon_{0}}}$ is the speed of light in vacuum (=$3\times 10^{8}$m/sec) with $\mu_{0}$, and $\epsilon_{0}$ representing the vacuum permeability, and vacuum permittivity, respectively, and $\bm{E}(\bm{r},t)$ denotes the electric field component of the electromagnetic wave. Equation~(\ref{eqn1}) can be derived by solving two of the Maxwell's equations for electromagnetism, i.e., $\bm{\nabla}\times\bm{E}(\bm{r},t)=-\frac{\partial \bm{B}(\bm{r},t)}{\partial t}$ and $\bm{\nabla}\times\bm{B}(\bm{r},t)=\mu_{0} \bm{J}(\bm{r},t)+\mu_{0}\epsilon_{0}\frac{\partial \bm{E}(\bm{r},t)}{\partial t}$ in vacuum, where there exists no charge ($\bm{\rho}(\bm{r},t)=0$) and current ($\bm{J}(\bm{r},t)=0$). It is interesting to note that we consider the electric field component in Eq.~(\ref{eqn1}) only, whereas, the magnetic field component is not taken into account. This is due to the fact that both electric and magnetic field components, separately, satisfy the wave equation and a time-varying electric (magnetic) field produces a time-varying magnetic (electric) field. Since Eq.~(\ref{eqn1}) is linear in the solution $\bm{E}(\bm{r},t)$, a linear superposition of any two solutions is also a solution to the wave equation.

For monochromatic fields (with high temporal coherence), it is often possible to separate the spatial part of the field from its temporal counterpart as follows:
\begin{eqnarray}
   \bm{E}(\bm{r},t)= \bm{u}(\bm{r}) \mathrm{e}^{i\omega t},
   \label{eqn2}
\end{eqnarray}
where $\omega=\frac{2\pi c}{\lambda}$ and $\lambda$ are the angular frequency and the wavelength of the monochromatic wave, respectively. The spatial part of Eq.~(\ref{eqn2}) i.e., $\bm{u}(\bm{r})$, is a solution to the well-known Helmholtz equation,
\begin{eqnarray}
    \left(\nabla^{2}+k^{2}\right)\bm{u}(\bm{r})=0,
    \label{eqn3}
\end{eqnarray}
where $k=\frac{\omega}{c}$ is the wavenumber. If we consider the $z$-direction to be the propagation direction of the wave, and apply the paraxial approximation i.e., $\left|\frac{\partial^2 u_{0}(\bm{r})}{\partial z^2}\right|<<\left|k\frac{\partial u_{0}(\bm{r})}{\partial z}\right|$ and $\left|\frac{\partial^2 u_{0}(\bm{r})}{\partial z^2}\right|<<\left|\nabla_{T}^{2}u_{0}(\bm{r})\right|$, we get:
\begin{eqnarray}
    \nabla_{T}^{2}u_{0}(\bm{r})+2ik\frac{\partial u_{0}(\bm{r})}{\partial z}=0,
    \label{eqn4}
\end{eqnarray}
where we have used $\bm{u}(\bm{r})=u_{0}(\bm{r}) \mathrm{e}^{ikz} \hat{z}$ as a solution to the Helmholtz equation in (3), with $u_{0}(\bm{r})$, and $\hat{z}$ representing the slowly varying envelop in $z$ (i.e., it changes slowly over a wavelength, $\lambda=2\pi/k$), and polarization direction of the wave, respectively. Additionally, $\nabla_{T}^{2}$ denotes the transverse Laplacian operator, which describes the diffracting behavior of the wave. It is also important to highlight that we drop the vectorial nature of $\bm{u}(\bm{r})$ while deriving Eq.~(\ref{eqn4}). This is because the polarization of the electromagnetic field becomes irrelevant while propagating in linear isotropic media. In cartesian coordinates, $\nabla_{T}^{2}=\frac{\partial^2}{\partial x^2}+\frac{\partial^2}{\partial y^2}$, whereas in cylindrical and spherical polar coordinates we can write $\nabla_{T}^{2}=\frac{1}{\rho}\frac{\partial}{\partial \rho}+\frac{\partial^2}{\partial \rho^2}+\frac{1}{\rho^2}\frac{\partial^2}{\partial \varphi^2}$ and $\nabla_{T}^{2}=\frac{1}{r^2}\left[\frac{1}{\sin \theta }\frac{\partial}{\partial \theta}\left(\sin \theta \frac{\partial}{\partial \theta}\right)+\frac{1}{\sin^{2}\theta}\frac{\partial^2}{\partial \varphi^2}\right]$, respectively. It is important to note that, from the perspective of geometric optics, the paraxial approximation holds true when the light ray travels closer to the optical axis i.e., when the divergence is extremely small. Similarly, from the wave optics view point, the paraxial approximation essentially corresponds to the situation when the light beam size is much larger than the wavelength of the beam.

The simplest solution to the Helmholtz equation, which one can think of, is a plane wave whose wavefront is basically characterized by $\mathrm{e}^{ikz}=C$, where $C$ is a constant, i.e., a solution that corresponds to planes transverse to the propagation direction $z$. The Poynting vector (mathematically, a cross product of the electric and magnetic field components), which qualitatively describes the direction of energy or momentum flow, points along the propagation direction in this case. Furthermore, the intensity profile of a plane wave is uniform at different propagation distances. However, realizing an idealized plane wave in practice is impossible due to its infinite transverse spatial extent. Therefore, it is necessary to figure out the practically realizable solutions of both the Helmholtz and the paraxial Helmholtz equations. We begin with the non-vortex solutions of the paraxial Helmholtz equation and subsequently address the vortex solutions. In particular, two practically realizable and fundamental solutions are the Gaussian and HG beams.

\subsection{Non-vortex solutions}
In this sub-section, we discuss some of the non-vortex solutions of the paraxial Helmholtz equation in different coordinate systems.

\subsubsection{Gaussian Beams}
\begin{figure}[h!]
\centering 
\includegraphics[width=1\linewidth]{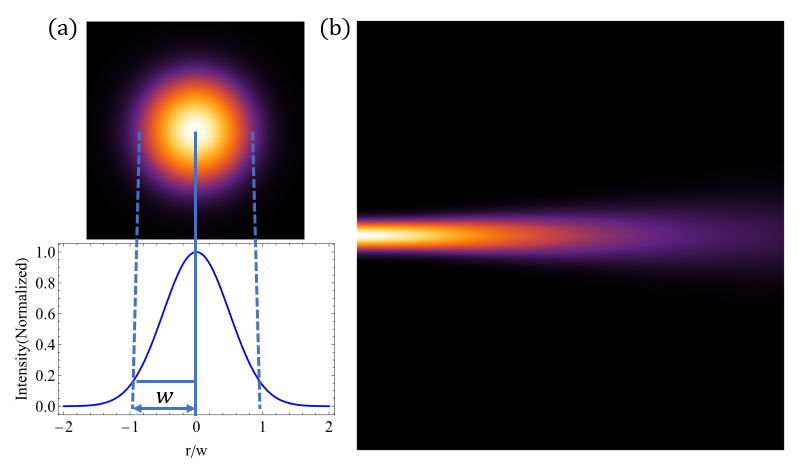}
\caption{Transverse (top) and line intensity (bottom) profiles of the fundamental Gaussian beam in (a), free-space propagation of the fundamental Gaussian beam in (b). In the line plot, we define $w$ as the beam width or beam radius i.e., the radial distance at which the intensity of the beam becomes $\frac{1}{e^2}$ of the peak intensity. In (b), the horizontal axis denotes the propagation distance and the vertical axis denotes one of the transverse coordinates (either $x$ or $y$).}
\label{Fig2}
\end{figure}

The complex field amplitude of Gaussian beams at a propagation distance, $z$, are typically expressed as:
\begin{eqnarray}
    u_{0,G}(\rho,z)=A_{0}\frac{w_{0}}{w(z)} \mathrm{e}^{-\frac{\rho^2}{w(z)^2}} \mathrm{e}^{\frac{ik\rho^2}{2R(z)}} \mathrm{e}^{i\Phi_{G,G}},
    \label{eqn5}
\end{eqnarray}
where $w_{0}$, $w(z)=w_{0}\sqrt{1+\left(\frac{z}{z_{R}}\right)^2}$, $R(z)=z\left[1+\left(\frac{z_{R}}{z}\right)^2\right]$, $\Phi_{G,G}=-\arctan\left(\frac{z}{z_{R}}\right)$, and $z_{R}=\frac{\pi w_{0}^2}{\lambda}=\frac{1}{2}kw_{0}^2$ represent the Gaussian beam waist size (i.e., the beam size at $z=0$), the beam size at a finite propagation distance $z$, the radius of curvature (inverse of the curvature) of the wavefront, the Gouy phase (that arises due to the spatial confinement of Gaussian beams), and the Rayleigh range (the distance at which the beam radius becomes $\sqrt{2}$ times the beam radius at the waist plane) of the beam, respectively. Furthermore, $A_{0}$ is a constant field amplitude, and $\rho(x,y)$ denotes the beam coordinate at a distance $z$. This Gaussian beam, shown in Fig.~\ref{Fig2}(a), is the intended output of most of the lasers used in laboratories. From Eq.~(\ref{eqn5}), it can be noted that at $z=0$, $R(z)=\infty$. Therefore, the wavefront is plane at $z=0$. In general, Gaussian beams do not have plane wavefronts. However, the radius of curvature of their wavefronts is much larger than the beam waist size. Therefore, for many practical applications, it is reasonable to approximate Gaussian beams as plane waves. Similarly, when $z\gg z_{R}$, both $w(z)$ and $R(z)$ scales linearly with $z$. It is interesting to highlight that the presence of the Gaussian envelope (the exponentially decaying term in Eq.~(\ref{eqn5})) is what constraints Gaussian beams to spatially \textit{diverge} at infinity and the term $\left(\frac{w_{0}}{w(z)}\right)$ stands for the energy (or, power) conservation during the course of beam propagation. In case of Gaussian beams, the maximum power is concentrated at the beam's center and goes away exponentially in the radial direction, thereby, giving the characteristic bright-lobed pattern at the center of the beam (see Fig.~\ref{Fig2}(a)). 

In Fig.~\ref{Fig2}(b), we show the longitudinal intensity distribution of a Gaussian beam propagating in the free-space by using Eq.~(\ref{eqn5}). It can be seen that the Gaussian beam maintains its beam size and peak intensity over a short propagation distance-up to its Rayleigh range-referred to as the near-field. However, as the beam propagates into the far field, i.e., at $z\gg z_{R}$, its size increases significantly while the peak intensity decreases. Therefore, we can safely consider that Gaussian beams are the diffracting solutions of the Helmholtz equation in the paraxial limit, and where the far-field divergence of Gaussian beams can be expressed as, $\beta_{\mathrm{div}}=\frac{\lambda}{\pi w_{0}}$. Despite the changes in the beam size and intensity in the far-field, the beam profile still remains Gaussian. This is due to the fact that the Fourier transform of a Gaussian function always remains Gaussian.

\subsubsection{Hermite-Gaussian Beams}
\begin{figure}[h!]
\centering 
\includegraphics[width=1\linewidth]{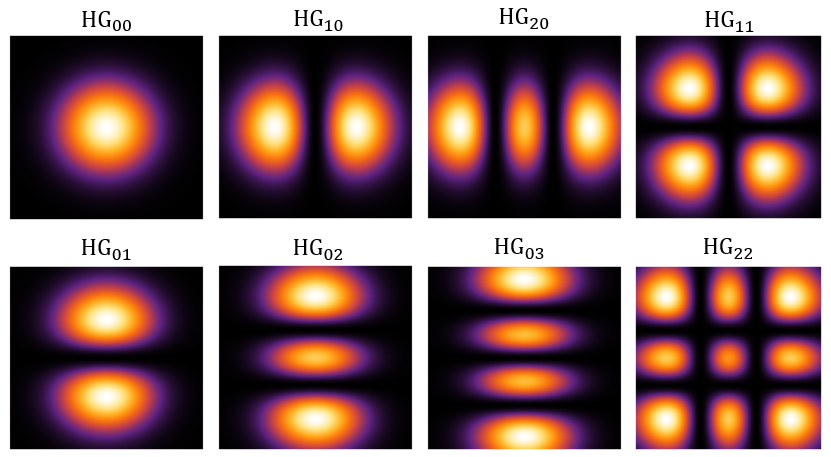}
\caption{Transverse intensity distributions of different HG modes at the source plane.}
\label{Fig3}
\end{figure}

Hermite-Gaussian (HG) beams are solutions to Eq.~(\ref{eqn4}) in cartesian coordinates. Their complex field amplitude at a propagation distance, $z$, is given by:
\begin{eqnarray}
    u_{0,HG}(x,y,z)=u_{mn} \frac{w_{0}}{w(z)} H_{m}\left(\frac{\sqrt{2}x}{w(z)}\right) H_{n}\left(\frac{\sqrt{2}y}{w(z)}\right) \mathrm{e}^{-\frac{x^2+y^2}{w(z)^2}} \mathrm{e}^{\frac{ik(x^2+y^2)}{2R(z)}} \mathrm{e}^{i\Phi_{G, HG}},
    \label{eqn6}
\end{eqnarray}
where $H_{m}(\ldots)$, and $H_{n}(\ldots)$ are the Hermite polynomials of order $m$, and $n$, respectively and $\Phi_{G,HG}=-(m+n+1)\arctan\left(\frac{z}{z_{R}}\right)$ denotes the Gouy phase for HG beams. Basically, $m$, and $n$ are some positive integers linked to the number of nodes (or, intensity nulls) along the horizontal ($x$-axis), and vertical ($y$-axis) directions of the HG beams, respectively, and directly determine the shape of the beam profile. $m=n=0$ corresponds to the fundamental TEM$_{00}$ mode (or, Gaussian mode with the highest beam quality), and $m,n>1$ corresponds to higher-order modes. It is also interesting to note that the Gouy phase of HG beams is ($m+n+1$) times the Gouy phase of Gaussian beams. Generally, HG beams show a lobed intensity pattern in the transverse dimensions, as can be seen from Fig.~\ref{Fig3}. Similar to the fundamental Gaussian mode, HG modes constitute stable propagation modes; that is, they maintain their characteristic intensity profiles during propagation, though their overall beam size scales with distance. Furthermore, HG modes can directly be excited from a laser cavity (i.e. resonator modes). Interestingly, HG modes form an orthonormal basis, i.e. any coherent paraxial beam can be expressed as a superposition of HG modes with different weights. Although we see intensity nulls along the horizontal and vertical axes, HG modes do not carry OAM (we will show this later). 

\begin{figure}[h!]
\centering 
\includegraphics[width=1\linewidth]{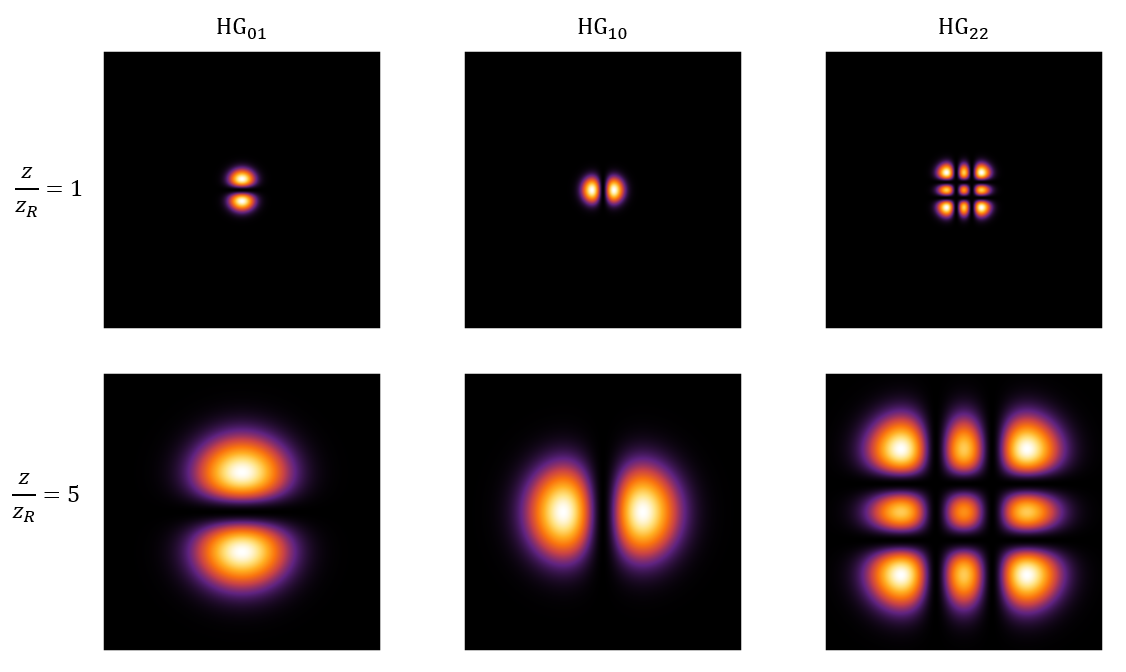}
\caption{Transverse intensity distributions of different HG modes propagating in free space.}
\label{Fig4}
\end{figure}
In Fig.~\ref{Fig4}, we plot the transverse intensity distributions of different HG modes, such as HG$_{01}$, HG$_{10}$, and HG$_{22}$ in free-space, at different propagation distances using Eq.~(\ref{eqn6}). In every case, the beam preserves its characteristic shape, even though its overall size changes. Therefore, like fundamental Gaussian beams, HG beams are also diffracting solutions to the paraxial Helmholtz equation. Furthermore, the divergence of HG modes in $x$, and $y$ directions are directly connected to the mode orders $m$, and $n$, respectively.

\subsection{Vortex solutions}

\begin{figure}[h!]
\centering 
\includegraphics[width=1\linewidth]{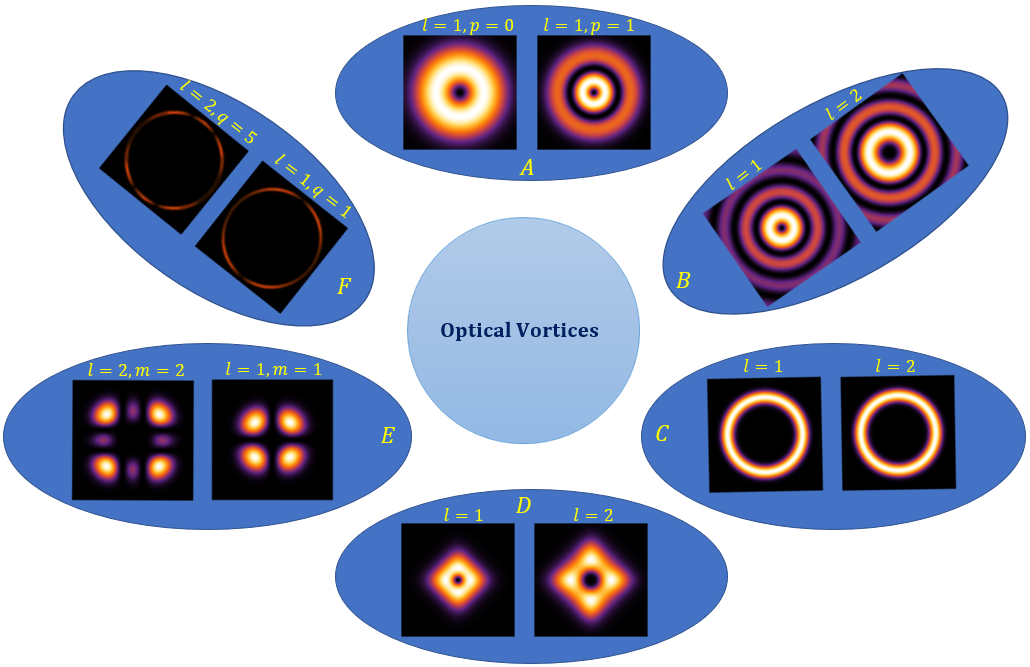}
\caption{Different types of vortex beams and their specific mode parameters. (A) Laguerre–Gaussian (LG) vortex beams are defined by their topological charge (TC) $l$ and radial index $p$. (B) Bessel–Gauss (BG) and (C) perfect optical vortex (POV) beams are primarily characterized by their TC $l$. (D) Lorentz–Gauss (LGa) beams are also defined by TC 
$l$. (E) Vortex Hermite–Cosh–Gaussian beams include TC 
$l$ and beam order $m$, and (F) perfect helical Mathieu beams include TC  $l$ and elliptical parameter $q$.}
\label{Fig5}
\end{figure}

We now turn our attention to the vortex solutions of the paraxial Helmholtz equation (Eq.~(\ref{eqn4})). Figure~\ref{Fig5} presents several representative families of vortex beams, each distinguished by specific mode parameters. LG beams include both a TC $l$ and a radial index $p$, while BG, POV, and Lorentz–Gauss beams are primarily defined by their TC, $l$. More complex structures, such as vortex Hermite–Cosh–Gaussian beams and perfect helical Mathieu beams, incorporate additional parameters (beam order $m$ and elliptical parameter $q$, respectively), reflecting the diversity of vortex beam profiles. We will describe the properties of several of these beam families in the following subsections.

\subsubsection{Laguerre-Gaussian Beams}
A complete set of solutions to Eq.~(\ref{eqn4}) are Laguerre-Gaussian (LG) beams, which can be written as~\cite{Alison}:
\begin{eqnarray}
    u_{l,p}(\rho,\varphi,z)=A_{0} \frac{w_{0}}{w(z)} \mathrm{e}^{-\frac{\rho^2}{w(z)^2}} L_{p}^{|l|}\left(\frac{2\rho^2}{w(z)^2}\right) \left(\frac{\sqrt{2}\rho}{w(z)}\right)^{|l|} \mathrm{e}^{il\varphi} \mathrm{e}^{\frac{ik\rho^2}{2R(z)}} \mathrm{e}^{i\Phi_{G,LG}},
    \label{eqn7}
\end{eqnarray}
where $L_{p}^{|l|}(\ldots)$ is the associated Laguerre polynomial with a real argument and $w(z)$, $R(z)$ are defined in the same way as for the Gaussian beams case. Furthermore, the Gouy phase for LG beams is defined as, $\Phi_{G,LG}=-(2p+|l|+1)\arctan\left(\frac{z}{z_{R}}\right)$. A closer look into the electric field distribution of LG beams in Eq.~(\ref{eqn7}), reveals that in addition to the parabolic phase (or, the radial phase) and the Gouy phase, LG beams also possess a phase that is dependent on the azimuthal angle i.e., the so-called helical phase $\exp(il\varphi)$). Light beams with helical or twisted phase are termed as \textit{vortex beams}. In case of vortex beams, the phase is indeterminate or undefined at the beam's center. In other words, a phase singularity exists at the center of the vortex beam, requiring the beam amplitude-and consequently its intensity-to vanish, thus producing the characteristic doughnut-shaped intensity distribution of vortex beams. In simple terms, in case of vortex beams, the energy is distributed over a (more) ring(s) with zero intensity at the center, which is clearly in contrast with Gaussian beams, where the energy is mostly distributed at the beam's center. Basically, LG beams are characterized by two discrete indices: (1) the azimuthal index or topological charge (TC) ($l$) and the radial index ($p$). The TC associated with the phase singularity, $\phi$, is defined as:
\begin{eqnarray}
    l=\frac{1}{2\pi}\times \oint_{C} \bm{\nabla} \phi(\bm{r}) \mathrm{d}r,
    \label{eqn8}
\end{eqnarray}
where $C$ is a tiny closed loop surrounding the singularity. In other words, the gradient of the phase circulation provides us the information about the TC carried by vortex beams. The TC of vortex beams can also be defined in many other ways, e.g: (1) The number of $2\pi$ phase jumps around the singularity along the azimuthal coordinate in a transverse plane, and (2) The number of helices per unit wavelength. The TC (proportional to the OAM of photons) only takes quantized values and it could be either an integer (resulting from the continuous phase variation along a closed loop) or fractional (manifested as a radial dark opening in the transverse intensity distribution during the beam propagation). Furthermore, positive and negative values of the TC indicate that the helical twisting of a vortex beam’s wavefronts occurs in opposite directions.
\begin{figure}[h!]
\centering 
\includegraphics[width=1\linewidth]{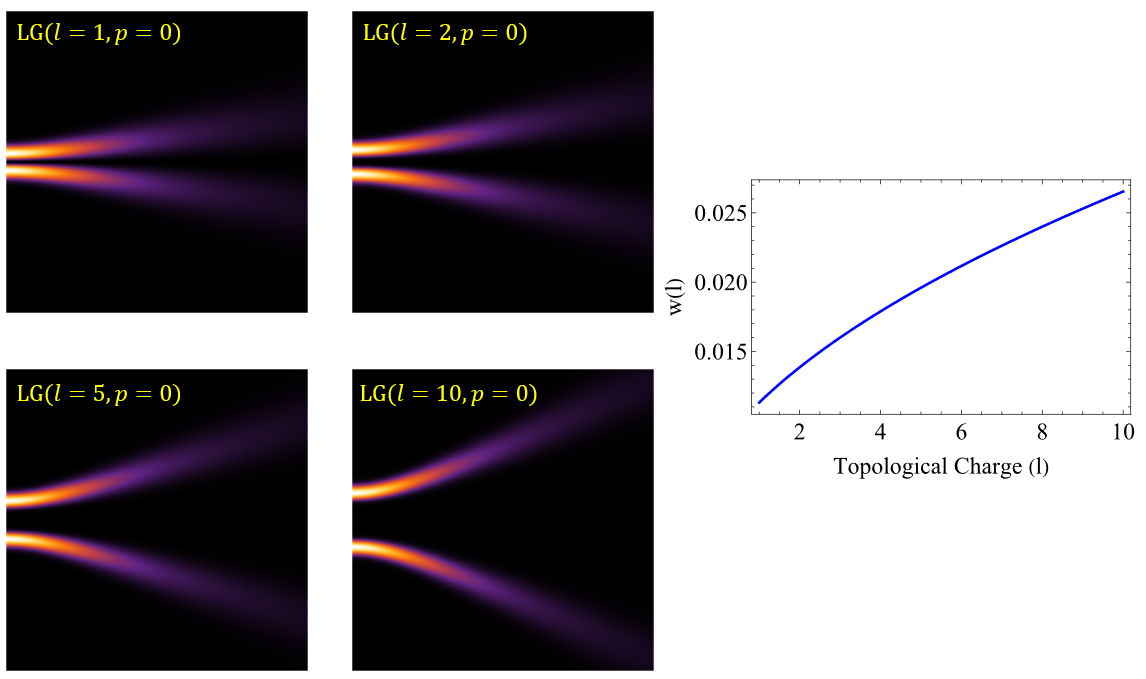}
\caption{(First and second columns) Longitudinal intensity distributions of LG modes propagating in free space, (third column) TC-dependent beam width of LG beams for $p=0$.}
\label{Fig6}
\end{figure}

An interesting property of LG vortex beams is the expansion of their central intensity null and the corresponding increase in the radius of maximum intensity (calculated from $\frac{\partial |u_{l,p}(\rho,\varphi,z)|^2}{\partial \rho}=0$) with the TC, which restricts the utilization of LG beams in many application scenarios, such as in optical trapping, fiber-based optical communication, and microscopy, to only name a few. On the other hand, the radial index, $p$, is quantitatively linked to the number of off-axis radial nodes (or, off-axis rings) and qualitatively to the hyperbolic momentum~\cite{Plick}. We show the transverse intensity distributions of LG beams for different values of $l$ and $p$ in Fig.~\ref{Fig5}(A). While the transverse phase profiles are not presented here, they can be straightforwardly derived from the argument of Eq.~(\ref{eqn7}), where the azimuthal coordinate is defined as $\varphi=\arctan\left(\frac{y}{x}\right)$. It is also important to highlight that in case of LG beams, the beam size, $w$, becomes dependent on both $l$, and $p$ through the relation $w(l,p)=w_{0}\sqrt{2p+|l|+1}$ (calculated from the definition of the second-moment width and shown in Fig.~\ref{Fig6} (third column)). This is completely different from the Gaussian beam case. Since the intensity is distributed over rings in vortex beams, the far-field divergence in such cases is defined as the angular spreading of the radius of the maximum intensity at the far-field. This is clearly in contrast with the case of Gaussian beams, where the divergence is defined as the far-field spreading of the maximum intensity lobe. Like HG beams, LG beams also form an orthonormal basis, i.e., any coherent paraxial beam can be decomposed into LG beams of different weights. In a simple way, it can be said that two LG beams with the same radial index but different TC values are orthogonal to each other. Likewise, the same definition goes for different radial indices. This is precisely why LG beams are employed as information carriers in optical communication, leading to significant enhancements in channel capacity and spectral efficiency. We will discuss more about this topic later in the ``Applications of vortex beams'' section (Section~\ref{applications}). Similar to Gaussian and HG beams, LG beams are also stable propagation modes and suffer from diffraction in the far-field. Experimentally, LG beams can be generated in the laboratory from fundamental Gaussian beams using various optical elements such as phase-only spatial light modulators~\cite{RuijianLi}, spiral phase plates~\cite{Sueda}, metasurfaces~\cite{Meifeng}, and digital micromirror devices~\cite{Yuxuanren}, or directly from lasers through intra-cavity modulation~\cite{Dunzhaowei}. In Fig.~\ref{Fig6}, we show longitudinal intensity distributions of different LG beams propagating in free-space (by using Eq.~(\ref{eqn7})). For simplicity, we set the radial index to $p=0$. From all the plots, it can be seen that LG beams typically preserve their beam size and intensity only in the near-field. However, in the far field, the beam size increases as a result of diffraction.

\subsubsection{Bessel and truncated Bessel Beams}
It is well-known that the Helmholtz equation governs the diffractive phenomena in every area of physics. In all the 
previous cases, we discuss about the diffracting solutions of the Helmholtz equation in the paraxial limit. Therefore, it is natural to ask the following question: Do non-diffracting (or diffraction-free) solutions to the Helmholtz equation exist? The answer to this question is indeed yes. Plane waves are the diffraction-free solutions to the Helmholtz equation as described earlier. However, realizing them in the laboratory is practically unfeasible. Therefore, plane waves are merely idealized mathematical constructs, and any light beam can be represented as a superposition of such waves. In 1987, Durnin et al., for the very first time, demonstrated that there exists exact non-singular solutions to the Helmholtz equation, which are diffraction-free, and possess narrow beam radii~\cite{Durnin}. More specifically, it was shown that an optical beam whose transverse intensity profile follows a zeroth-order Bessel function is immune to diffraction effects. They call those solutions Bessel beams (more specifically, zeroth-order Bessel beams). The complex field amplitude representing the zeroth-order Bessel beam can be expressed as~\cite{Durnin}:
\begin{eqnarray}
    u_{0,B}(x,y,z)=A_{0}J_{0}(k_{r}\rho) \mathrm{e}^{ik_{z}z},
    \label{eqn9}
\end{eqnarray}
with $k_{r}$, $k_{z}$, and $k=\sqrt{k_{r}^2+k_{z}^2}$ representing the magnitude of radial, longitudinal, and total wave vectors, respectively, $J_{0}(\ldots)$ denotes the zeroth-order Bessel function of the first kind and $\rho=\sqrt{x^2+y^2}$. Equation~(\ref{eqn9}) shows that for $k_{r}=0$, it will be a simple plane wave solution and for $0<k_{r}\leq \frac{\omega}{c}$, it represents a non-diffracting Bessel beam. If one calculates the intensity of the Bessel beam utilizing Eq.~(\ref{eqn9}), it can be found that the resulting intensity goes as $|J_{0}(k_{r}r)|^2$. Therefore, for a fixed value of $k_{r}$, the beam intensity remains constant in any plane perpendicular to the propagation axis (i.e., the $z$-axis in our description), which accounts for the non-diffracting nature of the Bessel beam. Transversely, one sees a bright lobe at the center surrounded by an infinite number of off-axis rings. It is important to clarify that when we say non-diffracting propagation of the zeroth-order Bessel beam, it actually refers to the non-diffracting spreading of the central bright lobe. Furthermore, the size of the central lobe is solely controlled by the radial wave vector. From a Fourier decomposition perspective, a Bessel beam can be thought of as a superposition of plane waves with their wave vectors lying on a cone whose axis of symmetry is the beam propagation axis i.e., the $z$-axis. Since an infinite number of rings are present in the transverse intensity profile of a Bessel beam, it is practically impossible to realize such a beam in the laboratory, as it would require an infinite amount of energy distributed over all rings (quantitatively, Bessel functions are not square-integrable). Therefore, a Gaussian envelope is typically included in the field amplitude of a Bessel beam to form a truncated Bessel or a Bessel-Gauss (BG) beam~\cite{GORI}. In this way, such beams can be readily generated in the laboratory from fundamental Gaussian beams using an axicon. An axicon is a conical lens defined by its cone angle, and the refractive index of the material used to design it. It is well-known that a converging lens is used to focus a light source into a small point (i.e., a point focus). However, an axicon generates a focal line along the optical axis as a result of interference. Within the region of beam overlap, it can replicate the properties of a Bessel beam. The distance until which the BG beam propagates without spreading is termed as the Bessel zone and is given by $z_{max}\approx \frac{w_{0}}{\theta}$ with $\theta\approx(n_{\mathrm{mat}}-1)\alpha$ assumed to be tiny, where $\theta$, $\alpha$, $n_{\mathrm{mat}}$ denote the half-cone angle, the half-apex angle, and the refractive index of the material of the axicon, respectively, and $w_{0}$ is the radius of the Gaussian beam. After the Bessel zone, the BG beam starts to spread (diffract) slowly, therefore, it is also called as quasi-nondiffracting Bessel beam. In general, BG beams can be regarded as a superposition of tilted Gaussian beams whose propagation axes lie on the surface of a cone with a half-angle $\theta$. The complex field amplitude of BG beams can be expressed as~\cite{GORI}:
\begin{eqnarray}
 u_{0,BG}(x,y,z)=A_{0}J_{0}(k_{r}\rho) \mathrm{e}^{-\frac{\rho^2}{w_{0}^2}} \mathrm{e}^{ik_{z}z}.
 \label{eqn10}
\end{eqnarray}
Similar to the zeroth-order solution, there also exist higher-order singular ($l>0$) solutions to the Helmholtz equation that carry OAM, with their wavefronts twisted around the propagation axis. The complex field amplitude of a BG beam carrying OAM can be expressed as~\cite{Vaity,ARLT1}:
\begin{eqnarray}
    u_{l,BG}(x,y,z)=A_{0}J_{l}(k_{r}\rho) \mathrm{e}^{-\frac{\rho^2}{w_{0}^2}} \mathrm{e}^{ik_{z}z} \mathrm{e}^{il\varphi},
    \label{eqn11}
\end{eqnarray}
where $J_{l}(\ldots)$ is the Bessel function of the first kind of order $l$. Since these higher-order BG beams carry OAM, their transverse intensity distribution looks like a central null at the center surrounded by a finite number of off-axis rings, as shown in Fig.~\ref{Fig5}(B).
\begin{figure}[h!]
\centering 
\includegraphics[width=1\linewidth]{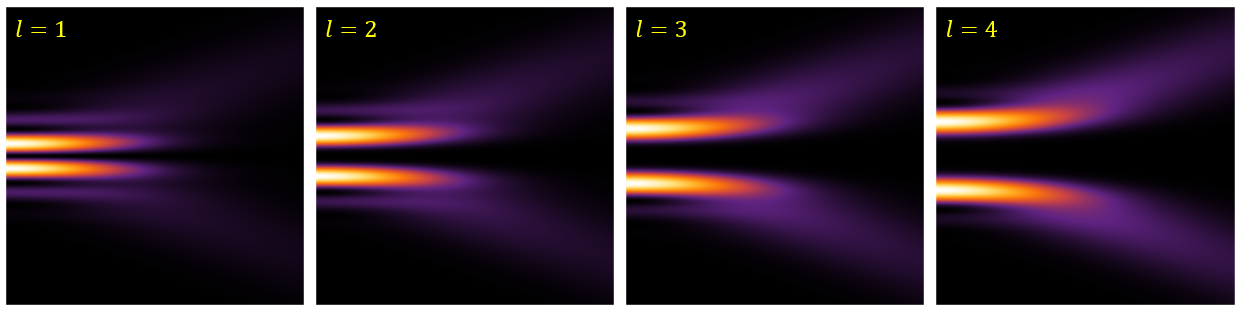}
\caption{Longitudinal intensity distributions illustrating the propagation behavior of BG beams in free space.}
\label{Fig7}
\end{figure}

Like LG vortex beams, the size of the central intensity null increases monotonically with the TC, $l$. In the language of singular optics, both LG and BG beams are termed as conventional vortex beams. Apart from the axicon~\cite{Vanitha}, BG beams can also be produced by using spatial light modulators~\cite{Chattrapiban}, coherent beam combining techniques~\cite{Taoyu}, vortex phase plates~\cite{Stoyanov}, to name a few. Owing to their quasi-nondiffracting nature, BG beams have found numerous applications, including particle trapping~\cite{Kotlyar}, optical communication~\cite{Mingjiancheng}, among others. In Fig.~\ref{Fig7}, we show the longitudinal intensity plots of BG vortex beams, carrying different TC values and propagating in free-space. It can be clearly seen that BG vortex beams maintain their non-diffracting nature only up to the Bessel zone. However, beyond the Bessel zone, they start to diffract slowly and in the far-field, they degrade into a single-ring annular intensity distribution.

\subsubsection{Perfect Optical Vortex beams}
As mentioned earlier, the size of the central dark core and the radius of the maximum intensity of conventional vortices such as LG and BG beams increase monotonically with the TC. Consequently, this intrinsic property imposes significant limitations on the use of them in various practical applications, including optical trapping and manipulation of microscopic particles, multiplexing of multiple OAM modes into optical fibers with fixed annular index profiles, and plasmonic structured illumination microscopy, among others. For instance, in optical trapping applications, it is often desirable to employ vortex beams with a small dark core and high TC to achieve an efficient confinement of microparticles—conditions that are difficult to realize with conventional LG and BG beams. Likewise, it is well known that OAM modes are mutually orthogonal, i.e., they are completely independent and do not interfere or couple with each other during propagation in free space or optical fibers. This orthogonality enables the use of distinct OAM modes as independent data channels—co-propagating at the same wavelength—to enhance data capacity and spectral efficiency in optical communication systems. However, coupling multiple OAM modes (for instance, LG modes) into an optical fiber with a fixed refractive index profile remains challenging for two primary reasons: (1) the beam size increases with increasing OAM, and (2) the side lobes beyond the main intensity ring introduce undesirable effects that must be carefully mitigated. A similar limitation arises in stimulated emission depletion (STED) microscopy—a super-resolution imaging technique—where a vortex beam serves as the depletion beam to minimize the effective focal spot size, thereby significantly enhancing lateral resolution (even down to a few nanometers). In this context, a vortex beam with a well-controlled dark core and beam size could further improve imaging performance. To address these challenges, Ostrovsky et al. theoretically introduced the concept of perfect optical vortex (POV) beams, which were subsequently realized experimentally using a spatial light modulator~\cite{Ostrovsky}. POV beams are characterized by a single bright intensity ring whose radius and width remain nearly invariant with increasing TC, thus overcoming the inherent limitations of conventional vortex beams~\cite{Ostrovsky,Vaity}. The complex field amplitude of POV beams at a propagation distance, $z$, is given by~\cite{Vaity}:
\begin{eqnarray}
    u_{0,POV}(\rho,\varphi,z)&=&A_{0} \frac{w_{0}}{w(z)} (-1)^{l} \mathrm{e}^{i(l\varphi+\Phi_{G, POV})} \mathrm{e}^{\frac{ik}{2R(z)}(\rho^2+\rho_{0}^2)} \mathrm{e}^{-\frac{1}{w(z)^2}\left(\rho^2-\left(\frac{\rho_{0}z}{z_{R}}\right)^2\right)} \nonumber\\
    &\times&I_{l}\left(\frac{2\rho_{0}\rho \mathrm{e}^{i\Phi_{G, POV}}}{w_{0}w(z)}\right)
    \label{eqn12}
\end{eqnarray}
where $w(z)$, $R(z)$, and $\Phi_{G,POV}$ denote the beam size at a distance $z$, the radius of curvature of the wavefront, and the Gouy phase, respectively, and are mathematically defined in the same way as the fundamental Gaussian beam. Here, $\rho_{0}$ represents the ring radius of the POV beam and $I_{l}(\ldots)$ is the modified Bessel function of the first kind of order $l$. 

First, let's examine the physical properties of POV beams at the source plane, i.e., at $z=0$. In a subsequent section we will discuss the propagation dynamics of POV beams in different media. At the source plane, the complex field amplitude of the POV beam can be expressed as~\cite{Vaity}:
\begin{eqnarray}
    u_{0,POV}(\rho',\varphi',z=0)=i^{l-1}\frac{w_{0}'}{w_{0}} \mathrm{e}^{il\varphi'} \mathrm{e}^{-\frac{\rho'^2+\rho_{0}^2}{w_{0}^2}} I_{l}\left(\frac{2\rho_{0}\rho'}{w_{0}^2}\right).
    \label{eqn13}
\end{eqnarray}
Here, ($\rho',\varphi'$) are the source plane coordinates, and $w_{0}'$ is the beam waist of the Gaussian beam, which is used to confine the BG beam. It is important to note that Eq.~(\ref{eqn13}) is derived by Fourier transforming the BG beam (physically, by using an optical lens). The form of the BG beam used for the Fourier transformation can be expressed as~\cite{Vaity}:
\begin{eqnarray}
    u_{0,BG}(r',\theta')=A_{0}J_{l}(k_{r}r') \mathrm{e}^{il\theta'}\mathrm{e}^{-\frac{r'^{2}}{w_{0}'^{2}}}.
    \label{eqn14}
\end{eqnarray}
The parameters characterizing POV and BG beams can be linked each other as follows: $\rho_{0}=\frac{k_{r}f}{k}$, and $w_{0}=\frac{2f}{kw_{0}'}$, where $\rho_{0}$, and $2w_{0}$ are the ring radius and full-ring width of the POV beam, respectively, and $f$ is the focal length of the lens used for the Fourier transformation. From Eq.~(\ref{eqn14}), one can clearly identify the exponentially decaying Gaussian envelope and the modified Bessel function of the first kind. The latter, for positive orders, exhibits an exponential growth behavior. Therefore, both functions intersect at $\rho' = \rho_{0}$, giving rise to the characteristic single-ring intensity profile of POV beams. In Fig.~\ref{Fig5}(C), the transverse intensity distributions of POV beams with different TC values are presented, clearly illustrating their stable and well-defined intensity structures.

An important question concerning POV beams is how \textit{perfect} they actually are. In particular, it is essential to examine whether the beam width remains perfectly invariant with respect to variations in the TC or exhibits a measurable dependence. To quantify this behavior, the beam width of POV beams can be evaluated using the second-moment definition:
\begin{eqnarray}
    w^2(l)=2\frac{\int_{0}^{\infty}\rho' \mathrm{d}\rho' \int_{0}^{2\pi}\rho'^{2}|u_{0,POV}(\rho',\varphi')|^2 \mathrm{d}\varphi'}{\int_{0}^{\infty}\rho' \mathrm{d}\rho' \int_{0}^{2\pi}|u_{0,POV}(\rho',\varphi')|^2 \mathrm{d}\varphi'}.
    \label{eqn15}
\end{eqnarray}
Now, by substituting Eq.~(\ref{eqn13}) into Eq.~(\ref{eqn15}) we obtain~\cite{Pinnell},
\begin{eqnarray}
    w^2(l)=w_{0}^2(l+1)+\rho_{0}^2 \left(1+\frac{I_{l+1}\left(\frac{\rho_{0}^2}{w_{0}^2}\right)}{I_{l}\left(\frac{\rho_{0}^2}{w_{0}^2}\right)}\right).
    \label{eqn16}
\end{eqnarray}
Equation~(\ref{eqn16}) indicates that the beam width is not entirely independent of the TC but exhibits a weak dependence on it. For relatively small TC values, this variation is negligible, whereas for very high ones, noticeable changes in the beam width can be observed. It is also worth noting that LG vortex beams show a similar dependence, with the beam width scaling as $w(l) \propto \sqrt{l}$. However, the variation is considerably slower in the case of POV beams, as discussed in Ref.~\cite{Pinnell}.

In the asymptotic limit, i.e., when $\rho_{0} \gg w_{0}$, the modified Bessel function in Eq.~(\ref{eqn13}) approaches an exponentially increasing form, such that $I_{l}\left(\tfrac{2\rho_{0}\rho'}{w_{0}^2}\right) \approx \exp\left(\tfrac{2\rho_{0}\rho'}{w_{0}^2}\right)$. Therefore, Eq.~(\ref{eqn13}) can be modified to~\cite{Vaity}:
\begin{eqnarray}
    u_{0,POV}(\rho',\varphi',z=0)=i^{l-1}\frac{w_{0}'}{w_{0}} \mathrm{e}^{il\varphi'} \mathrm{e}^{-\frac{(\rho'-\rho_{0})^2}{w_{0}^2}}.
    \label{eqn17}
\end{eqnarray}
Now, from Eq.~(\ref{eqn17}), it can be clearly seen that the amplitude of the beam is completely independent of the TC, and is fully controlled by $w_{0},w_{0}'$, and $\rho_{0}$. Therefore, plotting the transverse intensity distributions using Eq.~(\ref{eqn17}) yields TC–independent POV beam profiles. This form of the POV beam has been widely employed in numerous studies~\cite{Du,Xu,HONGYANWei}. Experimentally, POV beams can be generated using various techniques, including the Fourier transformation of BG beams~\cite{Vaity,Vanitha}, digital micromirror devices~\cite{Yue}, metasurfaces~\cite{Su}, planar Pancharatnam–Berry phase elements~\cite{Yachaoliuk}, and the Fourier transformation of LG beams with high radial indices~\cite{ZhenyuGuo}, among others. The use of POV beams has proven advantageous over conventional vortex beams in a variety of applications, such as optical trapping and manipulation~\cite{Mingzhou}, fiber-optic communication~\cite{Villalba}, plasmonic structured illumination microscopy~\cite{Chongleizhang}, and optical cryptography~\cite{Qingshuai}. For example, Ref.~\cite{Weishao} reported that multiplexing POV beams significantly enhances the system performance and reduces the bit error rate in free-space optical communication links. Likewise, Ref.~\cite{Chongleizhang} demonstrated that employing POV beams in plasmonic structured illumination microscopy not only increases surface plasmon excitation efficiency compared to LG beams but also substantially improves imaging resolution. Furthermore, in Ref.~\cite{Mingzhou}, Chen et al. demonstrated the ability of POV beams to efficiently trap and move multiple particles simultaneously in a ring trap.

\begin{figure}[h!]
\centering 
\includegraphics[width=1\linewidth]{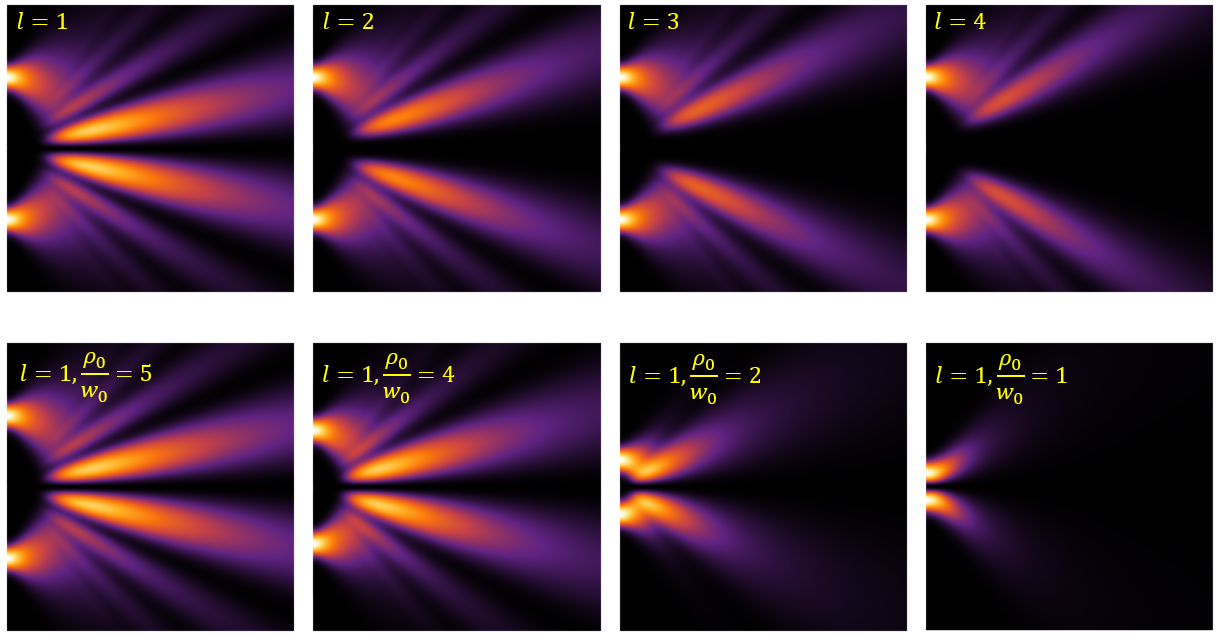}
\caption{Longitudinal intensity distributions illustrating the propagation dynamics of POV beams in free space.}
\label{Fig8}
\end{figure}

In Fig.~\ref{Fig8}, the longitudinal intensity distributions of POV beams propagating in free space are presented. As evident from the figure, the POV beam preserves its quasi–non-diffracting nature only up to the Rayleigh range. Beyond this region, several distinctive propagation features emerge. In the intermediate propagation distances, the beam undergoes self-focusing—that is, its transverse profile narrows significantly, accompanied by a pronounced increase in intensity. During this stage, multiple off-axis rings appear in the beam cross-section, indicating that the POV beam gradually transforms into a BG vortex beam. Upon further propagation, i.e., for $z \gg z_{R}$, the beam evolves into an ordinary vortex beam. Although the TC has only a minor influence on the beam characteristics within the non-diffracting region—for example, the effective propagation distance (determined by the Rayleigh range) increases slightly with higher TC values due to the charge-dependent beam width—the self-focusing behavior is strongly affected by the TC. As shown in Fig.~\ref{Fig8} (first row), the diameter of the central dark core expands with increasing TC in the self-focusing regime. Moreover, as the ratio of the beam radius ($\rho_{0}$) to the half-ring width ($w_{0}$) decreases, the self-focusing property gradually weakens: the self-focusing region contracts and the number of off-axis rings diminishes. When this ratio approaches unity, the self-focusing stage vanishes entirely, and the beam propagates similarly to a conventional vortex beam. This intrinsic self-focusing property of POV beams has been effectively utilized to enhance stability in optical communication systems~\cite{Shuailingwang}.

\subsubsection{Lorentz-Gauss vortex Beams}
With the growing demand for single-mode diode lasers in fiber-optic communication and optical sensing applications, the Lorentz–Gauss (LGa) vortex beam was introduced to model the output characteristics of such lasers. Unlike Gaussian, HG, or LG beams, LGa beams are not stable resonator modes. Instead, they can be produced by imparting a helical phase to the output of a single-mode diode laser, using refractive or diffractive optical elements, such as a spiral phase plate. An LGa vortex beam, at a propagation distance $z$, can be expressed as~\cite{Qusailah}:
\begin{eqnarray}
    u_{0,LGa}(\rho,z)&=&\left(\frac{\pi}{2w_{0x}w_{0y}}\right)\left(\frac{ik}{2\pi z}\right) \mathrm{e}^{-\frac{ik}{2z}(x^2+y^2)} \sum_{m=0}^{N}\sum_{n=0}^{N}\sigma_{2m}\sigma_{2n}  \nonumber\\
    &\times&\sum_{r=0}^{l}\frac{l!i^r}{r!(l-r)!}u(x,z)v(y,z),
    \label{eqn18}
\end{eqnarray}
where $\rho=\sqrt{x^2+y^2}$ is the transverse coordinate of the beam at a propagation distance $z$, ($w_{0x},w_{0y}$) are the beam widths of the Lorentz part of the LGa vortex beam in the $x$ and $y$ directions, respectively, and $w_{0}$ is the width of the Gaussian part. For the definition of $\sigma_{2m},\sigma_{2n},u(x,z)$, and $v(y,z)$, we refer the readers to Ref.~\cite{Qusailah}, where a detailed derivation of these quantities is presented. At the source plane (i.e., at $z=0$), the complex amplitude of the LGa vortex beam can be written as~\cite{Qusailah}:
\begin{eqnarray}
    u_{0,LG}(\rho',z=0)=\frac{w_{0x}w_{0y}}{(w_{0x}^2+x'^2)(w_{0y}^2+y'^2)} \mathrm{e}^{-\frac{x'^2+y'^2}{w_{0}^2}}(x'+iy')^l,
    \label{eqn19}
\end{eqnarray}
where $\rho'=\sqrt{x'^2+y'^2}$ is the transverse coordinate of the beam at the source plane. The term, $(x'+iy')^l$, translates to $\rho'^l \mathrm{e}^{il\varphi'}$, thereby, giving the characteristic vortex profile to the LGa beam. In Fig.~\ref{Fig5}(D), we show the transverse intensity distributions of the LGa vortex beam for different values of $l$. The basic vortex characteristic, such as the expansion of the vortex core with increasing values of $l$, can be clearly observed from the figure. 
\begin{figure}[h!]
\centering 
\includegraphics[width=1\linewidth]{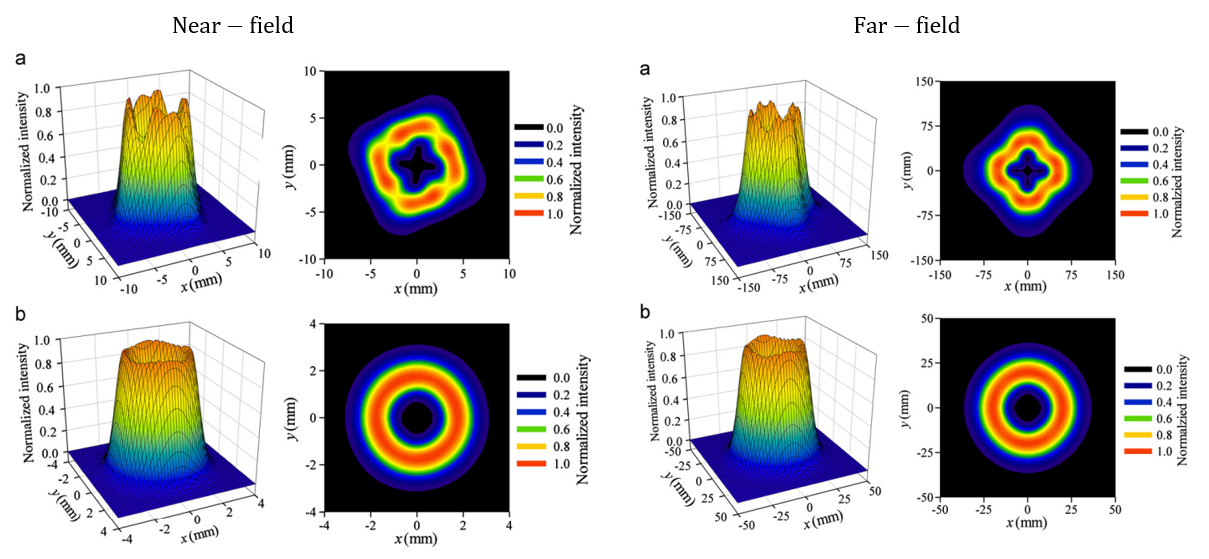}
\caption{Intensity distributions of LGa vortex beams propagating in free space. Reprinted with permission from Ref.~\cite{NI}.}
\label{Fig9}
\end{figure}
In Fig.~\ref{Fig9}, the near-field (left) and far-field (right) transverse intensity distributions of the LGa vortex beam propagating in free space are presented~\cite{NI}. Generally, the smaller of the beam width parameters—between the Gaussian component ($w_{0}$) and the Lorentz components ($w_{0x}, w_{0y}$)—dominates the overall profile of the LGa vortex beam. In the near-field regime, when the Lorentz contribution is stronger (i.e., $w_{0x}, w_{0y} < w_{0}$), the beam spot exhibits a rightward tilt, and the central dark region assumes a cross-like shape. Moreover, the bright ring surrounding the dark core becomes non-uniform, as shown in Fig.~\ref{Fig9}(left)(a). Conversely, when the Gaussian part dominates (i.e., $w_{0} < w_{0x}, w_{0y}$), the beam spot closely resembles that of a Gaussian vortex beam, with a uniformly distributed intensity around the central null [Fig.~\ref{Fig9}(left)(b)]. In the far-field regime, when the Lorentz component dominates, the beam tilt nearly vanishes. Although the intensity distribution remains somewhat non-uniform, it is significantly improved compared to the near-field case [Fig.~\ref{Fig9}(right)(a)]. When the Gaussian component dominates, the beam achieves a more uniform intensity distribution, as illustrated in Fig.~\ref{Fig9}(right)(b).

For comprehensive discussions of additional vortex beam families—such as vortex Hermite–Cosh–Gaussian, Airy vortex, Mathieu vortex, and Ince–Gaussian beams—readers are referred to Refs.~\cite{Hricha,ELHALBA,Lazrek1,Lazrek2,Xiaojin,Xinguang,Senhua,Ruihuang,Shijiechen,Dongye,Deng}. A detailed treatment of all existing vortex beam types lies beyond the scope of this review.
\begin{figure}[h!]
\centering 
\includegraphics[width=1\linewidth]{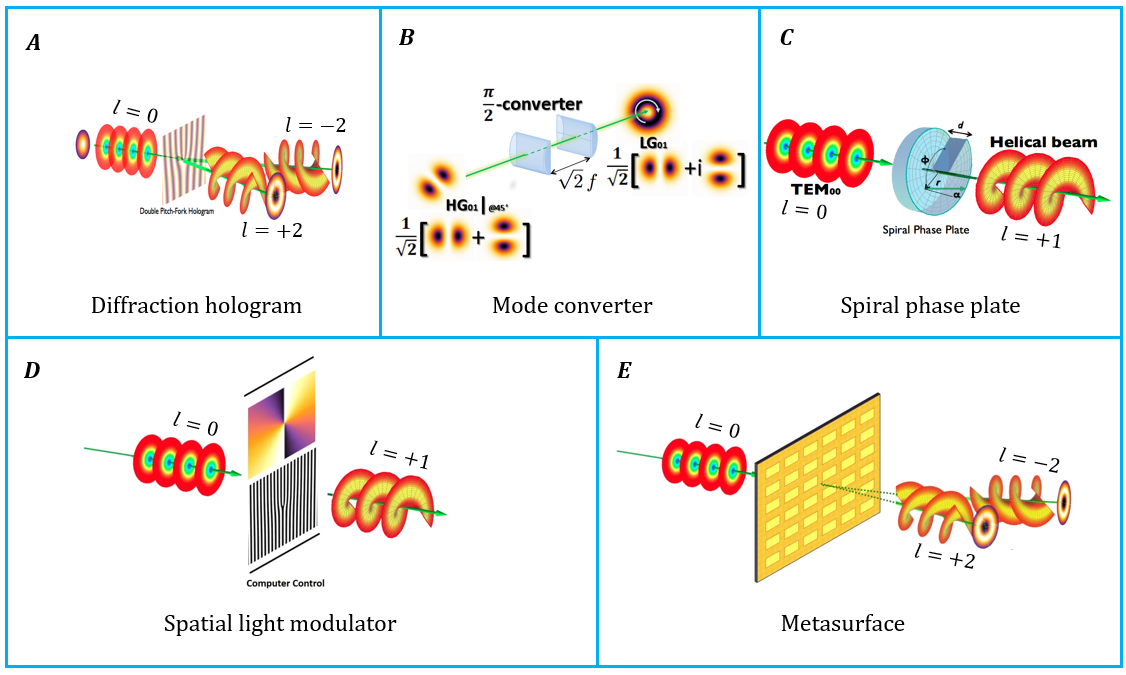}
\caption{Generation methods of vortex beams. (A) A forked hologram, formed due to the interference of a spiral phase and a tilted plane wave phase, imprints OAM on the input Gaussian beam, (B) Two cylindrical lenses, placed apart at a specific distance, convert a HG beam into a LG beam carrying OAM, (C) A dielectric plate, whose thickness varies with angle, imparts OAM to a Gaussian beam, (D) A liquid crystal device composed of a pixelated matrix of liquid crystals, whose molecules are birefringent, can modify the parameters of the incident beam and its phase, creating a twisted beam, and (E) An ultra-thin metamaterial can change the phase of an input Gaussian beam to imprint an OAM on it.}
\label{Fig10}
\end{figure}

\section{Generation methods of vortex beams and detection of their OAM}\label{vortex generation and detection}
\subsection{Generation of vortex beams}
\label{vortexgeneration}
Over the past three decades, numerous techniques have been developed for the generation of optical vortex beams. In this section, we provide an overview of several representative and widely adopted approaches that have become standard in laboratory implementations. These encompass both conventional optical components and modern digital devices, including diffraction holograms, spiral phase plates, cylindrical lens mode converters, spatial light modulators, and metasurfaces (see Fig.~\ref{Fig10}). In addition, alternative generation schemes—such as photon sieves, q-plates, photonic integrated structures, Fresnel diffraction from circular apertures, and flat nanostructured gradient-index vortex phase masks—have also been proposed and experimentally realized.

\subsubsection{Diffraction holograms}
\label{diffholog}
\begin{figure}[h!]
\centering 
\includegraphics[width=1.0\linewidth]{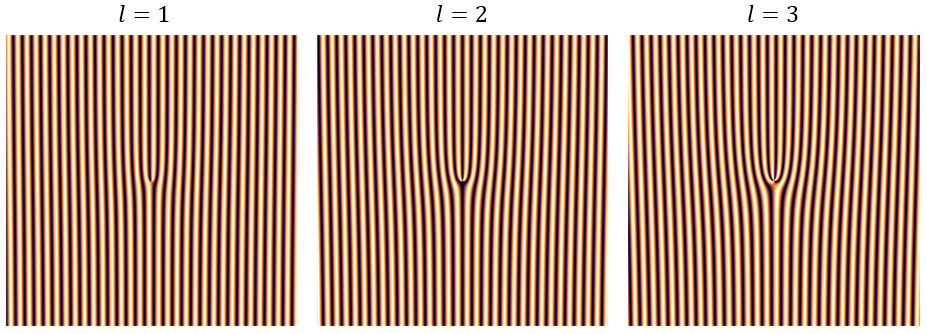}
\caption{The computer generated holograms with sinusoidal transmission function for generating optical vortices with TC 1 (left), 2 (middle), and 3 (right).}
\label{Fig11}
\end{figure}

Holography is a technique based on interference phenomena that enables the recording and reconstruction of three-dimensional images, commonly referred to as holograms. In 1971, the Nobel Prize in Physics was awarded to Dennis Gabor for his pioneering contributions to the development of holographic methods. In general, a hologram represents an interference pattern formed by the superposition of an object wave and a reference wave~\cite{BRYNGDAHL}. The use of computer-generated holograms (CGHs) provides a simple yet powerful approach for the generation of optical vortex beams~\cite{Bazhenov,Heckenberg}. In this method, the first step involves numerically computing the two-dimensional interference pattern resulting from the superposition of an object wave (the desired optical vortex beam) and a reference wave (a tilted plane wave, in this case). To generate a vortex beam carrying a TC $l$, one can interfere the spiral phase term $\mathrm{e}^{il\varphi}$ with a tilted plane wave $\mathrm{e}^{ik_{\text{tilt}}x}$, where $k_{\text{tilt}}$ denotes the spatial frequency of the interference fringes and is inversely proportional to their spatial period. Physically, $k_{\text{tilt}}$ can be adjusted by varying the tilt angle between the object and reference beams. For simplicity, the amplitude of the object beam is typically set to unity. The resulting interference signal can then be expressed as a term proportional to $\cos\left[l\arctan\left(\frac{y}{x}\right) - k_{\text{tilt}}x\right]$. This interference pattern, readily obtained using computer software, exhibits a characteristic forked-dislocation structure (see Fig.~\ref{Fig11}). 
The difference in the number of fringes above and below the dislocation line corresponds to the TC value, while the fork orientation indicates its sign. In practice, the computed pattern is printed onto a transparent substrate using a high-resolution printer. When illuminated with a laser beam—typically with a Gaussian intensity profile (see Fig.~\ref{Fig10}(A))—the hologram produces vortex beams of both positive and negative TC values in the first diffraction order, while the zeroth order (central beam) retains its Gaussian profile. This holographic technique offers several advantages for vortex beam generation: (1) it is compatible with both continuous-wave and pulsed laser sources, and (2) it can be implemented across a wide range of wavelengths with minimal modification. Although the holographic approach appears relatively straightforward for generating vortex beams, it presents several limitations. (1) The CGH pattern cannot be reconfigured once it is fabricated using photolithographic techniques. (2) Care must be taken to suppress unwanted diffraction orders, as they may introduce noise and interfere with applications such as optical communication and quantum information processing.

\subsubsection{Mode converters}
\begin{figure}[h!]
\centering 
\includegraphics[width=1.0\linewidth]{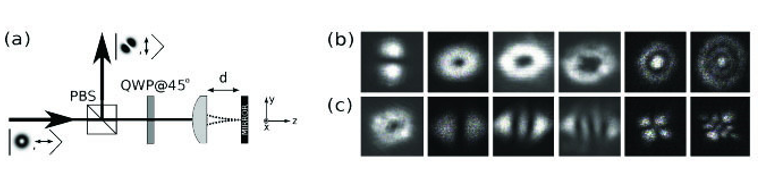}
\caption{Experimental setup demonstrating a single cylindrical-lens-based mode-conversion mechanism. Reprinted with permission from Ref.~\cite{Souza}.}
\label{Fig12}
\end{figure}

As the name suggests, this approach is employed to convert one mode into another (see Fig.~\ref{Fig10}(B)). It was Allen and his collaborators who demonstrated that a pair of cylindrical lenses with identical focal lengths ($f$), placed at a specific separation, can transform a superposition of HG modes into an LG mode~\cite{BEIJERSBERGEN}. The operation principle of such a system relies on the Gouy phase and the phenomenon of astigmatism. As discussed earlier in Section~\ref{solutions}, an LG mode can always be expressed as a superposition of HG modes, since the latter form an orthonormal basis. The working of a mode converter can be understood as follows: a superposition of HG modes (say, HG$_{01}$ and HG$_{10}$, aligned parallel and perpendicular to the cylindrical lens, respectively) is incident on one of the lenses. The separation between the two lenses determines the phase difference induced between the HG modes. For instance, when the distance between the lenses is $\sqrt{2}f$, a phase shift of $\frac{\pi}{2}$ is introduced, and the setup functions as a $\frac{\pi}{2}$-converter. Likewise, a separation of $2f$ induces a phase difference of $\pi$, resulting in a $\pi$-converter. In this sense, a $\frac{\pi}{2}$-converter (or $\pi$-converter) acts analogously to a half-wave (or full-wave) plate. Generally, a $\frac{\pi}{2}$ ($\pi$)-converter is used to generate LG modes with positive (negative) TC values, i.e., $+1$ ($-1$) in our current notation. The radial index of the generated LG mode is given by $p = \textrm{min}(m,n)$, where $m$ and $n$ are the orders of the Hermite polynomials as introduced earlier.
Moreover, by replacing one of the two cylindrical lenses with a flat mirror and adjusting the distance between the remaining lens and the mirror, one can also realize the mode conversion operation (see Fig.~\ref{Fig12})~\cite{Souza}. 
This modified setup significantly relaxes the alignment constraints encountered in the conventional two-lens configuration. It is well known that HG modes are solutions to the paraxial Helmholtz equation in Cartesian coordinates and exhibit rectangular symmetry, while LG modes are solutions in cylindrical coordinates and exhibit circular symmetry. Therefore, the mode converter effectively transforms the beam symmetry from rectangular to circular. It is noteworthy that such mode converters are also employed to retrieve OAM vortex beams (as will be discussed later). Although the setup appears straightforward for generating LG modes with TC $l = \pm 1$, the production of higher-order modes ($|l| > 1$) requires superpositions of higher-order HG modes, thereby increasing the optical system’s complexity. Furthermore, careful alignment of the cylindrical lenses (or the lens–mirror configuration) is crucial, as deviations from the optimal separations of $\sqrt{2}f$ or $2f$ can lead to mode contamination.

\subsubsection{Spiral phase plates}
A spiral phase plate (SPP) is a refractive optical element made of a transparent dielectric material, capable of directly imparting a spiral or helical phase to a coherent light beam (see Fig.~\ref{Fig10}(C))~\cite{BEIJERSBERGEN1}. 
\begin{figure}[h!]
\centering 
\includegraphics[width=1.0\linewidth]{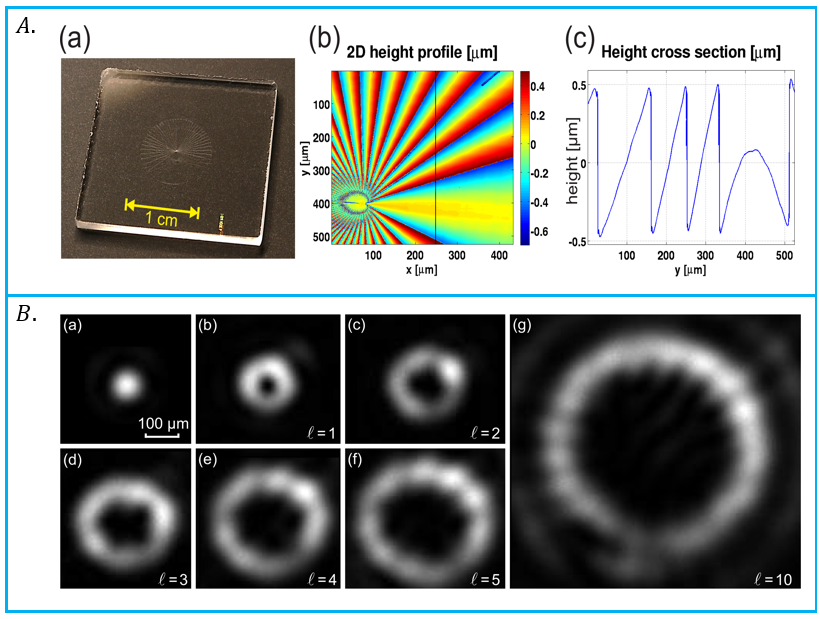}
\caption{(A): (a) Photograph of one of the fabricated Moiré diffractive optical elements (MDOEs). (b) Height measurement of its phase profile in the central region by white light interferometry, and (c) Phase profile along the vertical line indicated in (b). (B): (a) Unmodified fundamental Gaussian beam and (b)-(g) Generation of LG-like vortex beam carrying different TC values. Reprinted with permission from Ref.~\cite{Harm}.}
\label{Fig13}
\end{figure}
The thickness of an SPP varies azimuthally around its center while remaining uniform along the radial direction. Typically, an SPP is designed for a specific wavelength, meaning that the difference between its maximum and minimum optical thickness must be carefully adjusted to correspond to that wavelength. When an optical beam with a flat phase front, such as a Gaussian beam, passes through the SPP, the device introduces a spiral phase to the beam’s wavefront. Consequently, the initially flat phase front evolves azimuthally from $0$ to $2\pi$, generating a vortex beam with a TC of $+1$. By simply reversing the orientation of the phase plate, one can produce a vortex beam carrying a TC of $-1$. It is important to note that an SPP is usually fabricated for a particular wavelength and a fixed TC. Therefore, optimal performance is achieved when the illuminating laser beam matches the design wavelength of the plate. Any deviation from this wavelength can lead to noticeable distortions in the output beam profile. For instance, as discussed in Ref.~\cite{Stoyanov1}, the phase plate can generate vortices with \textit{fractional} TC values when the illumination wavelength differs from the design wavelength-an effect that becomes increasingly pronounced as the deviation grows. To overcome these limitations, the concept of an adjustable SPP was theoretically proposed and later experimentally demonstrated~\cite{Harm}. This device can generate multiple TC values and operate efficiently across a range of wavelengths. A representative design, known as the Moiré diffractive SPP, is shown in Fig.~\ref{Fig13}(A). Using this configuration, LG-like vortex beams with variable TC values were successfully produced (see Fig.~\ref{Fig13}(B)). Moreover, it was shown that these adjustable SPPs are polarization-insensitive and exhibit a conversion efficiency exceeding 90\%.

In addition to the azimuthal variation in thickness, a spiral phase can also be introduced by modulating the refractive index, and thus the optical path length, with the azimuthal angle. This principle was realized through the design of a flat SPP~\cite{Wenbingwu}. One of the notable advantages of using an SPP for vortex beam generation is its high conversion efficiency, typically around 80\%, which can reach nearly 100\% with proper anti-reflection coating. Furthermore, an SPP can withstand the high intensities of powerful laser beams, making it one of the few optical components suitable for generating vortex beams with ultrashort laser pulses.

\subsubsection{Spatial light modulators}

\begin{figure}[h!]
\centering 
\includegraphics[width=1\linewidth]{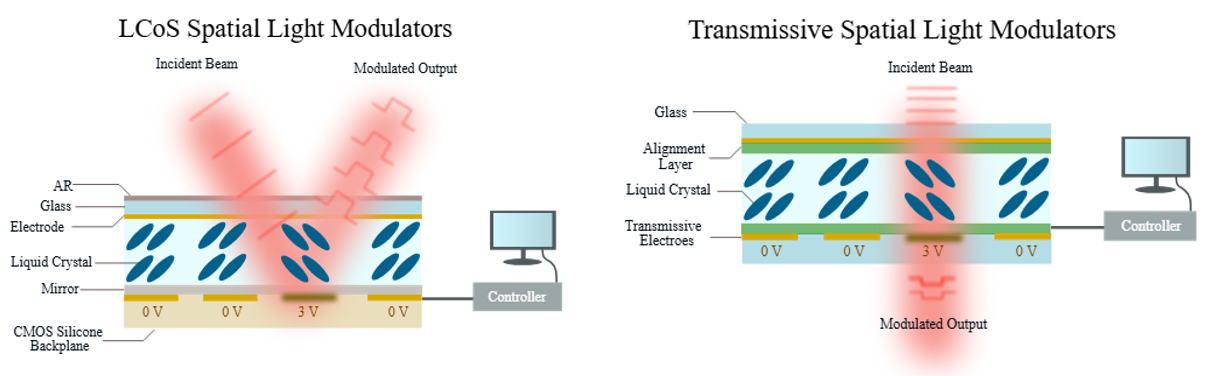}
\caption{Different types of SLMs. (Left) Reflective, (Right) Transmissive SLMs to enable the modulation of amplitude, phase and polarization of incident light.}
\label{Fig14}
\end{figure}

All the methods discussed above for vortex beam generation can be regarded as static. For instance, the CGH method can generate a vortex beam with a fixed TC, which is often impractical in experiments. Therefore, when it comes to flexible (or dynamic) generation of vortices, there is essentially no competition between the spatial light modulator (SLM) and other optical or digital devices. An SLM belongs to the class of programmable optical devices used for vortex beam generation, and its operating principle can be understood as follows: an SLM typically consists of a pixelated matrix of liquid crystals, whose optical and electrical anisotropy are controlled by voltages applied across individual pixels. When a desired phase distribution is uploaded to the SLM, the phase of each pixel is converted into a corresponding voltage across the liquid-crystal cell. The modulation of the laser beam's phase consequently leads to a modulation of its amplitude or intensity. For instance, to generate a vortex beam carrying a TC of $l$, the phase pattern displayed on the SLM is given by $\phi_{SLM}(x,y) = l\varphi$, where $\phi_{SLM}(x,y)$ is the phase delay at each pixel coordinate $(x,y)$, and $\varphi = \arctan\left(\frac{y}{x}\right)$ is the azimuthal angle around the center of the SLM screen. This procedure generates a phase pattern resembling a spiral staircase. Then, a Gaussian beam with a flat phase front is expanded (for example, using a telescope configuration) and directed onto the active area of the SLM. The liquid-crystal pixels behave as tiny, adjustable wave plates, each retarding the phase of the reflected Gaussian beam by the precise amount specified in the CGH. As the beam reflects from different azimuthal regions of the spiral staircase pattern, it acquires the corresponding helical phase, thus forming a vortex beam. The resolution of the generated vortex beam largely depends on the waist size of the input Gaussian beam and the active area of the SLM. A larger beam waist makes better use of the active area, resulting in a higher-resolution vortex beam. However, if the waist is too large, part of the incident beam will miss the SLM surface, thereby reducing the overall quality of the generated beam.
In practice, displaying only a spiral phase pattern on the SLM can sometimes deflect the vortex beam at an undesirable angle. To avoid this, it is advantageous to add a blazed (linear) grating phase to the spiral phase (see Fig.~\ref{Fig10}(D)). The combination of the spiral and blazed grating phases forms a forked diffraction grating, i.e., a grating with a fork-like dislocation at the center. The number of fringes above and below the dislocation corresponds to the TC of the generated vortex beam. This composite phase serves two primary purposes: (1) it separates the first-order diffracted vortex beam (which contains the desired mode) from the zeroth and higher diffraction orders, and (2) it ensures that the vortex beam is generated on-axis in the first-order diffraction, facilitating its isolation and use.
As mentioned earlier, the key advantages of SLMs for vortex beam generation are: (1) their flexibility in generating single vortices, vortex arrays, and composite vortex beams, (2) the real-time control of the TC, (3) the high-quality beam generation with high-resolution SLMs, and (4) the compatibility with a wide range of wavelengths. Generally, SLMs are categorized as either reflective-type, where liquid-crystal displays are mounted on a Si substrate, or transmissive-type, where transparent liquid-crystal displays are used (see Fig.~\ref{Fig14}). It is important to note that SLMs can be used not only to generate vortex beams but also to detect their OAM states, as will be discussed later.

\subsubsection{Metasurfaces}
Metamaterials are a class of artificially engineered periodic structures that exhibit electromagnetic responses generally not found in nature~\cite{Jingbo}. For instance, it has been shown that these materials can display magnetic behavior at terahertz (THz) frequencies-an effect absent in naturally occurring materials~\cite{Fan}. Furthermore, it is possible to observe different phenomena such as the negative refraction~\cite{PADILLA}, and hyperbolic light dispersions~\cite{Drachev}, which are not so natural. Metamaterials are usually designed from multiple materials (could be metals or plastics), which are arranged in repeating patterns, at scales that are smaller than the wavelengths of the phenomena they influence. Depending on their shape, size, geometry, and orientation, metamaterials can be used to manipulate light and acoustic waves. Metamaterials have found a wide range of applications, including optical filters, medical devices, ultrasonic sensors, super-lenses capable of imaging beyond the traditional diffraction limit, earthquake-shielding structures, and even invisibility cloaks, among others. A metasurface, on the other hand, represents the two-dimensional counterpart of a three-dimensional metamaterial, consisting of an ultra-thin, planar array of subwavelength building blocks known as meta-atoms. By suitably engineering the meta-atoms, one can precisely tailor the optical properties of metasurfaces—an aspect that is particularly crucial when dealing with ultrashort pulses carrying OAM. It is well-known that metasurfaces can control the amplitude, phase, frequency, and polarization of light at a sub-wavelength scale. 
\begin{figure}[h!]
\centering 
\includegraphics[width=1.0\linewidth]{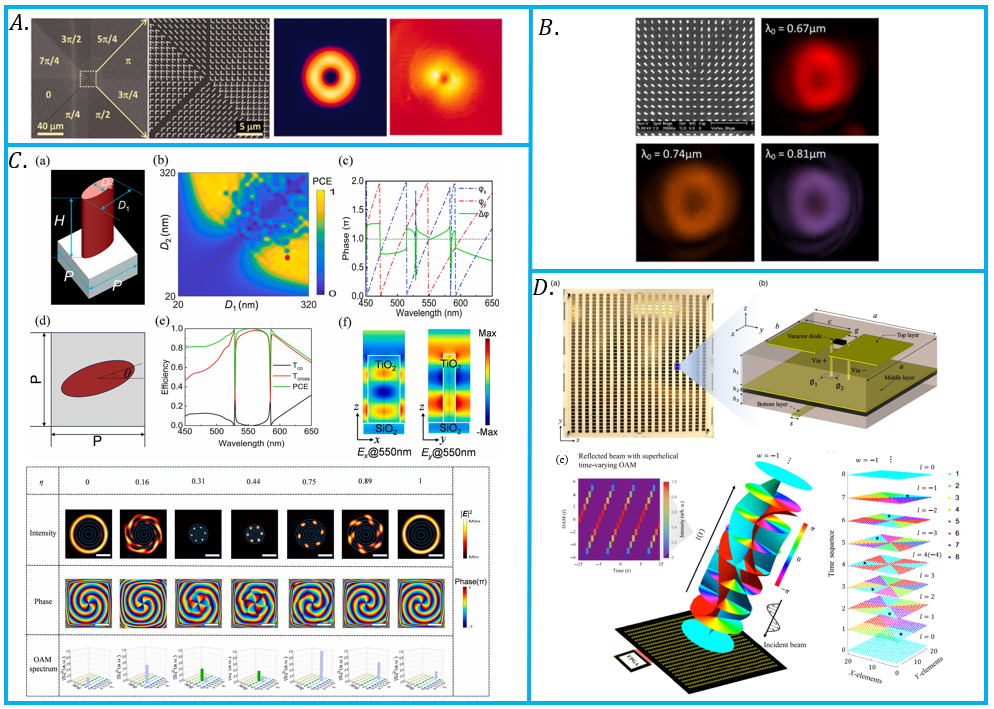}
\caption{Metasurface-based vortex beams generation. (A): (from left to right): Panels 1 and 2: Scanning electron microscopy (SEM) images of a fabricated plasmonic metasurface that generates an optical vortex. Panels 3 and 4: Simulated and measured far-field intensity pattern of an optical vortex with $l=1$, respectively, taken from Ref.~\cite{Nanfang}. (B): (top left) SEM image of the fabricated gold nanopillars-based optical vortex generator. Measured intensity distribution of the vortex
beam patterns at 670 nm (top right), 740 nm (bottom left) and 810 nm (bottom right), taken from Ref.~\cite{Lingling}. (C): (a) Schematic illustration of the unit cell on the CGM, (b) Calculated polarization conversion efficiency (PCE) as a function of nanopillars major axis of D1 and minor axis of D2 at a wavelength of 550 nm, (c) Distribution of the calculated phase for the $x-$ and $y-$polarized light, as well as their phase difference, (d) Top view of the unit cell with dimension parameters of a nominal lattice constant of $P = 350$ nm, (e) Distribution of transmittance of cross-polarization and co-polarization components, as well as PCE, as a function of wavelength, (f) Electric field distributions at 550 nm excited by $x-$polarized (left) and $y-$polarized incident light (right), taken from Ref.~\cite{Chenyangwang}. (D): Scheme to generate time-varying OAM beams, taken from Ref.~\cite{JingxinZhang}. Reprinted with permission from Refs.~\cite{Nanfang,Lingling,Chenyangwang,JingxinZhang}}
\label{Fig15}
\end{figure}
One of the fascinating applications of metasurfaces is their ability to generate optical vortices (see Fig.~\ref{Fig10}(E)). In 2011, Yu et al. for the first time demonstrated that a plasmonic metasurface composed of V-shaped nanoantennas can induce a helical phase shift upon interference with an incident linearly polarized plane wave, thereby generating an optical vortex carrying a TC of $l=1$ (see Fig.~\ref{Fig15}(A))~\cite{Nanfang}. Furthermore, in Ref.~\cite{Lingling}, it was shown that a metasurface, consisting of an array of metallic nanorods with the same geometry but spatially varying orientations, can generate vortices from an initial circularly polarized plane wave at visible and near-infrared (NIR) wavelengths (see Fig.~\ref{Fig15}(B)). In Ref.~\cite{ShanJiang}, Jiang et al. demonstrated that a four-layer azimuthally symmetric achromatic metasurface with broad bandwidth can efficiently convert a left-handed (right-handed) circularly polarized incident plane wave to a right-handed (left-handed) circularly polarized vortex beam. In Ref.~\cite{Fengmei}, to overcome the limitations in tuning vortex modes in compact metasurface devices—stemming from the difficulty of realizing dynamic meta-units—Mei et al. proposed a cascaded-metasurface-based system that enables dynamic switching between high-quality vortex modes by adjusting only a small number of meta-units. In Ref.~\cite{Chenyangwang}, the authors demonstrated that a single composite geometric metasurface (CGM), with ellipsoidal TiO$_{2}$ nanopillars as unit cells, can generate composite vortex beams (CVBs) with petal-like intensity distributions (see Fig.~\ref{Fig15}(C)). Moreover, it was shown that the proposed CGM can also generate two POV beams with different TC values. For the dynamic generation of vortex beams, an optically controlled programmable metasurface design was proposed. In Ref.~\cite{Ma}, the authors proposed a single-layer metallic porous metasurface design with V-shaped base elements, and demonstrated that by leveraging the photonic spin-orbit interaction and optimizing the geometric phase of the design, vortex beams with variable TC values can be produced. Additionally, in Ref.~\cite{Nadi}, Nadi et al. proposed a 1-bit programmable metasurface antenna to generate highly efficient dynamic multi-mode vortex beams. In contrast to the generation of time-varying OAM beams (self-torque of light) via the highly nonlinear process of high-order harmonic generation (HHG), a simple space–time-coding digital metasurface operating in the microwave regime has been proposed to produce time-varying OAM beams (see Fig.~\ref{Fig15}(D))~\cite{JingxinZhang}. Moreover, to eliminate the bulky experimental setups traditionally required for producing different types of perfect optical vortex (POV) beams, metasurface-based approaches have also been developed~\cite{Xiujuanliu}.

As noted at the beginning of this subsection, we have reviewed only a selection of the most widely used methods for generating vortex beams. For a more comprehensive discussion of additional techniques employed in the generation of optical vortices, we refer readers to Refs.~\cite{Keli,Shujunzheng,Weikezhao,Peiliang}.

\subsection{Detection of OAM of vortex beams}
Since vortex beams are widely employed in real-world applications such as optical trapping, optical communication, and quantum information technologies, accurately determining their TC (or equivalently, their OAM) is of fundamental importance. A broad variety of OAM detection schemes have been developed, including those based on diffraction, interference, geometric coordinate transformations, and spectroscopy (for instance, coherent anti-Stokes Raman scattering spectroscopy). Other notable approaches rely on deep learning, metasurfaces, surface plasmon polaritons, the rotational Doppler effect, or optical elements such as cylindrical and tilted convex lenses. In this subsection, we briefly review some of the most representative methods.

\subsubsection{Diffraction}
In Sec.~\ref{diffholog}, we discussed how diffraction holograms can be employed to generate vortex beams. Here, we illustrate how diffraction phenomena can likewise be exploited to determine the TC of vortex beams.

\textit{Case 1- Single-slit diffraction:}

\begin{figure}[h!]
\centering 
\includegraphics[width=1\linewidth]{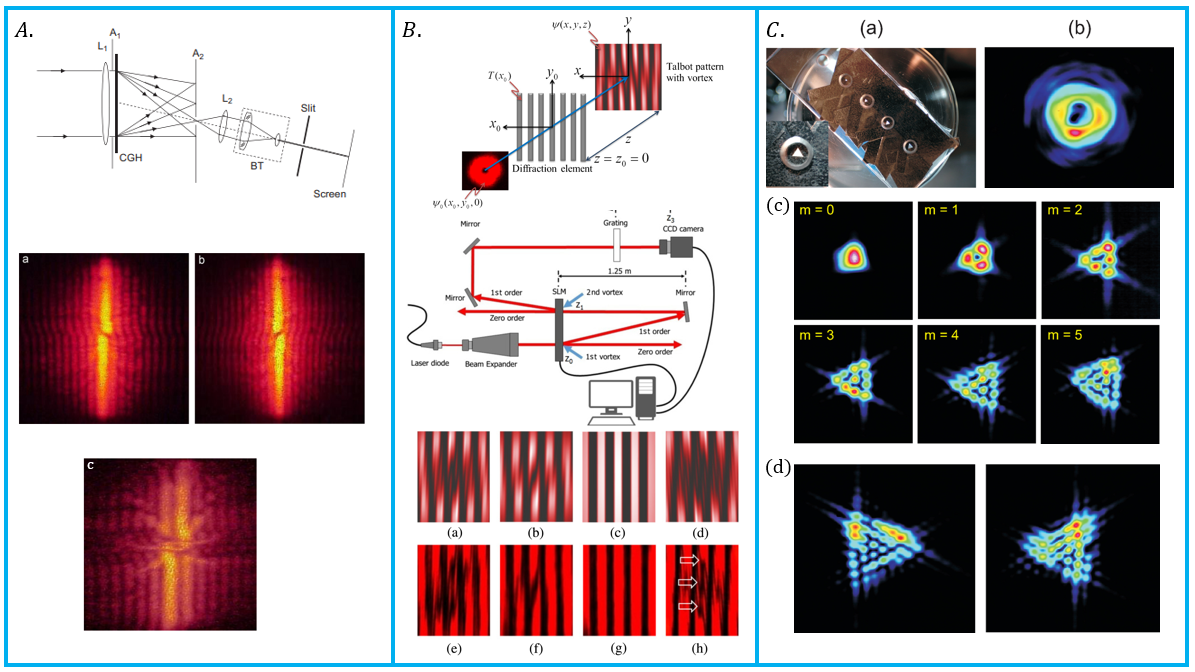}
\caption{Detection of OAM of vortex beams based on diffraction. (A) Single-slit diffraction of vortex beams, taken from Ref.~\cite{GHAI}. (B) Diffraction of vortex beams through a periodic diffraction grating (Talbot effect), taken from Ref.~\cite{Panthong}. (C) Diffraction of vortex beams through a triangular aperture, taken from Ref.~\cite{deAraujo}. Reprinted with permission from Refs.~\cite{GHAI,Panthong,deAraujo}.}
\label{Fig16}
\end{figure}

Diffraction of a plane wave through a single slit is a well-understood phenomenon. Typically, it produces an intensity distribution on a screen placed behind the slit that follows a $\mathrm{sinc}^2$ profile. Qualitatively, one observes a bright central maximum with the highest intensity, flanked by rapidly diminishing side lobes. A natural question then arises: What happens when a vortex beam illuminates a single slit? Can information about the TC—its magnitude and sign—be extracted from the resulting diffraction pattern? Moreover, how does the central dark core characteristic of vortex beams manifest in single-slit diffraction? To address these questions, Ghai et al. were, to the best of our knowledge, the first to experimentally study the single-slit diffraction of vortex beams, allowing the dark core of the incident beam to pass directly through the slit (see Fig.~\ref{Fig16}(A), top, for the experimental setup)~\cite{GHAI}. It is worth noting that earlier works had employed diffraction primarily to manipulate the TC of vortex beams; however, the distinctive dark core of the vortex was not visible in those diffraction patterns because in all previous configurations it was blocked~\cite{Sztul,MASAJADA}. In Ref.~\cite{GHAI}, it was demonstrated that: (1) Each diffraction fringe exhibits a noticeable bend near the center—a clear signature of vortex-beam diffraction; (2) Reversing the sign of the input beam’s TC reverses the bending direction; (3) For an incident vortex beam with $l=1$, the bright (dark) fringes in the upper half of the diffraction pattern bend and align with the dark (bright) fringes in the lower half. In contrast, for $l=2$, the bright (dark) fringes in the upper half align with the corresponding bright (dark) fringes in the lower half. These characteristic features are clearly visible in Figs.~\ref{Fig16}(A) (middle and bottom).

\textit{Case 2- The Talbot effect:}

In 1836, Henry Fox Talbot discovered a remarkable diffraction phenomenon, later termed the \textit{Talbot effect}~\cite{Talbot}. It can be described as follows: when a plane wave illuminates a periodic diffraction grating, self-images of the grating are formed at regular intervals along the propagation axis. These intervals correspond to the \textit{Talbot length} ($z_{T}$), given approximately by $z_{T}\approx\frac{2a^2}{\lambda}$, where $a$ denotes the grating period and $\lambda$ the wavelength of the incident light. The recurring replicas of the grating are referred to as \textit{Talbot images} or \textit{self-images}. It is important to note that this expression for 
$z_T$ is valid under the condition $\lambda\ll a$. In \textit{Case 1}, we discussed how the single-slit diffraction of vortex beams reveals both the magnitude and the sign of their TC. Now, what happens if we replace the single slit with a periodic diffraction grating? In other words, can the Talbot effect be utilized to extract information about the TC of vortex beams? The answer is affirmative. In Ref.~\cite{Panthong}, the authors investigated several configurations---including single-slit, double-slit, multiple-slit, and grating diffraction---and concluded that the grating setup, exploiting the near-field Talbot effect, provides the most effective means for detecting the TC of vortex beams (see Fig.~\ref{Fig16}(B)). For instance, a closer examination of Fig.~\ref{Fig16}(B) (bottom) shows that: (1) the number of dark stripes (indicated by arrow marks) and (2) their orientation within the diffraction pattern encode, respectively, the magnitude and the sign of the TC.

\textit{Case 3- Diffraction through a triangle aperture:}

One of the simplest and most reliable methods to determine both the magnitude and sign of the TC of vortex beams is to diffract them through an equilateral triangular aperture. In Ref.~\cite{deAraujo}, the authors diffracted vortex beams carrying various TC values and recorded their far-field intensity distributions, which resemble triangular lattices, to extract the corresponding TC information (see Fig.~\ref{Fig16}(C)). The authors derived several interesting relations, such as:
(1) $l = N - 1$, where $l$ denotes the TC of the input vortex beam and $N$ is the number of bright spots along one edge of the far-field intensity pattern;
(2) $l = \frac{\sqrt{1 + 8N_{t}} - 3}{2}$, where $N_{t}$ represents the total number of bright spots in the far-field intensity distribution;
(3) if the input vortex beam carries a negative TC, the resulting triangular lattice in the far-field pattern is rotated by $180^{\circ}$ relative to that of a vortex with a positive TC. It is important to emphasize that this method performs efficiently with both continuous-wave and pulsed laser sources and can be readily applied to the characterization of ultrashort vortex beams.

\subsubsection{Interference}

The principle of superposition forms the foundation of interferometry. In essence, it involves combining waves in such a manner that their superposition produces an interference pattern, whose characteristics-such as the number of fringes, their shape, and relative orientation-can be analyzed to reconstruct properties of the original waves. From a fundamental understanding of interference, it is known that when two waves of the same frequency combine, the resulting intensity pattern depends critically on their relative phase difference: constructive (destructive) interference occurs when the waves are in phase (out of phase). Likewise, an intermediate interference pattern appears when the waves are only partially in phase. Consequently, such interference patterns can be exploited to determine the relative phase difference between the waves~\cite{Adhikari}. Extending this idea, one can infer that interference phenomena can be used to retrieve the phase information-or equivalently, the TC—of vortex beams, since they exhibit characteristic nonuniform phase distributions. Several straightforward interferometric techniques have been demonstrated for this purpose:
(1) The interference of a vortex beam (whose TC is to be determined) with a plane wave produces a spiral-shaped pattern, where the number and orientation of the spirals indicate the magnitude and sign of the TC, respectively~\cite{BASISTIY1,Soskin}.
(2) The interference of a vortex beam with its mirror image, typically generated using a Dove prism, which inverts the handedness of the beam when the prism is traversed an odd number of times. The resulting interference pattern exhibits a petal-like structure, and the TC of the vortex beam can be obtained by dividing the number of petals by two~\cite{Vickers}.

In this subsection, we review several advanced interferometric approaches for characterizing vortex beams, including the improved multipoint interferometer (IMI) and self-referenced interference of laterally displaced vortex beams.

$\textit{Case 1- Improved multipoint interferometer (IMI):}$

\begin{figure}[h!]
\centering 
\includegraphics[width=1\linewidth]{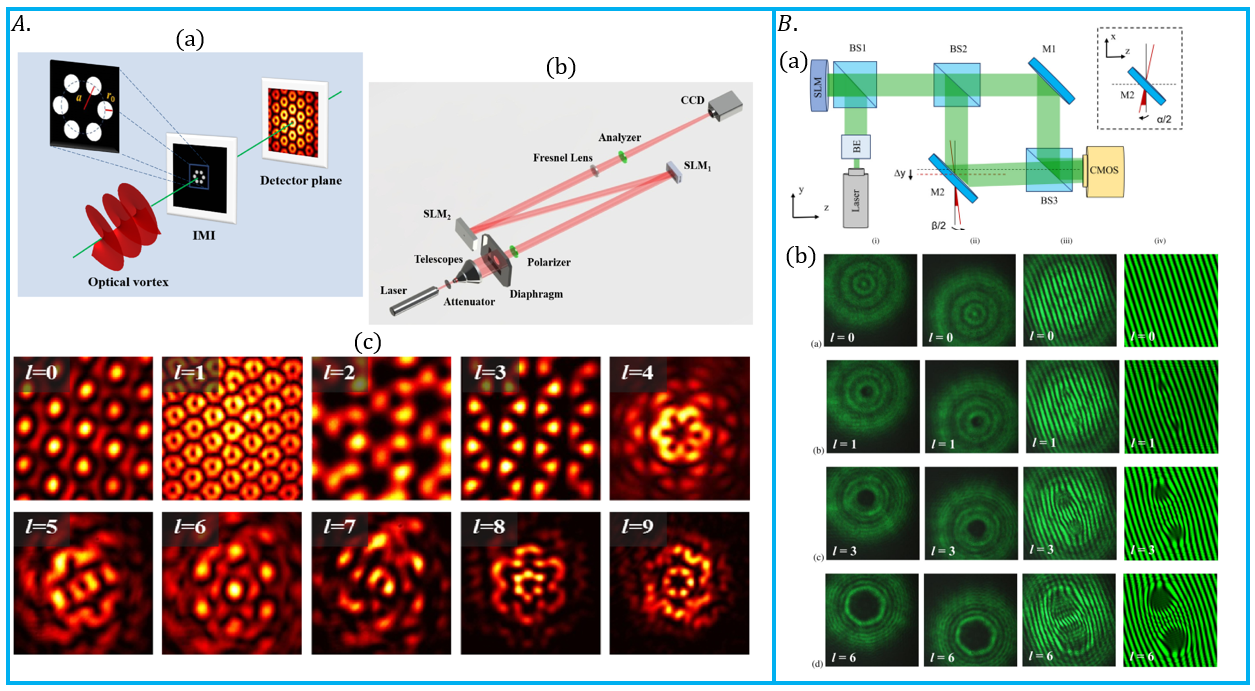}
\caption{Detection of OAM of vortex beams based on interference. (A) Improved multipoint interferometer, taken from Ref.~\cite{QiZhao}. (B) Self-referenced interference of laterally displaced vortex beams, taken from Ref.~\cite{Praveen}. Reprinted with permission from Refs.~\cite{QiZhao,Praveen}.}
\label{Fig17}
\end{figure}

In 2008, Berkhout et al. proposed that point-like pinholes, uniformly distributed along a circular ring, can be effectively employed to determine the TC of vortex beams of arbitrary size~\cite{Berkhout}. In this method—based on a multipoint or multipinhole interferometer (MI)—the TC of a vortex beam is identified from the fact that beams carrying different TC produce distinct far-field intensity distributions after passing through a multipinhole plate, in which the pinholes are evenly arranged along a ring. Although this approach is simple, robust, and remarkably successful in characterizing optical vortices—even those originating from astronomical sources (a feat unattainable by diffraction-based methods)—it has a few limitations:
(1) It can only characterize vortex beams carrying relatively low TC values, since the far-field intensity patterns repeat periodically for beams with higher TC values;
(2) While the magnitude of the TC can be determined for any number of pinholes, extracting its sign becomes problematic when the number of pinholes is even—vortex beams with charges $+l$ and $-l$ yield identical far-field intensity patterns, making them indistinguishable. To address these issues, Zhao et al. introduced a modified version of the MI, referred to as the improved multipoint interferometer (IMI) (see Fig.~\ref{Fig17}(A))~\cite{QiZhao}. In this enhanced design, each pinhole is replaced by a small circular aperture. It was shown that, unlike in the original MI, the far-field intensity patterns no longer repeat periodically for vortex beams with higher TC values. In other words, enlarging the pinholes (or circular apertures in the IMI) extends the measurable range of TC values, enabling the detection of high-$l$ vortex beams. 
 
$\textit{Case 2- Self-referenced interference of laterally displaced vortex beams:}$

Self-referenced interferometry has long served as a powerful technique for characterizing optical beams. Based on either wavefront splitting or amplitude splitting, this method produces an interferogram that encodes information about the original beam. Importantly, self-referenced interferometry is applicable not only to conventional Gaussian beams but also to vortex beams. Both wavefront-splitting and amplitude-splitting interferometers can be effectively employed for this purpose. One such example, discussed earlier, involves the interference of two vortex beams carrying equal but opposite TC values, resulting in a distinctive petal-like intensity pattern.

In this case, we discuss how the interference between a vortex beam and its laterally shifted copy can be used to determine, in a simple and direct manner, both the magnitude and sign of the TC. In Ref.~\cite{Praveen}, the authors employed a self-referenced interferometric technique using a Mach–Zehnder interferometer and observed interferograms featuring conjoined fork-like structures (see Fig.~\ref{Fig17}(B)). The essence of the method is as follows: the interference occurs between the original vortex beam (whose TC magnitude and sign are to be determined) and a modified version of the same beam—that is, a vortex beam carrying the same TC but with a slightly altered propagation direction. This modification is achieved by introducing controlled tilts and lateral displacements in one of the beams. It was shown that introducing a small tilt angle between the two interfering beams produces interference fringes containing a pair of fork-like structures. A lateral displacement between the beams further separates these forks. Moreover, when both tilt and lateral displacement are introduced along orthogonal (same) directions, the resulting interferogram exhibits conjoined (disjoined) fork structures. The number of tines observed in each fork structure directly corresponds to the magnitude of the TC, while the way in which the forks are connected reveals its sign: if the fork structures are connected by their handles (tines), the TC is negative (positive). This approach is simple, robust, and highly versatile, making it a reliable method for extracting the TC information of vortex beams.

\subsubsection{Geometric coordinate transformation}

Sorting of OAM modes through optical geometric coordinate transformation is another efficient technique for detecting vortex beams, typically relying on the conversion of OAM into linear momentum~\cite{Berkhout1,Berkhout2}. In this approach, diffractive optical elements perform a Cartesian ($x, y$) to log-polar ($u, v$) coordinate transformation, effectively converting the helical phase structure of the incident light beam—associated with its OAM—into a transverse phase gradient. A subsequent lens then focuses input vortex beams with different TC values to distinct transverse positions, thereby enabling their spatial separation and identification.

\begin{figure}[h!]
\centering 
\includegraphics[width=1\linewidth]{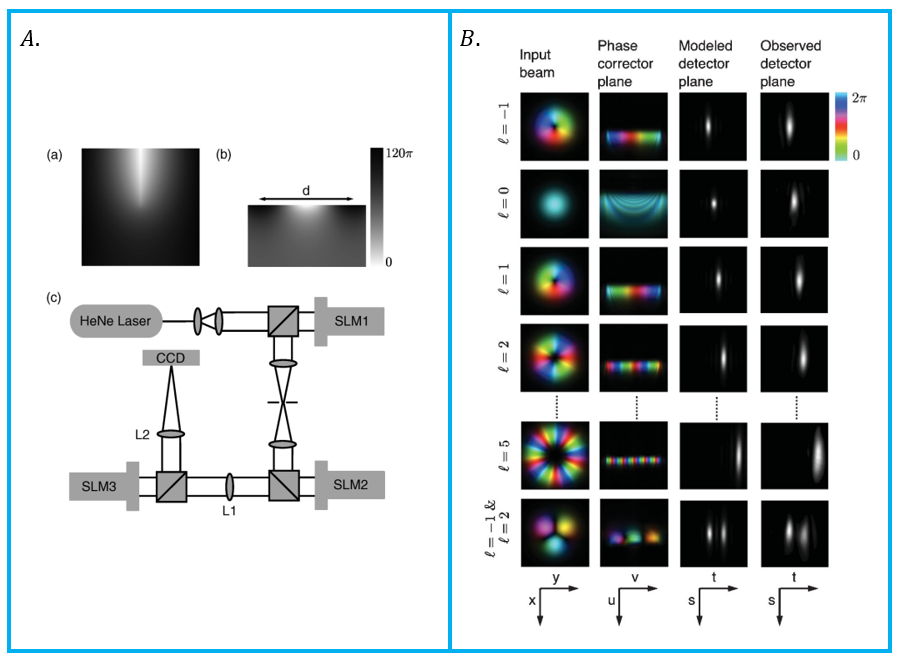}
\caption{Illustrates OAM detection via optical geometric coordinate transformation. (A) Experimental set-up. (B) Single and superposed vortex beams carrying different OAM at the input plane, and their sorting in the modeled (theoretical) and observed (experimental) detection planes.  Reprinted with permission from Ref.~\cite{Berkhout1}.}
\label{Fig18}
\end{figure}

The experimental setup used to realize such a coordinate transformation is illustrated in Fig.~\ref{Fig18}(A). Spatial light modulators (SLM1, SLM2, and SLM3) are employed sequentially to: (1) generate vortex beams with different TC values, (2) convert the azimuthal coordinate in the Cartesian frame into the transverse position in log-polar coordinates, and (3) compensate for the phase distortion introduced by the geometric transformation. Lenses L1 and L2 are used to perform the Fourier transformation and to focus the transformed light beams. It was further demonstrated that this geometric transformation can, in principle, be achieved using a single optical element, provided that the mapping between the Cartesian and log-polar coordinates is conformal. Experimental results obtained using this approach are shown in Fig.~\ref{Fig18}(B). It is clearly observed that different OAM modes focus at distinct transverse positions on the detection plane. This method not only enables the measurement of the OAM associated with a single vortex mode but also allows for the sorting of multiple OAM modes present in a superposition state (see the last row of Fig.~\ref{Fig18}(B), where the input beam—a superposition of modes with $l=-1$ and $l=+2$—produces two separate spots at the detector plane). Furthermore, such a mapping converts concentric rings in the input plane into parallel lines in the output plane. Despite its simplicity and effectiveness, this method presents one notable limitation: partial overlap of the focused spots corresponding to adjacent OAM states, leading to cross-talk between neighboring modes. However, this issue can be mitigated by introducing an additional phase grating to the transforming optical elements.

\subsubsection{Cylindrical and convex lenses}

In the “generation” section of vortex beams (Subsection~\ref{vortexgeneration}), we discussed in detail how a simple arrangement of two cylindrical lenses can convert a superposition of HG modes into a LG mode. A natural question then arises: Can the same setup be used to detect the TC of vortex beams? The answer is straightforward—yes. To achieve this, one simply reverses the mode conversion process. Specifically, an LG input beam carrying a nonzero TC $l$ is directed into the same mode converter, consisting of two cylindrical lenses separated by a distance of $\sqrt{2}f$, where $f$ is the focal length of each lens. Under this configuration, the mode converter transforms the input LG beam into an HG mode oriented at an angle of $45^{\circ}$ relative to the axes of the lenses. By identifying the resulting HG mode and counting the number of minima (or intensity nulls) between the bright lobes provides the TC information of the original LG vortex beam~\cite{Allen,BEIJERSBERGEN}.  

\begin{figure}[h!]
\centering 
\includegraphics[width=1\linewidth]{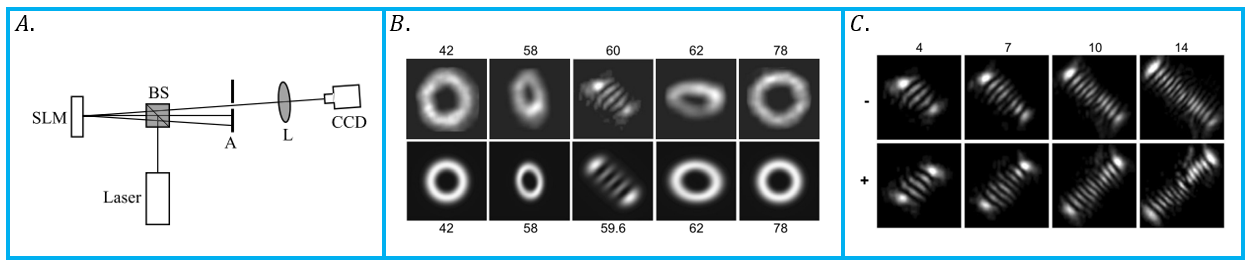}
\caption{Determination of TC values of vortex beams using a tilted convex lens. (A) Experimental set-up. (B) Experimental (top row) and theoretical (bottom row) intensity distributions at different propagation distances (in cm), measured from the tilted lens, for a TC of +4. (C) Experimentally measured intensity distributions at a propagation distance of 60 cm from the tilted lens, for vortices carrying TC values of -4, -7, -10, and -14 (top row) and +4, +7, +10, and +14 (bottom row). Reprinted with permission from Ref.~\cite{VAITY1}.}
\label{Fig19}
\end{figure}

Following the measurement of TC values using two cylindrical lenses, a natural question arises: can the same be achieved with a single lens? Doing so would render the detection setup more compact and less susceptible to aberrations. In 2008, Vaity et al. proposed a remarkably simple method to determine the TC of vortex beams by illuminating a single convex lens—slightly tilted in one of the transverse directions—with an LG mode (see Fig.~\ref{Fig19}(A))~\cite{VAITY1}. They observed that, at a specific distance beyond the lens, a tilted HG-like mode appears, as the Hermite polynomial component in the complex field amplitude of the propagating beam becomes dominant at that plane. By counting the number of intensity minima between the bright lobes of this pattern, one can directly infer the TC of the input vortex beam (see Fig.~\ref{Fig19}(B)). This technique is simple, robust, and versatile, allowing the detection of arbitrary TC values for a wide range of vortex beams.

\subsubsection{Rotational Doppler effect}

The Doppler effect is a well-known phenomenon in both optics and acoustics. In general, it describes the frequency shift of a wave relative to an observer moving with respect to the source. Specifically, from the observer’s point of view, the frequency of the wave increases (decreases) when the source approaches (recedes from) the observer. These descriptions pertain to the translational (or linear) motion of the source or observer. However, when the motion involved is rotational, it can also influence the properties of a beam carrying OAM during their interaction. In this case, the frequency of light is altered through what is known as the rotational Doppler shift~\cite{Martin,Courtial1}, which can be understood as follows: 
\begin{figure}[h!]
\centering 
\includegraphics[width=1\linewidth]{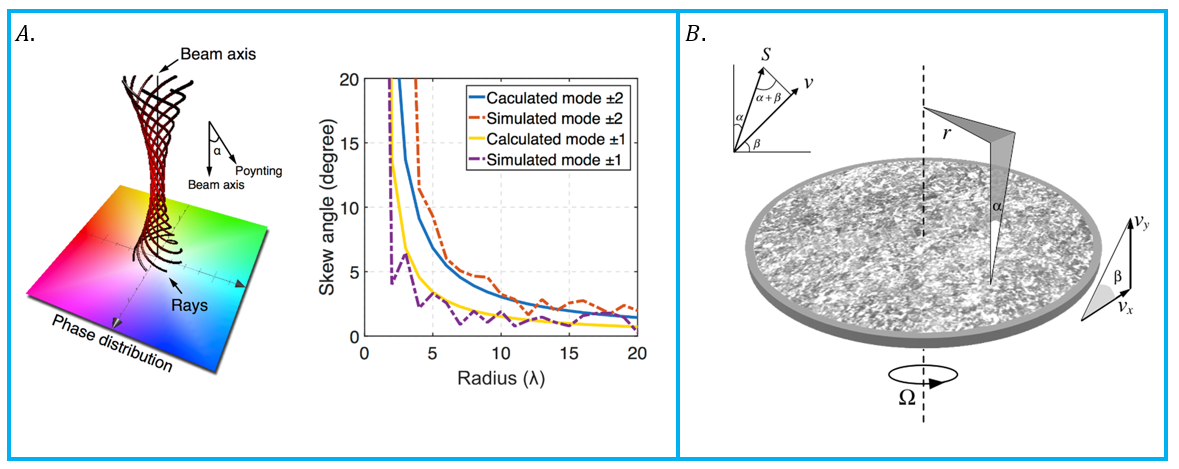}
\caption{Illustration of the skew angle of an OAM beam. (A) Skewing of rays and phase distribution of an OAM beam. (B) Scattering of an OAM beam from a rough surface undergoing rotational and translational motion, leading to multiple Doppler effects. Reprinted with permission from Ref.~\cite{Baiyangliu}.}
\label{Fig20}
\end{figure}
In an OAM-carrying light beam, the helical phase front corresponds to a local skew angle (denoted as $\alpha$, as illustrated in Fig.~\ref{Fig20}(A)) of the Poynting vector $\left( \bm{S} = \frac{1}{\mu_{0}} \bm{E} \times \bm{B} \right)$. This angle is given by $\alpha=\frac{l}{kr}=\frac{l\lambda}{2\pi r}$, where $l$, $k = \frac{2\pi}{\lambda}$, and $\lambda$ represent the TC (OAM), wavenumber, and wavelength of the vortex beam, respectively, while $r$ denotes the radial distance from the beam axis. This relationship implies that an OAM beam possesses an azimuthal component of momentum at every point across its cross-section, and this component is fundamentally responsible for the rotational Doppler effect. In Fig.~\ref{Fig20}(B), we illustrate a rough surface capable of undergoing both transverse rotation and longitudinal translation. When an OAM beam illuminates such a surface, the reflected beam experiences both rotational and linear Doppler shifts, such that the frequency shift of the reflected beam is expressed as~\cite{Baiyangliu}:
\begin{eqnarray}
    \Delta f=\frac{\Delta l \;\Omega}{2\pi}+ \frac{v_{y}f_{0}}{c},
    \label{eqn20}
\end{eqnarray}
where $f_{0}$ is the original frequency of the incident OAM beam, $\Omega$ is the transversal rotation angular frequency, $v_{y}$ is the vertical velocity (as shown in Fig.~\ref{Fig20}(B)), and $c$ is the speed of light. Note that the horizontal component of the linear velocity is, $v_{x}=\Omega r$ and Eq.~(\ref{eqn20}) is derived under the condition $\alpha \ll 1$. It is also important to note that: (1) The first term in Eq.~(\ref{eqn20}) represents the frequency shift of the reflected OAM beam due to the rotational Doppler effect, created by the transversal rotation of the rough surface, which depends only on the OAM variation between the incident and reflected OAM beams i.e., $\Delta l$, and the angular frequency $\Omega$. Furthermore, this component of the frequency shift is independent of the incident OAM beam’s optical frequency, meaning that every spectral component of the reflected beam undergoes the same magnitude of frequency shift. (2) The second term in Eq.~(\ref{eqn20}) corresponds to the linear Doppler effect arising from the longitudinal translation of the surface, and its magnitude depends on the optical frequency of the incident OAM beam.

Furthermore, Eq.~(\ref{eqn20}) is derived under the assumption that the incident OAM beam is linearly polarized. However, when a circularly polarized OAM beam is used, an additional term appears in the expression for the frequency shift in Eq.~(\ref{eqn20}), to account for the change in spin angular momentum between the incident and reflected beams. The rotational Doppler shift has also been observed for white-light beams carrying OAM after backscattering from a rotating object, where it was found that even a broadband (white-light) source can produce a single-valued frequency shift~\cite{Lavery}. Moreover, when the incident beam is a superposition of multiple OAM states, the reflected spectrum exhibits multiple frequency sidebands corresponding to the different OAM components.

\section{Applications of vortex beams}
\label{applications}

Owing to their characteristic doughnut-shaped transverse intensity profile and inherent ability to carry OAM, vortex beams have found widespread applications across numerous fields. These include optical trapping and rotation of microscopic particles, optical communication, microscopy, image processing, high-precision metrology, cryptography, quantum entanglement, quantum logic operations, sensing, chiral studies of magnetic materials, and even astrophysics, among others. In this section, we briefly review some of these representative applications.

\subsection{Optical trapping and rotation of microscopic particles}

\begin{figure}[h!]
\centering 
\includegraphics[width=1\linewidth]{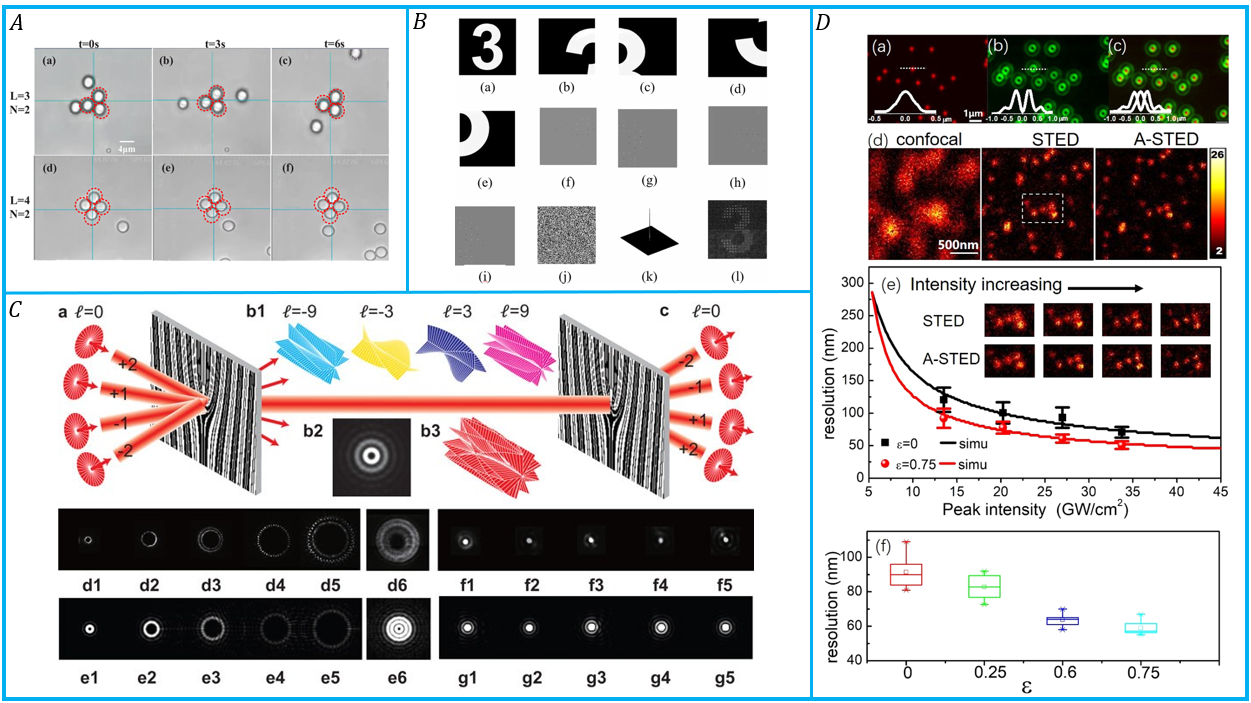}
\caption{Applications of OAM beams in different domains. (A) Optical trapping of multiple particles using OAM beams, taken from Ref.~\cite{Zihengwu}. (B) Optical encryption with OAM beams, taken from Ref.~\cite{Li_2025}. (C) OAM based free-space optical communications, taken from Ref.~\cite{Lei2015}. (D) Improving lateral resolution of images using OAM beams in stimulated emission depletion (STED) microscopy- a super resolution microscopy technique, taken from Ref.~\cite{BinWang}.  Reprinted with permission from Refs.~\cite{Zihengwu,Li_2025,Lei2015,BinWang}.}
\label{Fig21}
\end{figure}

Generally, the size of an optical beam lies in the micrometer range, which coincides with the characteristic scale of microscopic particles. This correspondence makes the study of light–matter interactions more accessible. In particular, when the optical beam is a vortex beam, the transfer of its OAM to microscopic particles becomes especially efficient. In 1995, He et al. demonstrated that absorptive particles trapped within the central dark core of a vortex beam can be set into rotation through the transfer of OAM from the photons to the particles~\cite{HeH}. Moreover, the chirality (or handedness) of the helical wavefront determines the direction of particle rotation. It was also shown that, depending on the particle’s shape and size, and for laser powers of only a few milliwatts, the rotational frequency typically ranges between 1–10 Hz. In Ref.~\cite{Dasgupta}, Dasgupta et al. showed that LG vortex modes with increasing TC values can be used to achieve controlled orientation and rotation of red blood cells (RBCs). They further observed that LG beams with TC values greater than 15 could drive RBCs as natural micro-rotors. In Ref.~\cite{Leimingzhou}, Zhou et al. investigated the sensitivity of displacement detection for large spherical particles using vortex beams and found that, with an appropriate choice of TC, the sensitivity could be enhanced by at least an order of magnitude compared to Gaussian beams. In Ref.~\cite{Bobkova}, Bobkova et al. designed an “optical grinder” by combining two LG beams of different radii and opposite OAM to trap silica spheres of various diameters in the micrometer range. This optical grinder was found to not only trap particles but also sort (or spatially separate) them according to their size. Furthermore, in Ref.~\cite{Zihengwu}, multiple particles were simultaneously trapped using a rotationally symmetric power-exponent-phase vortex beam (RSPEPVB), establishing a one-to-one correspondence between the number of trapped particles and the TC of the RSPEPVB (see Fig.~\ref{Fig21}(A)). It was also demonstrated that the trapped particles do not rotate around the circular light intensity distribution. In Ref.~\cite{Jisenwen}, two spherical microparticles were simultaneously trapped using a pair of vortex beams, and it was shown that the separation distance between the trapped particles could be tuned by adjusting the beam parameters. Finally, perfect vortex beams have also been used to trap microscopic particles arranged in a necklace-like configuration and to induce their continuous rotation~\cite{Mingzhou}.

\subsection{Optical encryption}

In today’s world, data serves as the essential fuel driving innovation and efficiency across all sectors. It empowers scientists to make groundbreaking discoveries, enables governments to enhance public services, and helps businesses better understand customer needs. In other words, data is one of the most valuable assets for any organization or individual. Consequently, securing data over unprotected networks has become a critical necessity. Optical encryption is a powerful tool that addresses this challenge by overcoming the key bottlenecks and vulnerabilities faced by conventional electronic or digital encryption methods. It is not merely about strengthening encryption algorithms mathematically, but about enhancing the physical process of encryption itself—making it faster, more efficient, and inherently more secure for high-demand applications. In simple terms, optical encryption can be defined as a technique for securing data by encoding it within the physical properties of a light beam—such as its amplitude, polarization, phase, wavelength, or spatial distribution. The encoded signal is then transmitted through free space or optical fibers and can only be decrypted by a receiver possessing the correct key.

Optical encryption, in its inception stages, was primarily based on the use of Gaussian beams. However, this approach has several limitations:
(1) Using Gaussian beams, data can only be encoded in a limited set of parameters such as amplitude, wavelength, and polarization.
(2) Because a Gaussian beam’s intensity profile appears identical to an unauthorized observer, the encrypted data is not resilient against simple eavesdropping. Contrariwise, the use of vortex beams offers significant advantages in optical encryption over conventional Gaussian beams for the following reasons:
(1) It enables data encoding in the OAM degree of freedom, which can be a single OAM state or a superposition of multiple states.
(2) Since different OAM states are mutually orthogonal (i.e., no crosstalk between modes), multiple independent data channels can be transmitted simultaneously on the same wavelength, with each channel assigned a distinct OAM state. This greatly enhances the overall channel capacity. Moreover, the encrypted message and key can be distributed across different OAM channels.
(3) Vortex beam-based encryption schemes are inherently more resilient to eavesdropping compared to their Gaussian counterparts.

Due to the lack of selectivity for TC values in the Bragg diffraction condition, OAM—despite being an independent physical degree of freedom—was not initially employed as an information carrier in holography. Fang et al.~\cite{XinyuanFang} addressed this limitation with an insightful idea: OAM beams with different TC values exhibit distinct spatial-frequency distributions in Fourier space, with higher TC values corresponding to larger light rings (for example, in the case of LG beams, the radius of maximum intensity scales as the square root of the TC). The authors exploited this property to design a sampling array specific to each OAM state, with the array period matched to the spatial-frequency distribution of the incident OAM beam. This innovation enabled the realization of OAM-dependent holography. Until 2020, most OAM holography systems relied solely on phase-only metasurfaces. However, Ren et al. were the first to demonstrate a complex-amplitude metasurface-based OAM holography technique capable of multiplexing up to 200 independent OAM channels—making it highly promising for large-capacity OAM-based encryption~\cite{HaoranRen}. In 2021, Zhu et al. further advanced conventional OAM holography by introducing radial modulation as an additional degree of freedom~\cite{Zhu:21}. They demonstrated that: (1) this new degree of freedom significantly enhances both the holographic capacity and image fidelity, and (2) the space–bandwidth product efficiency improves by a factor of 20 compared to traditional phase-only metasurface-based OAM holography. In Ref.~\cite{Li_2025}, Li et al. proposed an optical encryption system that integrates OAM holography with a nonlinear authentication mechanism, showing excellent performance for secure data transmission (see Fig.~\ref{Fig21}(B)). It is well known that the annular intensity profiles of perfect vortex beams remain nearly unchanged with increasing TC in their true (asymptotic) field representation. Leveraging this unique property, Yang et al. achieved ultra-secure image encryption by tightly focusing perfect vortex beams—demonstrating potential applications in multiplexed data storage, optical communication, and beyond~\cite{Qingshuai}.    

\subsection{Optical communication}

Optical communication, in its broadest sense, dates back to ancient times, when people used smoke signals or mirrors to convey information across distances. However, in such early methods, the amount of information that could be transmitted was inherently limited. Over the centuries, continuous efforts were made to increase the volume and speed of data transmission. A major breakthrough came in 1792, when Claude Chappe introduced the concept of transmitting mechanically coded messages over long distances using intermediate relay stations—a system he called the optical telegraph~\cite{Chappe}. In these systems, light was used to make the coded signals visible so they could be relayed from station to station. Nevertheless, the achievable bit rate (i.e., the number of bits processed per unit time) was only about 1 bit/s. Later, the development of coaxial cables and microwave communication systems improved data transmission rates to nearly 100 Mbit/s. However, the advent of optical waves for communication brought an unprecedented leap forward—offering exceptionally high bandwidth due to their high carrier frequency, minimal transmission loss when propagated through optical fibers, and strong immunity to electromagnetic interference.

As discussed earlier, LG modes form an orthonormal basis. Owing to this property, it was predicted that they could introduce a new form of diversity in optical telecommunications. It is well known that there is no theoretical limit to the TC of vortex beams; in other words, the TC can range from $-\infty$ to $\infty$ for a given optical frequency. Consequently, an infinite number of modes can, in principle, coexist at the same frequency. Since these modes are mutually orthogonal, each can serve as an independent communication channel, allowing—at least theoretically—an infinite number of multiplexed modes to dramatically increase the system’s channel capacity. In practice, however, the number of usable or multiplexed modes is limited by an inherent characteristic of conventional vortex beams, such as LG and BG modes: their beam size intrinsically depends on the TC. To overcome this constraint in communication systems, perfect vortex beams are employed, as their size remains independent of the TC.

Owing to their ability to carry OAM, vortex beams are known to exhibit greater resilience to atmospheric turbulence compared to conventional Gaussian beams~\cite{Gbur}. In Ref.~\cite{Krenn}, Krenn et al. demonstrated that the relative phase of superposed OAM modes remains largely unaffected by atmospheric disturbances. The authors successfully distinguished between 16 different OAM mode superpositions with an exceptionally low error rate, thereby establishing the feasibility of conducting long-distance quantum experiments using OAM beams. In Ref.~\cite{Mirhosseini_2015}, the authors utilized the OAM degree of freedom of light for information encoding and showed that the quantum key distribution (QKD) transmission rate can be significantly enhanced in such systems. They also discussed how QKD protocols based on spatial-mode encoding offer improved resistance to eavesdropping attacks. From the fundamental principles of nonlinear optics and quantum information science, it is known that entanglement between two photons in the OAM degree of freedom naturally arises in the process of spontaneous parametric down-conversion, due to the momentum conservation—specifically, transverse momentum. By exploiting this intrinsic OAM entanglement, quantum teleportation can be successfully realized~\cite{Wangxilin}. In Ref.~\cite{Lei2015}, the authors proposed a Dammann optical vortex grating and demonstrated its capability for multiplexing a large number of OAM channels, enabling individual channel modulation and simultaneous detection of all channels (see Fig.~\ref{Fig21}(C)). As a proof-of-concept, they demonstrated the simultaneous detection of 10 OAM channels, with an OAM interval of 6 between the consecutive channels to reduce the cross-talk between them.

\subsection{Microscopy and imaging}

Microscopy is the practice of observing objects—or regions within objects—that are invisible to the naked eye, using a microscope. Fundamentally, the goal of microscopy is twofold: to magnify the object and to improve resolution, i.e., the ability to distinguish two closely spaced points as separate. For more than a century, it was considered a fundamental principle of physics that a conventional light microscope (such as those commonly used in school laboratories) could not resolve objects smaller than roughly half the wavelength of the light used for illumination. This restriction is known as the Abbe diffraction limit, expressed as
$\left(d=\frac{\lambda}{2 a}\right)$, where $\lambda$ is the wavelength of light and $a$ is the numerical aperture of the objective lens. This expression represents the theoretical maximum resolution achievable by any optical imaging system. A direct consequence of this diffraction limit is that a laser beam cannot be focused to a spot smaller than about half its wavelength. For example, visible light has wavelengths in the range of 400–700 nm, implying that structures smaller than approximately 200 nm cannot be resolved using traditional optical microscopy. This poses a serious challenge in biological imaging, where many critical structures—such as proteins, viruses, and synapses between neurons—fall within the 1–120 nm size range. Over time, various near-field and far-field techniques have been developed to enhance lateral resolution. Among far-field approaches, methods such as 4Pi microscopy~\cite{Hell:92}, I5M~\cite{Gustafsson}, and structured illumination microscopy (SIM)~\cite{Gustafsson1} have been shown to improve resolution by roughly a factor of two. Although these techniques extend the Abbe limit, they still remain fundamentally constrained by its principles. Similarly, near-field methods such as total internal reflection fluorescence (TIRF)~\cite{AMBROSE1956} and near-field scanning optical microscopy (NSOM)~\cite{Eric} have achieved sub-diffraction resolution, but their applicability is limited—for instance, they can probe only near-surface regions of biological samples and not the interior of cells.

A major breakthrough in this field occurred in 1994, when Stefan W. Hell and Jan Wichmann developed stimulated emission depletion (STED) microscopy~\cite{Hell:94}, which was later experimentally demonstrated in 1999~\cite{Klar:99}. STED microscopy is one of several super-resolution imaging techniques, and its underlying principle can be understood as follows: the method employs two laser beams—one to excite fluorescent molecules in a sample, and a second, doughnut-shaped STED beam (also called the depletion beam, which is essentially a vortex beam) to immediately de-excite them. The depletion beam suppresses fluorescence everywhere except at the very center of the doughnut. By effectively reducing the periphery of the excitation spot, the technique confines the region of fluorescence emission to a volume much smaller than that permitted by the conventional diffraction limit. Scanning this tiny, sub-diffraction excitation spot across the sample produces an image of extraordinary sharpness and detail, revealing structures far smaller than those resolvable by standard optical microscopes. Notably, STED microscopy can be applied to samples containing both ensembles of fluorophores and single fluorescent molecules. For this groundbreaking achievement, Eric Betzig, Stefan W. Hell, and W. E. Moerner were jointly awarded the 2014 Nobel Prize in Chemistry “for the development of super-resolved fluorescence microscopy,” as announced by the Royal Swedish Academy of Sciences.

In Ref.~\cite{YanLu}, Yan et al. proposed an all-fiber, compact STED microscopy system based on vortex fiber modes, demonstrating excellent resilience to perturbations. Using their setup, they successfully imaged fluorescent bead samples and achieved a lateral resolution of 103 nm. In Ref.~\cite{BinWang}, Wang et al. employed an annular mask to precisely reshape the doughnut-shaped depletion beam. By compressing the effective point spread function of the STED microscope through a tighter depletion beam profile, they achieved a significant improvement in lateral resolution—down to 40 nm (see Fig.~\ref{Fig21}(D)). It is well known that maintaining consistent lateral resolution deep within a specimen is highly challenging in STED microscopy, primarily due to specimen-induced aberrations and scattering distortions. To overcome these limitations, Yu et al. demonstrated deep-imaging STED microscopy using a Gaussian beam as the excitation source and a hollow BG beam as the depletion beam~\cite{Wentaoyu}. With this approach, the imaging depth was extended up to 115~$\mu$m in both solid agarose and polydimethylsiloxane samples. In Ref.~\cite{Wei:15}, sub-100-nm resolution was achieved in plasmonic structured illumination microscopy (PSIM)—another super-resolution technique—using vortex beams with fractional TC values. The authors further highlighted the potential of their method for low-cost, dynamic biological imaging applications. In Ref.~\cite{Chongleizhang}, perfect vortex beams were employed in PSIM, and it was found that they enhance plasmon excitation efficiency by nearly a factor of six compared to conventional LG beams. In this configuration, a lateral resolution better than 200 nm was achieved.

\section{Propagation of vortex beams in different material media} \label{Propagation of vortex beams}

Understanding the propagation behavior of vortex beams is as crucial as their generation and detection. This is because propagation characteristics hold the key to unlocking their groundbreaking applications in fields such as communications, microscopy, and quantum computing. In Section~\ref{solutions}, we discussed in detail the free-space propagation properties of several types of vortex beams, including LG, BG, POV, and LGa beams. Owing to their distinctive transverse spatial structures, vortex beams interact with matter in ways that fundamental Gaussian beams do not. Consequently, monitoring how their properties evolve during propagation through various media enables us to probe intrinsic material characteristics. In this section, we focus on the paraxial propagation of LG, BG, and POV beams in different material environments—specifically gradient-index (GRIN) and chiral media—using the Huygens–Fresnel diffraction integral, which can be expressed as~\cite{Qusailah}:
\begin{eqnarray}
    u_{0}(\rho,\varphi,z)=\frac{i}{\lambda B}\mathrm{e}^{ikz} \int_{0}^{\infty} \int_{0}^{2\pi}u_{0}(\rho',\varphi') \mathrm{e}^{-\frac{ik}{2B}\left[A\rho'^2-2\rho \rho' \cos(\varphi-\varphi')+D\rho^2\right]} \rho' \mathrm{d}\rho' \mathrm{d}\varphi',
    \label{eqn21}
\end{eqnarray}
where $A,B,C,$ and $D$ are elements of the ray transfer matrix representing the physical optical system.

\subsection{Propagation of LG, BG and POV beams in GRIN media}

A GRIN medium is a material whose refractive index varies continuously in space, typically following a quadratic profile. Owing to its inherent self-focusing ability, the GRIN medium has found wide applications in areas such as optical communication and optical sensing~\cite{Song:98,Tuchida}. Depending on the spatial distribution of the refractive index gradient, GRIN media are generally categorized into radial, axial, and spherical types. In practice, a graded-index optical fiber often serves as a convenient realization of a GRIN medium. To analyze the propagation behavior of various vortex beams within such a medium, we consider the refractive index in the following form~\cite{Bikash3}:
\begin{eqnarray}
    \mu(\rho)=\mu_{0}\left(1-\frac{1}{2}\beta^2 \rho^2\right), \beta^2 \rho^2 \ll 1,
    \label{eqn22}
\end{eqnarray}
where $\beta$ is the gradient-index parameter, $\mu_{0}$ is the base refractive index (along the symmetry axis i.e., the $z$-axis in our case). Basically, Eq.~(\ref{eqn22}) represents a radial-type GRIN medium. The ray transfer matrix of a GRIN medium is typically written as~\cite{Bikash3}:
\begin{eqnarray}
\label{grinmat}
    \begin{pmatrix}
        A & B\\
        C & D
    \end{pmatrix}
    =
    \begin{pmatrix}
        \cos(\beta z) & \frac{\sin(\beta z)}{\mu_{0}\beta}\\
        -\mu_{0}\beta \sin(\beta z) & \cos(\beta z)
    \end{pmatrix}.
    \label{eqn23}
\end{eqnarray}
Next, we employ the Huygens–Fresnel diffraction integral, Eq.~(\ref{eqn21}), in conjunction with the ray transfer matrix, Eq.~(\ref{grinmat}), to investigate the propagation features of various types of vortex beams.

\subsubsection{LG beams propagating in a radial GRIN medium}
The complex field amplitude of an LG beam (with a zero radial index) at the source plane i.e., at $z=0$, can be expressed as:
\begin{eqnarray}
    u_{0,LG}(\rho',\varphi')=A_{0} \left(\frac{\sqrt{2}\rho'}{w_{0}}\right)^{|l|} \mathrm{e}^{-\frac{\rho'^2}{w_{0}^2}} \mathrm{e}^{il\varphi'},
    \label{eqn24}
\end{eqnarray}
where all the beam parameters are explained in detail in Section~\ref{solutions}. Now, after substituting Eq.~(\ref{eqn24}) in Eq.~(\ref{eqn21}) and evaluating the radial and angular integrals analytically, we obtain:
\begin{eqnarray}
 u_{0,LG-GRIN}(\rho,\varphi,z)&=&\frac{i^{l+1}A_{1}k}{B} \mathrm{e}^{il\varphi} \mathrm{e}^{ikz} \mathrm{e}^{-\frac{ikD\rho^2}{2B}} \frac{\left(\frac{k\rho}{B}\right)^l \Gamma\left(\frac{l+|l|+2}{2}\right)}{2^{l+1}\left(\frac{1}{w_{0}^{2}}+\frac{ikA}{2B}\right)^{\left(\frac{l+|l|+2}{2}\right)} \Gamma(l+1)} \nonumber \\
 &\times&
 {_1F_1}\left[\frac{l+|l|+2}{2}; l+1;-\frac{\left(\frac{k\rho}{B}\right)^2}{4\left(\frac{1}{w_{0}^{2}}+\frac{ikA}{2B}\right)}\right],
 \label{eqn25}
\end{eqnarray}
where $\Gamma(\ldots)$ and ${_1F_1}(\ldots)$ denote the Gamma and the confluent hypergeometric functions, respectively, and all constant factors are absorbed into $A_{1}$. Using Eq.~(\ref{eqn25}), we can readily visualize both the intensity ($|u_{0,LG-GRIN}(\rho,\varphi,z)|^2$) and phase (Arg[$u_{0,LG-GRIN}(\rho,\varphi,z)$]) distributions of the LG beam propagating through a GRIN medium. To capture the essential propagation characteristics of the LG beam, we focus only on the longitudinal intensity distributions ($y$ vs $z$) by setting $x=0$. 
\begin{figure}[h!]
\centering 
\includegraphics[width=1\linewidth]{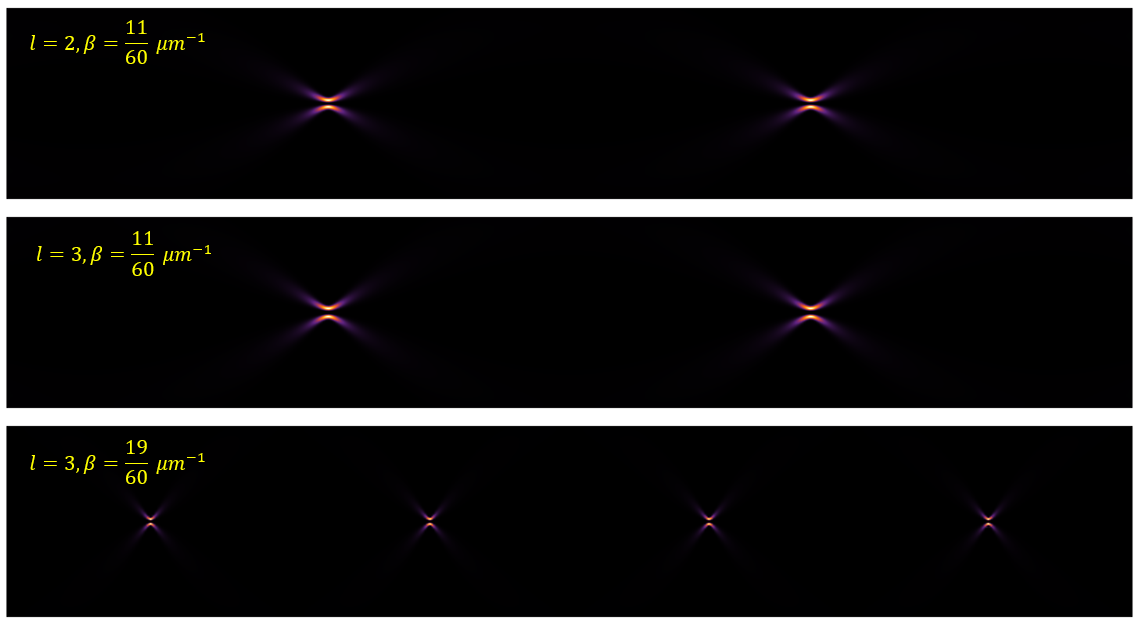}
\caption{Propagation of an LG beam in a GRIN medium for different values of the TC ($l$) and gradient-index parameter ($\beta$). (Top) $l=2$, $\beta=11/60$~$\mu m^{-1}$, (Middle) $l=3$, $\beta=11/60$~$\mu m^{-1}$, (Bottom) $l=3$, $\beta=19/60$~$\mu m^{-1}$}
\label{Fig22}
\end{figure}
The corresponding results are presented in Fig.~\ref{Fig22}. As shown in Fig.~\ref{Fig22} (Top), the LG beam exhibits periodic refocusing behavior in the GRIN medium for a TC $l=2$ and a gradient-index parameter $\beta=11/60$~$\mu$m$^{-1}$. The beam size reaches a minimum at the focusing region, accompanied by a corresponding enhancement in beam intensity. The periodicity of this focusing, however, is determined solely by the gradient-index parameter. To substantiate this, we increase the TC from $l=2$ to $l=3$ while keeping $\beta$ fixed. As seen in Fig.~\ref{Fig22} (Middle), the dark core expands and the radius of maximum intensity increases, yet the focusing periodicity remains unchanged. We then vary the gradient-index parameter to explore its influence on beam propagation. Keeping the TC fixed at $l=3$, we increase $\beta$ from $11/60$~$\mu$m$^{-1}$ to $19/60$~$\mu$m$^{-1}$. As illustrated in Fig.~\ref{Fig22} (Bottom), the focusing periodicity decreases markedly, revealing an inverse relationship between the periodicity and the gradient-index parameter. Notably, the LG beam preserves its single-ring intensity distribution throughout propagation in the GRIN medium, a feature that can be particularly advantageous for applications such as fiber-optic communications and optical beam guiding.

\subsubsection{BG beams propagating in a radial GRIN medium}

\begin{figure}[h!]
\centering 
\includegraphics[width=1\linewidth]{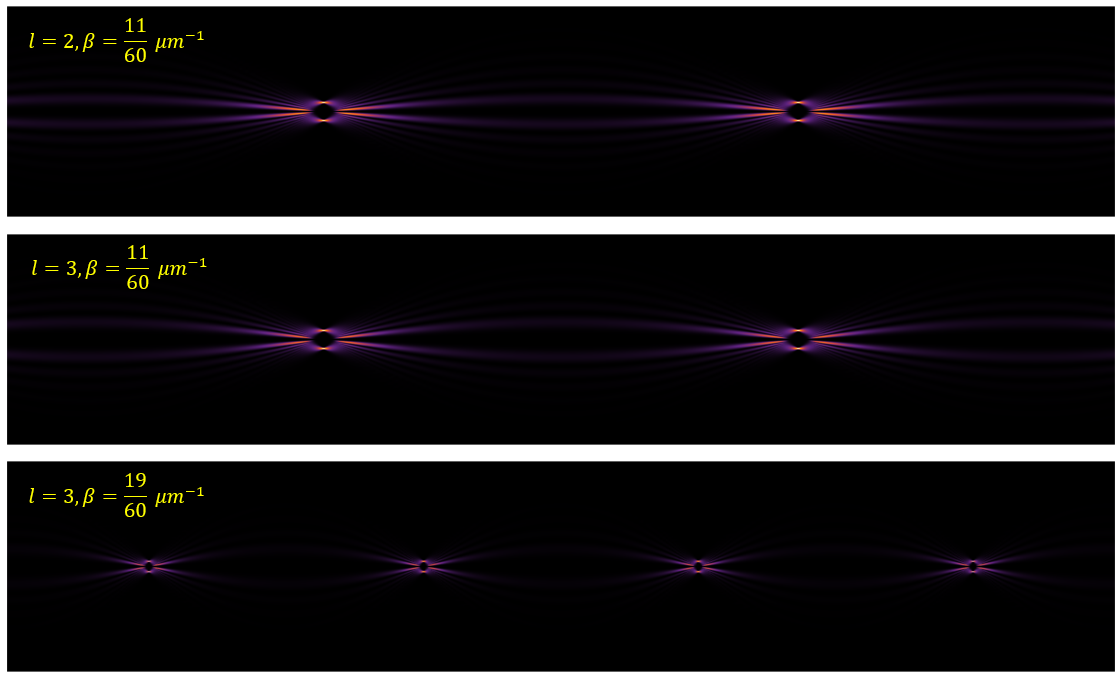}
\caption{Propagation of an BG beam in a GRIN medium for different values of the TC ($l$) and gradient-index parameter ($\beta$). (Top) $l=2$, $\beta=11/60$~$\mu m^{-1}$, (Middle) $l=3$, $\beta=11/60$~$\mu m^{-1}$, (Bottom) $l=3$, $\beta=19/60$~$\mu m^{-1}$.}
\label{Fig23}
\end{figure}

At the source plane, the complex field amplitude of the BG beam can be expressed as:
\begin{eqnarray}
 u_{0,BG}(\rho',\varphi')=A_{0} J_{l}(k_{r}\rho') \mathrm{e}^{-\frac{\rho'^2}{w_{0}^2}} \mathrm{e}^{il\varphi'},
 \label{eqn26}
\end{eqnarray}
where $J_l(\ldots)$ is the Bessel function of the first kind of order $l$. Note that a detailed description on the beam parameters of the BG beam is presented in Section~\ref{solutions}. After substituting Eq.~(\ref{eqn26}) in Eq.~(\ref{eqn21}) and evaluating the radial and angular integrals, we obtain the following expression for the BG beam propagating in the GRIN medium:
\begin{eqnarray}
    u_{0,BG-GRIN}(\rho,\varphi,z)&=& \frac{i^{l+1}A_{0}k}{B} \mathrm{e}^{il\varphi} \mathrm{e}^{ikz} \mathrm{e}^{-\frac{ikD\rho^2}{2B}} \frac{1}{2\left(\frac{1}{w_{0}^{2}}+\frac{ikA}{2B}\right)} \mathrm{e}^{-\frac{k_{r}^2+\left(\frac{k\rho}{B}\right)^2}{4\left(\frac{1}{w_{0}^{2}}+\frac{ikA}{2B}\right)}} \nonumber \\
    &\times&
    I_{l}\left[\frac{k_{r}k\rho/B}{2\left(\frac{1}{w_{0}^{2}}+\frac{ikA}{2B}\right)}\right],
    \label{eqn27}
\end{eqnarray}
 where $I_l(\ldots)$ is the modified Bessel function of the first kind
of order $l$. Using Eq.~(\ref{eqn27}) and by setting $x=0$, we obtain the longitudinal intensity distributions ($|u_{0,BG-GRIN}(\rho,\varphi,z)|^2$) shown in Fig.~\ref{Fig23}.

From Fig.~\ref{Fig23} (Top), it is evident that the BG beam also exhibits a periodic refocusing behavior during its propagation through the GRIN medium. However, several key differences emerge when compared with its LG counterpart: (1) During propagation, the BG beam forms a single-ring intensity pattern at the focusing region, whereas its initial profile at $z=0$ displays multiple off-axis rings. In contrast, the LG beam maintains a single-ring intensity pattern both at $z=0$ and at the focusing region. (2) For the LG beam, the radius of the maximum-intensity ring at the focusing region increases monotonically with the TC, while for the BG beam, this radius remains nearly invariant with respect to the TC (see Figs.~\ref{Fig23} (Top) and (Middle)). It is well-known that BG and POV beams are Fourier pairs. Hence, the emergence of a single-ring intensity pattern in the focusing region during the propagation of the BG beam in the GRIN medium can be attributed to the far-field transformation of the BG beam into its Fourier counterpart, namely, the POV beam. This interpretation is further supported by the observation that the radius of the maximum-intensity ring remains unchanged as the TC increases—a distinctive characteristic of the POV beam. Meanwhile, the overall size of the BG beam profile expands with increasing TC in regions away from the focusing point. Similar to the LG beam case, the periodicity of the focusing behavior is governed solely by the gradient-index parameter (see Figs.~\ref{Fig23} (Middle) and (Bottom)). 

\subsubsection{POV beams propagating in a radial GRIN medium}

\begin{figure}[h!]
\centering 
\includegraphics[width=1\linewidth]{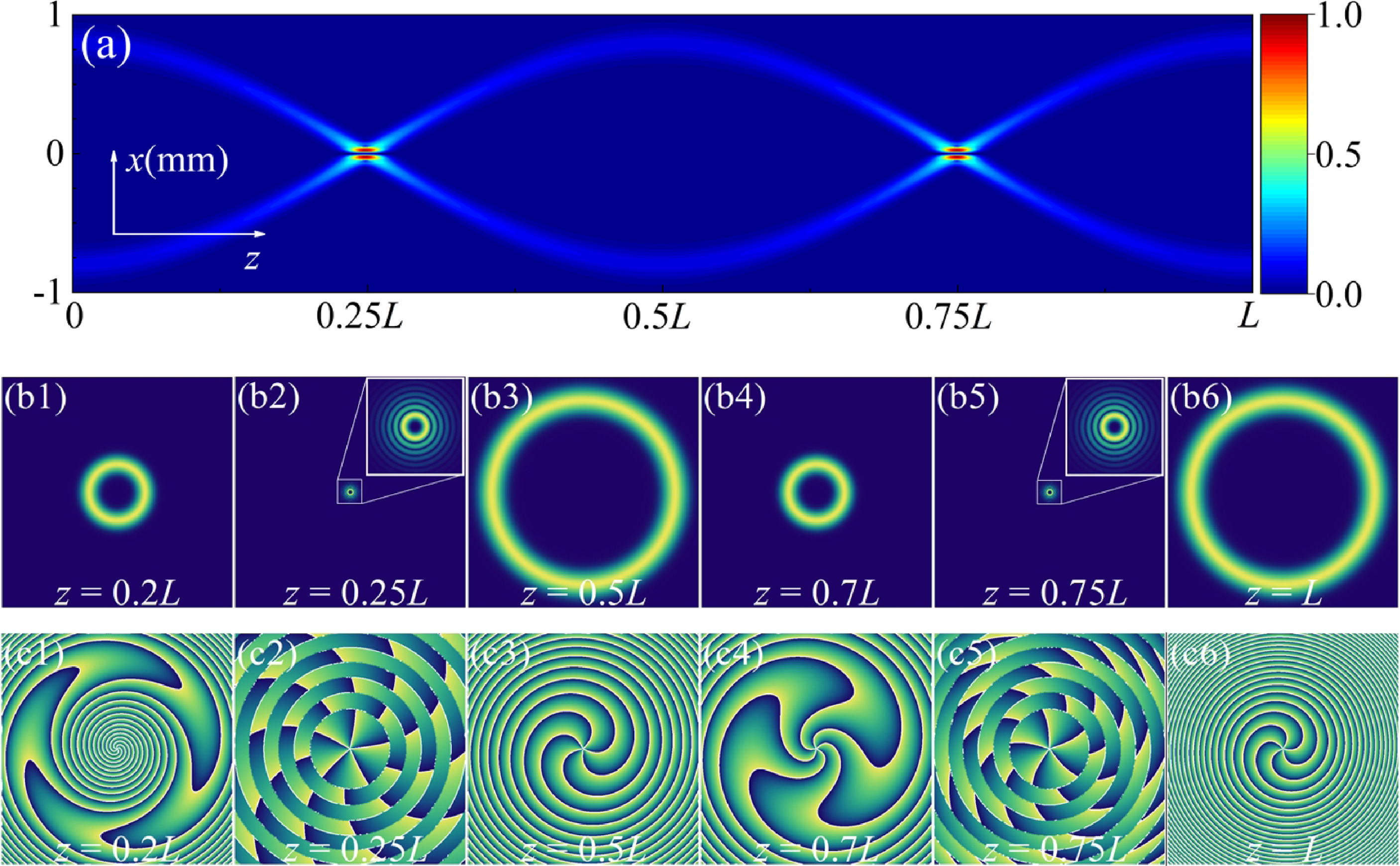}
\caption{Propagation of a POV beam in a GRIN medium. (a) Longitudinal intensity, (b1)-(b6) Transverse intensity, (c1)-(c6) Transverse phase distributions of the POV beam propagating in the GRIN medium. Reprinted with permission from Ref.~\cite{Hanghang}.}
\label{Fig24}
\end{figure}

As already defined in Section~\ref{solutions} (see Eq.~(\ref{eqn13})), the complex field amplitude of the POV beam at the source plane can be written as:
\begin{eqnarray}
    u_{0,POV}(\rho',\varphi')=A_{POV} \mathrm{e}^{-\frac{\rho'^2}{w_{0}^2}} I_{l}\left(\frac{2 \rho_{0}\rho'}{w_{0}^2}\right) \mathrm{e}^{il\varphi'},
    \label{eqn28}
\end{eqnarray}
where $A_{POV}$ absorbs all constant prefactors. Substituting Eq.~(\ref{eqn28}) into Eq.~(\ref{eqn21}) and evaluating of the radial and angular integrals results in:
\begin{eqnarray}
    u_{0,POV-GRIN}(\rho,\varphi,z)&=&\frac{i^{l+1}A_{1}k}{B} \mathrm{e}^{il\varphi} \mathrm{e}^{ikz} \mathrm{e}^{-\frac{ikD\rho^2}{2B}} \frac{1}{2\left(\frac{1}{w_{0}^{2}}+\frac{ikA}{2B}\right)} \mathrm{e}^{\frac{\left(\frac{2\rho_{0}}{w_{0}^2}\right)^2-\left(\frac{k\rho}{B}\right)^2}{4\left(\frac{1}{w_{0}^{2}}+\frac{ikA}{2B}\right)}} \nonumber \\
    &\times&
    J_{l}\left[\frac{\left(\frac{2\rho_{0}}{w_{0}^2}\right) \left(\frac{k\rho}{B}\right)}{2\left(\frac{1}{w_{0}^{2}}+\frac{ikA}{2B}\right)}\right].
    \label{eqn29}
\end{eqnarray}
Basically, Eq.~(\ref{eqn29}) is used to plot the longitudinal intensity, transverse intensity, and transverse phase distributions of the POV beam propagating in the GRIN medium (see Fig.~\ref{Fig24}). It can be seen from Fig.~\ref{Fig24}(a) that the POV beam also undergoes periodic refocusing in the GRIN medium just like LG and BG beams. Therefore, it can be safely concluded that any optical beam (with or without having OAM) propagating in the GRIN media shows periodic refocusing behavior. Furthermore, in the region of focusing, the propagating beam exhibits a multi-ring intensity pattern i.e., the POV degrades into the BG beam.
 This can be clearly seen from Figs.~\ref{Fig24} (a), (b2) and (b5). However, after the focusing region, the beam shows a single-ring intensity pattern with a gradual enhancement of the beam profile size. In this case, the gradient-index parameter solely decides the periodicity of refocusing as well, just like the LG and BG beams cases. It is important to highlight that a unique propagation property of POV beams is that they undergo self-focusing or auto-focusing (i.e., inward diffraction of POV beams) without showing any periodic behavior, even when propagating in free-space. It is the GRIN medium which is responsible for their periodic focusing. However, LG and BG beams do not show a self-focusing effect naturally (let's say in free-space) unless some material media like the GRIN medium modify their propagation dynamics and induce this focusing behavior. From the free-space propagation of POV beams, we know that the ratio between the ring radius an the half-ring width ($\rho_{0}/w_{0}$) of the POV beam controls its self-focusing behavior. When this ratio is much larger than one, the self-focusing effect is clearly observed along with the formation of a multi-ring intensity pattern. However, when ($\rho_{0}/w_{0}$) approaches unity, the self-focusing effect gradually diminishes and completely disappears when the ratio becomes unity (i.e., the propagated beam behaves just like an ordinary LG vortex beam with one single-ring in the intensity distribution). Therefore, in the propagation of POV beams in the GRIN medium, it can be expected that the multi-ring patterns in the focusing regions remain intact as long as the ratio, ($\rho_{0}/w_{0}$), is well above one.       

\subsection{Propagation of LG, BG and POV beams in chiral media}

An object is said to be chiral if it cannot be made to coincide with its mirror image through any combination of translation or rotation. To understand this concept more intuitively, consider a simple example: when you look into a mirror, you see your reflection—your mirror image. Now imagine that this mirror image becomes a real three-dimensional (3D) object. One might ask, “Can this mirror image be superposed on the original you such that every feature of the reflection matches perfectly with the original?” The answer is simply no. You and your mirror image are non-superposable. For instance, if you wear a ring on the ring finger of your left hand, your mirror image will have it on the ring finger of the right hand. In other words, you and your reflection act as two distinct objects. This, in essence, is the idea of chirality. Chiral objects are widespread in nature—examples include DNA, proteins, and numerous drugs such as ibuprofen and penicillamine, among others.

Understanding the propagation dynamics of optical beams—with or without OAM—is essential, as it paves the way for numerous applications in fields such as optics, biochemistry, and medicine. It is well known that linearly polarized light, when propagating through a chiral medium, undergoes polarization rotation due to the medium’s optical activity. Consequently, the incident beam splits into left- (LCP) and right-circularly polarized (RCP) components. These components propagate with different phase velocities and follow distinct trajectories because they experience different effective refractive indices within the chiral medium. In this review, we explore how various types of optical vortex beams behave during propagation through a chiral medium. In particular, we focus on the propagation dynamics of LG, BG, and POV beams, while acknowledging that a vast body of literature also exists on the propagation of other vortex beam types in chiral media.

The ABCD transfer matrix of the optical system in the chiral medium can be described as~\cite{Bikash2}:
\begin{eqnarray}
    \begin{pmatrix}
    A_{L} & B_{L}\\
    C_{L} & D_{L}
    \end{pmatrix}
    =
    \begin{pmatrix}
    1 & \frac{z}{n_{L}}\\
    0 & 1
    \end{pmatrix}, \nonumber\\
    \begin{pmatrix}
    A_{R} & B_{R}\\
    C_{R} & D_{R}
    \end{pmatrix}
    =
    \begin{pmatrix}
    1 & \frac{z}{n_{R}}\\
    0 & 1
    \end{pmatrix},
    \label{eqn30}
\end{eqnarray}
where the subscripts $L$, and $R$ denote the LCP, and RCP beams, respectively. Here, $n_{L}=\frac{n_{0}}{1+n_{0}k\gamma}$, and $n_{R}=\frac{n_{0}}{1-n_{0}k\gamma}$ represent the effective refractive indices of the LCP, and RCP beams, respectively. Finally, $n_{0}$ denotes the base refractive index of the chiral medium, and $\gamma$ is the chiral parameter.

\subsubsection{LG beams propagating in a chiral medium}
The total complex field of the LG beam propagating in a chiral medium can be expressed as:
\begin{eqnarray}
    u_{0,LG-Chiral-Total}(\rho,\varphi,z)&=& u_{0,LG-Chiral-Left}(\rho,\varphi,z) \nonumber \\
    &+& u_{0,LG-Chiral-Right}(\rho,\varphi,z),
    \label{eqn31}
\end{eqnarray}
\begin{figure}[h!]
\centering 
\includegraphics[width=1\linewidth]{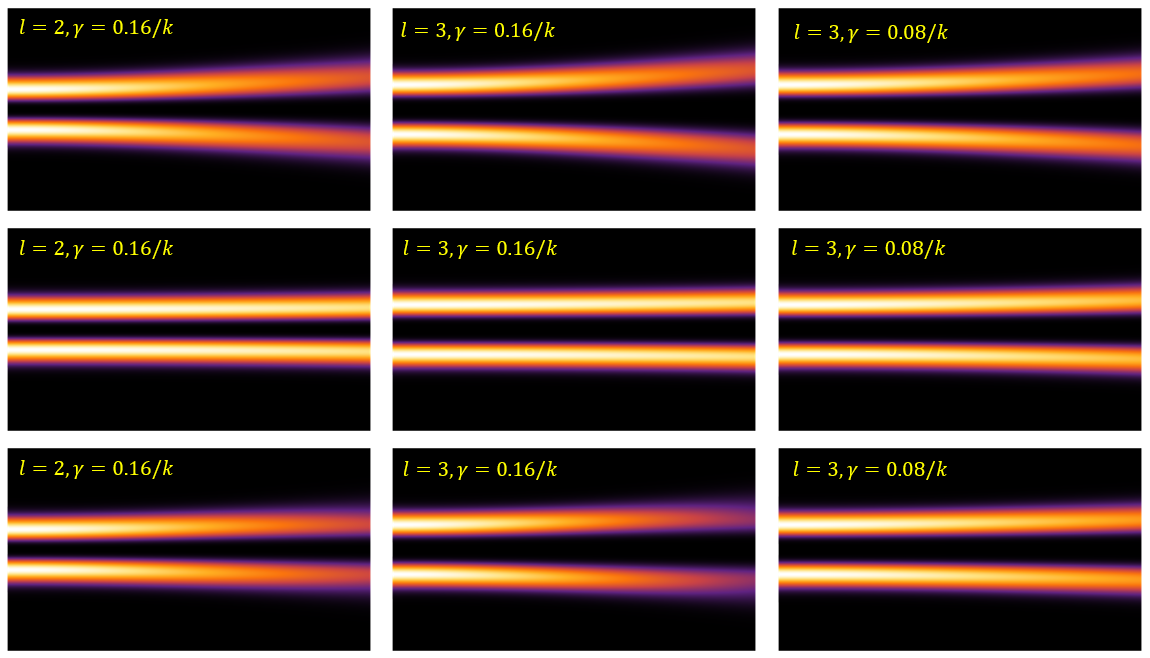}
\caption{Propagation of Lg beams in a chiral medium. (First row) Longitudinal intensity distributions of the LCP LG beam. (Second row) Longitudinal intensity distributions of the RCP LG beam. (Third row) Longitudinal intensity distributions of the total LG beam}
\label{Fig25}
\end{figure}
where,
\begin{eqnarray}
   u_{0,LG-Chiral-Left}(\rho,\varphi,z)&=&\frac{i^{l+1}A_{1}k}{B_{L}} \mathrm{e}^{il\varphi} \mathrm{e}^{ikz} \mathrm{e}^{-\frac{ikD_{L}\rho^2}{2B_{L}}} \frac{\left(\frac{k\rho}{B_{L}}\right)^l \Gamma\left(\frac{l+|l|+2}{2}\right)}{2^{l+1}\left(\frac{1}{w_{0}^{2}}+\frac{ikA_{L}}{2B_{L}}\right)^{\left(\frac{l+|l|+2}{2}\right) \Gamma(l+1)}} \nonumber \\
 &\times&
 {_1F_1}\left[\frac{l+|l|+2}{2}; l+1;-\frac{\left(\frac{k\rho}{B_{L}}\right)^2}{4\left(\frac{1}{w_{0}^{2}}+\frac{ikA_{L}}{2B_{L}}\right)}\right],
 \label{eqn32}
\end{eqnarray}
and
\begin{eqnarray}
   u_{0,LG-Chiral-Right}(\rho,\varphi,z)&=&\frac{i^{l+1}A_{1}k}{B_{R}} \mathrm{e}^{il\varphi} \mathrm{e}^{ikz} \mathrm{e}^{-\frac{ikD_{R}\rho^2}{2B_{R}}} \frac{\left(\frac{k\rho}{B_{R}}\right)^l \Gamma\left(\frac{l+|l|+2}{2}\right)}{2^{l+1}\left(\frac{1}{w_{0}^{2}}+\frac{ikA_{R}}{2B_{R}}\right)^{\left(\frac{l+|l|+2}{2}\right) \Gamma(l+1)}} \nonumber \\
 &\times&
 {_1F_1}\left[\frac{l+|l|+2}{2}; l+1;-\frac{\left(\frac{k\rho}{B_{R}}\right)^2}{4\left(\frac{1}{w_{0}^{2}}+\frac{ikA_{R}}{2B_{R}}\right)}\right].
 \label{eqn33}
\end{eqnarray}
To obtain Eqs.~(\ref{eqn31})–(\ref{eqn33}), we substitute Eq.~(\ref{eqn24}) into Eq.~(\ref{eqn21}) and solve it for the left-, right-, and total LG beam components. Now using Eq.~(\ref{eqn31}), the intensity of the total LG beam can be calculated as:
\begin{eqnarray}
    I_{0,LG-Chiral-Total}(\rho,\varphi,z)&=&|u_{0,LG-Chiral-Left}(\rho,\varphi,z)|^2 \nonumber \\
    &+& |u_{0,LG-Chiral-Right}(\rho,\varphi,z)|^2 \nonumber \\
    &+& \left[u_{0,LG-Chiral-Interference}(\rho,\varphi,z)\right],
    \label{eqn34}
\end{eqnarray}
where the last term in Eq.~(\ref{eqn34}) defines an interference term defined as:
\begin{eqnarray}
   u_{0,LG-Chiral-Interference}(\rho,\varphi,z)&=& [u_{0,LG-Chiral-Left}(\rho,\varphi,z)\nonumber \\ &\times& u_{0,LG-Chiral-Right}^{*}(\rho,\varphi,z)] \nonumber \\
   &+&
   [u_{0,LG-Chiral-Left}^{*}(\rho,\varphi,z) \nonumber \\
   &\times&u_{0,LG-Chiral-Right}(\rho,\varphi,z)].
   \label{eqn35}
\end{eqnarray}
Using Eq.~(\ref{eqn35}), we plot the longitudinal intensity distributions of the LCP (first row), RCP (second row), and total (third row) LG beams propagating in the chiral medium by setting $x=0$ (see Fig.~\ref{Fig25}). From these plots, it is evident that, in all cases, the beam maintains its non-diffracting nature over a certain propagation distance before gradually diffracting away from the beam center. The propagation dynamics of the LCP component can be summarized as follows:
(1) As the TC increases, the non-diffracting propagation distance extends (see Fig.~\ref{Fig25}, first row). This behavior arises from the fact that, for a zero radial index, the width of the LG beam scales with the square root of the TC. Consequently, a higher TC leads to a broader beam. It is well known that the Rayleigh range ($z_R$) determines the non-diffracting propagation distance of an optical beam and scales with the square of the beam width for a fixed wavelength and refractive index of the medium. Therefore, an increase in the TC enhances the beam width and, in turn, extends the Rayleigh range.
(2) A decrease in the chiral parameter results in an increase in the non-diffracting propagation distance. This can be understood as follows: for the LCP component, the effective refractive index ($n_L$) increases as the chiral parameter ($\gamma$) decreases, assuming a fixed base refractive index ($n_0$) and beam wavelength. Since the Rayleigh range scales linearly with the effective refractive index, a smaller chiral parameter leads to a larger Rayleigh range and hence a longer non-diffracting propagation distance. 

Similarly, the propagation characteristics of the RCP component can be summarized as follows: (1) An increase in the TC leads to a longer non-diffracting propagation distance, following the same reasoning discussed for the LCP component.
(2) A decrease in the chiral parameter results in a reduction of the non-diffracting distance (see Fig.~\ref{Fig25}, second row). This occurs because, for a fixed base refractive index and beam wavelength, the effective refractive index of the RCP component ($n_{R}$) decreases as the chiral parameter ($\gamma$) decreases. Consequently, the Rayleigh range of the propagating beam diminishes, leading to a shorter non-diffracting propagation distance. It is also worth noting that, for a given set of beam and medium parameters, the RCP component consistently exhibits a longer non-diffracting distance compared to its LCP counterpart. This can be attributed to the fact that the effective refractive index of the RCP component is always greater than that of the LCP component (see Fig.~\ref{Fig25}, first and second rows).

Since the LCP and RCP components propagate at different phase velocities in a chiral medium—owing to their distinct effective refractive indices—the resulting relative phase difference between them leads to interference effects. These interference effects cause the total beam to exhibit propagation trajectories that differ markedly from those of the individual LCP and RCP components (see Fig.~\ref{Fig25}, third row).

\subsubsection{BG beams propagating in a chiral medium}

\begin{figure}[h!]
\centering 
\includegraphics[width=1\linewidth]{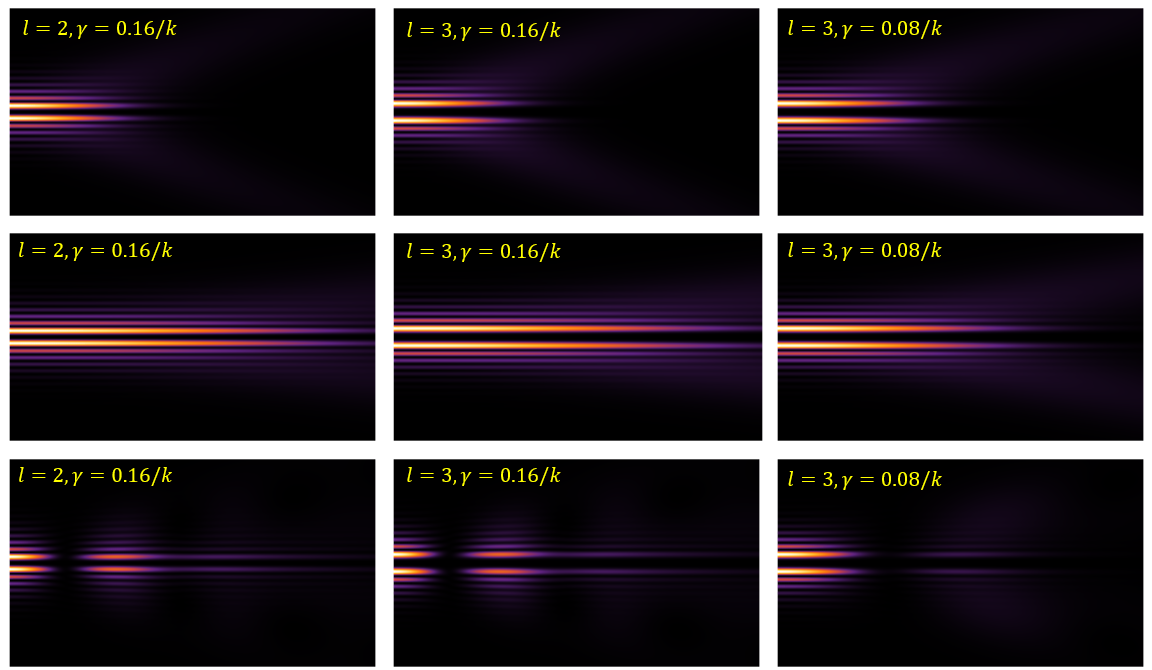}
\caption{Propagation of BG beams in a chiral medium. (First row) Longitudinal intensity distributions of the LCP BG beam. (Second row) Longitudinal intensity distributions of the RCP BG beam. (Third row) Longitudinal intensity distributions of the total BG beam.}
\label{Fig26}
\end{figure}

Following the same procedure outlined for the LG beam in the previous subsection, we derive the analytical expression for the complex field amplitude of the BG beam propagating through a chiral medium:
\begin{eqnarray}
    u_{0,BG-Chiral-Total}(\rho,\varphi,z)&=& u_{0,BG-Chiral-Left}(\rho,\varphi,z) \nonumber \\
    &+& u_{0,BG-Chiral-Right}(\rho,\varphi,z),
    \label{eqn36}
\end{eqnarray}
where
\begin{eqnarray}
    u_{0,BG-Chiral-Left}(\rho,\varphi,z)&=& \frac{i^{l+1}A_{0}k}{B_{L}} \mathrm{e}^{il\varphi} \mathrm{e}^{ikz} \mathrm{e}^{-\frac{ikD_{L}\rho^2}{2B_{L}}} \frac{1}{2\left(\frac{1}{w_{0}^{2}}+\frac{ikA_{L}}{2B_{L}}\right)} \mathrm{e}^{-\frac{k_{r}^2+\left(\frac{k\rho}{B_{L}}\right)^2}{4\left(\frac{1}{w_{0}^{2}}+\frac{ikA_{L}}{2B_{L}}\right)}} \nonumber \\
    &\times&
    I_{l}\left[\frac{k_{r}k\rho/B_{L}}{2\left(\frac{1}{w_{0}^{2}}+\frac{ikA_{L}}{2B_{L}}\right)}\right],
    \label{eqn37}
\end{eqnarray}
and
\begin{eqnarray}
    u_{0,BG-Chiral-Right}(\rho,\varphi,z)&=& \frac{i^{l+1}A_{0}k}{B_{R}} \mathrm{e}^{il\varphi} \mathrm{e}^{ikz} \mathrm{e}^{-\frac{ikD_{R}\rho^2}{2B_{R}}} \frac{1}{2\left(\frac{1}{w_{0}^{2}}+\frac{ikA_{R}}{2B_{R}}\right)} \mathrm{e}^{-\frac{k_{r}^2+\left(\frac{k\rho}{B_{R}}\right)^2}{4\left(\frac{1}{w_{0}^{2}}+\frac{ikA_{R}}{2B_{R}}\right)}} \nonumber \\
    &\times&
    I_{l}\left[\frac{k_{r}k\rho/B_{R}}{2\left(\frac{1}{w_{0}^{2}}+\frac{ikA_{R}}{2B_{R}}\right)}\right],
    \label{eqn38}
\end{eqnarray}
where $I_{l}(\ldots)$ represents the modified Bessel function of the first kind of order $l$. Using Eqs.~(\ref{eqn36})–(\ref{eqn38}), we compute the longitudinal intensity distributions ($|u_{0,BG-Chiral-Total}(\rho,\varphi,z)|^2$) of the BG beam propagating through the chiral medium, and the corresponding results are shown in Fig.~\ref{Fig26}. The main observations can be summarized as follows: (1) Both LCP and RCP BG beams exhibit non-diffracting behavior over a finite propagation distance—determined by their effective refractive indices, the chiral parameter of the medium, and the TC-dependent beam widths—beyond which both components begin to diffract.
(2) As the TC increases, the non-diffracting propagation distance extends for both components, owing to the corresponding increase in beam width with TC.
(3) A reduction in the chiral parameter leads to an increase in the non-diffracting distance for the LCP BG beam, whereas it causes a decrease for the RCP component.
(4) Between the two, the RCP component consistently exhibits a longer non-diffracting distance than the LCP component, a consequence of its higher effective refractive index. (5) The total beam, resulting from the interference between the LCP and RCP components, displays intricate longitudinal intensity distributions.

\subsubsection{POV beams propagating in a chiral medium}
\begin{figure}[h!]
\centering 
\includegraphics[width=1\linewidth]{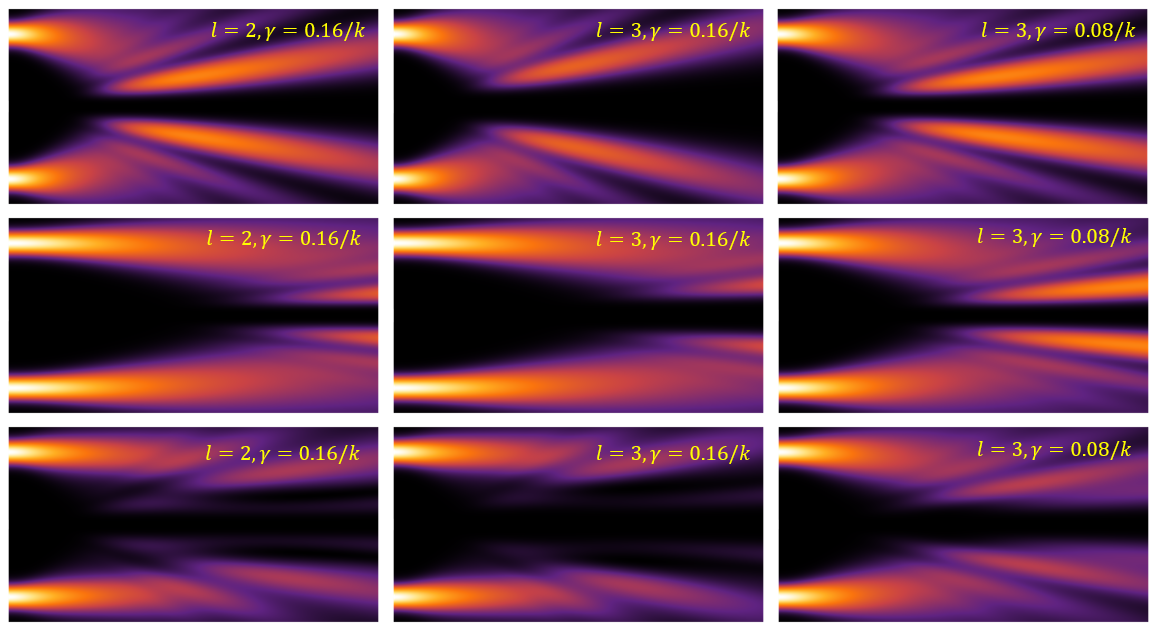}
\caption{Propagation of POV beams in a chiral medium. (First row) Longitudinal intensity distributions of the LCP POV beam. (Second row) Longitudinal intensity distributions of the RCP POV beam. (Third row) Longitudinal intensity distributions of the total POV beam.}
\label{Fig27}
\end{figure}

The complex amplitude of the total POV beam propagating in a chiral medium can be expressed as:
\begin{eqnarray}
    u_{0,POV-Chiral-Total}(\rho,\varphi,z)&=& u_{0,POV-Chiral-Left}(\rho,\varphi,z) \nonumber \\
    &+& u_{0,POV-Chiral-Right}(\rho,\varphi,z),
    \label{eqn39}
\end{eqnarray}
where
\begin{eqnarray}
    u_{0,POV-Chiral-Left}(\rho,\varphi,z)&=&\frac{i^{l+1}A_{1}k}{B_{L}} \mathrm{e}^{il\varphi} \mathrm{e}^{ikz} \mathrm{e}^{-\frac{ikD_{L}\rho^2}{2B_{L}}} \frac{1}{2\left(\frac{1}{w_{0}^{2}}+\frac{ikA_{L}}{2B_{L}}\right)} \mathrm{e}^{\frac{\left(\frac{2\rho_{0}}{w_{0}^2}\right)^2-\left(\frac{k\rho}{B_{L}}\right)^2}{4\left(\frac{1}{w_{0}^{2}}+\frac{ikA_{L}}{2B_{L}}\right)}} \nonumber \\
    &\times&
    J_{l}\left[\frac{\left(\frac{2\rho_{0}}{w_{0}^2}\right) \left(\frac{k\rho}{B_{L}}\right)}{2\left(\frac{1}{w_{0}^{2}}+\frac{ikA_{L}}{2B_{L}}\right)}\right],
    \label{eqn40}
\end{eqnarray}
and
\begin{eqnarray}
    u_{0,POV-Chiral-Right}(\rho,\varphi,z)&=&\frac{i^{l+1}A_{1}k}{B_{R}} \mathrm{e}^{il\varphi} \mathrm{e}^{ikz} \mathrm{e}^{-\frac{ikD_{R}\rho^2}{2B_{R}}} \frac{1}{2\left(\frac{1}{w_{0}^{2}}+\frac{ikA_{R}}{2B_{R}}\right)} \mathrm{e}^{\frac{\left(\frac{2\rho_{0}}{w_{0}^2}\right)^2-\left(\frac{k\rho}{B_{R}}\right)^2}{4\left(\frac{1}{w_{0}^{2}}+\frac{ikA_{R}}{2B_{R}}\right)}} \nonumber \\
    &\times&
    J_{l}\left[\frac{\left(\frac{2\rho_{0}}{w_{0}^2}\right) \left(\frac{k\rho}{B_{R}}\right)}{2\left(\frac{1}{w_{0}^{2}}+\frac{ikA_{R}}{2B_{R}}\right)}\right],
    \label{eqn41}
\end{eqnarray}
where $J_{l}(\ldots)$ represents the Bessel function of the first kind of order $l$. To obtain Eqs.~(\ref{eqn39})–(\ref{eqn41}), we substitute Eq.~(\ref{eqn28}) into Eq.~(\ref{eqn21}) and solve for the LCP, RCP, and total POV beam components. Using these equations, we then compute the longitudinal intensity distributions of the POV beam, with the results shown in Fig.~\ref{Fig27}. 

The main findings from the propagation of POV beams in a chiral medium can be summarized as follows: (1) Both LCP and RCP components of the POV beam exhibit non-diffracting and self-focusing behavior, each occurring at distinct propagation distances. (2) For the LCP component, the non-diffracting propagation distance increases with increasing TC, whereas for the RCP component, it decreases. This behavior arises because, for a POV beam, the beam width scales with the TC according to 
\begin{equation}
 w^2(l)=w_{0}^2(l+1)+\rho_{0}^2 \left(1+\frac{I_{l+1}\left(\frac{\rho_{0}^2}{w_{0}^2}\right)}{I_{l}\left(\frac{\rho_{0}^2}{w_{0}^2}\right)}\right).
 \label{eqn42}
 \end{equation}
Hence, an increase in the TC leads to a larger beam width and, consequently, a longer Rayleigh range.
(3) The RCP component propagates over longer distances without significant changes in its beam characteristics compared to the LCP component.
(4) A decrease in the chiral parameter leads to an increase (decrease) in the non-diffracting propagation distance for the LCP (RCP) component.
(5) When the ratio of the ring radius to the half-ring width ($\rho_{0}/w_{0}$) equals unity, the self-focusing stage vanishes completely from the intensity distributions for both LCP and RCP components.
(6) The total POV beam exhibits intricate longitudinal intensity patterns resulting from the interference between its LCP and RCP components.

Up to this point, our discussion has focused exclusively on vortex beams carrying longitudinal OAM, also known as spatial vortex beams. We now turn our attention to a distinct class of vortex beams, known as spatiotemporal vortex beams, which are characterized by transverse orbital angular momentum and an intrinsic space–time coupling. These unique properties have attracted growing interest within the optics and photonics community, opening promising avenues for novel applications. In the following section, we briefly introduce the fundamental concepts underlying spatiotemporal vortices.

\section{Spatiotemporal optical vortex and pulsed vortex beams}\label{STOV PVB}
This section provides a brief overview of the so-called spatiotemporal optical vortex (STOV) beams, covering their fundamental concept, common generation techniques, and propagation dynamics in various media. The discussion then extends to the broader context of pulsed vortex beams.

\subsection{Transition from longitudinal OAM to transverse OAM}
In Sections~\ref{solutions} -~\ref{Propagation of vortex beams}, the description of vortex beams has been limited to those carrying longitudinal OAM, where the OAM density vector ($\bm{L}_{OAM}=\bm{r}\times\bm{p}$) remains parallel to the beam's propagation direction ($\bm{k}_{z}$), if we assume the $z$-direction to be the beam propagation direction. Here, $\bm{r}$ represents the position vector and $\bm{p}$ is the linear momentum density vector, which is linked to a local wave vector $\bm{k}_{l}$. From the basic definition of the vector product, it is easy to figure out that the direction of $\bm{L}_{OAM}$ is perpendicular to the plane containing $\bm{r}$ and $\bm{p}$. Therefore, $\bm{L}_{OAM}$ remains perpendicular to $\bm{k}_{l}$ as well. Since $\bm{L}_{OAM} \parallel \bm{k}_{z}$ for longitudinal OAM beams, the direction of $\bm{k}_{l}$ remains perpendicular to the direction of $\bm{k}_{z}$. In this way, there exists no coupling between these components. Such optical beams are characterized by an azimuthal phase that depends solely on the transverse spatial coordinates and are well-explained within the paraxial approximation. 

\begin{figure}[h!]
\centering 
\includegraphics[width=1\linewidth]{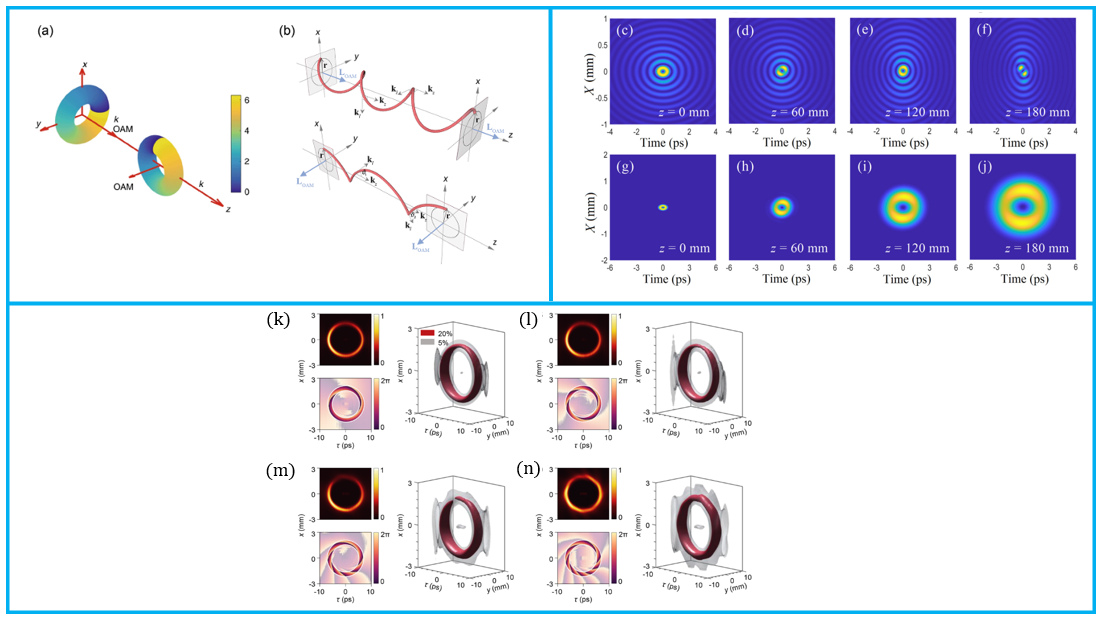}
\caption{Illustration of various STOV beams. (a)-(b) 3D profiles of longitudinal and transverse OAM beams. The colorbar shows the phase. (c)-(f) Propagation of BeSTOV in an anomalous dispersive medium. (g)-(j) Propagation of Gaussian STOV in an anomalous dispersive medium. (k)-(n) Spatiotemporal intensity, phase and 3D iso-intensity profiles of PSTOV with topological charges $l=2,4,6,$ and $8$, respectively. Reprinted with permission from Refs.~\cite{ChenWei,CAO2022133,Shuoshuozhang}.}
\label{FigSTOV}
\end{figure}

A more recent and intriguing development is the class of STOVs. In these optical beams, the spatial and temporal degrees of freedom are intrinsically coupled (due to the breaking of the orthogonal relationship between $\bm{k}_{l}$ and $\bm{k}_{z}$), which is typically manifested in their vortical phase structure. Consequently, STOVs possess transverse OAM, meaning that their OAM density is oriented in a direction perpendicular to the propagation axis (see Fig.~\ref{FigSTOV}). While longitudinal vortices possess on-axis phase singularities, STOVs manifest their singularities within the space-time domain. The optical field for such transverse OAM beams can be expressed as:
\begin{eqnarray}
  E_{STOV}(x,y,z;\tau)=E_{0}(x,y,z;\tau) \mathrm{e}^{il\varphi_{STOV}},
  \label{eqn43}
\end{eqnarray}
where $E_{0}(x,y,z;\tau)$ denotes the field amplitude and $\varphi_{STOV}$ is the azimuthal phase of the STOV in the $x-\tau$ plane, with $\tau=t-\frac{z}{v_{g}}$, and $v_{g}$ representing the local time coordinate of the pulse, and its associated group velocity, respectively. Various types of STOVs have been proposed in the past, based on the amplitude distribution ($E_{0}(x,y,z;\tau)$) used to define their optical fields. For instance, the generation of Gaussian, Bessel STOV (BeSTOV), and perfect STOV (PSTOV)- spatiotemporal counterparts of Gaussian, Bessel, and perfect spatial vortices- were thoroughly investigated~\cite{ChenWei,CAO2022133,Shuoshuozhang}.

In 2016, Jhajj et al. provided the first experimental evidence of STOVs~\cite{Jhajj}. In particular, they demonstrated that STOVs arise naturally as a consequence of an arrested self-focusing collapse during the nonlinear interaction of intense optical pulses with air. Furthermore, based on a spatiotemporal spectral modulation technique, STOVs were realized in free space~\cite{Hancock,Chong}. Various other strategies have been adopted in the past to generate STOVs~\cite{WeinerAM,Mirando:21,HuangJunyi22,WangHaiwen}. For instance, in Ref.~\cite{WangHaiwen}, it was demonstrated that STOVs can be generated with arbitrary spatiotemporal tilt by utilizing specially designed photonic crystals. In Ref.~\cite{ZangMirandoChong}, the authors showed that focusing of STOVs with cylindrical lens results in the tilting of the pulse. In Ref.~\cite{Hancock}, it was demonstrated that free space propagation of STOVs conserves angular momentum in space-time. We suggest readers to go through Refs.~\cite{ChenWei,Bekshaev} for a detailed understanding on the generation, characterization, and propagation dynamics of STOVs. Various other interesting studies have been conducted in the context of STOV. For instance, in Ref.~\cite{CHEN202544}, the authors investigated on the orbit-orbit coupling between longitudinal and transverse OAMs in the tight focusing of 3D STOVs. The authors proved that a ring-shaped trace of the phase singularity forms at the central part of the focused wavepacket due to such coupling. 

\subsection{Pulsed vortex beams}
The ability to generate high-peak-power ultrashort vortex pulses—spanning picosecond (ps) to femtosecond (fs) durations—unlocks significant potential in fields like optical trapping, strong-field physics, and nonlinear optics. Here, we provide a concise overview of the recent progress toward the design of compact and efficient sources for such pulsed vortex beams.

\subsubsection{Picosecond vortex beams}
In 2009, Yuichi et al., for the very first time, demonstrated the production of intense vortex pulses with a temporal duration of $4.5$ ps and a peak power of approximately 12.5 kW from a stressed large-mode-area fiber amplifier~\cite{Tanaka:09}. The optical efficiency (ratio of the output power to the input power) achieved in their study was nearly $29$ \%. Additionally, it was demonstrated that the sign of the TC of the generated vortices can be controlled by varying the strength and direction of the applied stress.  In Ref.~\cite{Koyama:11}, the authors generated ps vortex pulses in a stressed Yb-doped fiber amplifier seeded by a ps mode-locked Nd:YVO$_{4}$ laser. The achieved peak power, and temporal pulse width of the vortex output were $34.2$ kW, and $8.2$ ps, respectively. A maximum optical efficiency of $47.9$ \% was obtained in this case. Furthermore, in Ref.~\cite{Liang:09}, the authors reported the generation of mode-locked vortex pulses with temporal widths falling in $20-25$ ps regime and a repetition rate of $3.5$ GHz. In Ref.~\cite{Huang:18}, Huang et al. reported the generation of $8.5$ ps vector vortex beams from a mode-locked fiber laser by controlling geometric phases inside the laser cavity to map polarization to OAM. In Ref.~\cite{TONG2022108396}, the authors demonstrated the generation of high-power ps vortex arrays in an all-solid-state laser. The maximum output power and pulse duration of the two-vortex array were found to be $3.7$ W and $16.2$ ps, whereas, for the four-vortex array, these values were $3$ W and $17$ ps. In Ref.~\cite{ZhouWei}, the authors utilized a unique self-consistent intracavity mode conversion scheme to generate high-power ps vortex pulses from a passively mode-locked Nd:YVO$_{4}$ laser. For an LG mode of TC $1$, the achieved maximum output power, repetition rate, and pulse duration were $3.51$ W, $133.28$ MHz, and $10.2$ ps, respectively. Similarly, for an LG mode of TC $16$, the maximum output power and pulse duration were $3.16$ W, and $10.8$ ps, respectively.

\subsubsection{Femtosecond vortex beams}
Generally, fs pulse generation requires more stringent and extreme conditions such as broader gain bandwidth and specialized gain media, high-precision mode-locking, dispersion management, and tightly focused pumping spot, when compared to ps pulse generation. In Ref.~\cite{Zhuang:13}, Zhuang et al. reported a high-power fs monolithic self-mode-locked Yb:KGW laser with a repetition rate up to tens of GHz. It was further demonstrated that the length of the laser crystal strongly influences the repetition rate and temporal duration of the generated pulses. In Ref.~\cite{Chang:16}, the authors demonstrated the direct generation of fs vortex beams by selective pumping. Additionally, it was shown that the output power can reach up to $1.45$ W for a pump power of $10$ W. In Ref.~\cite{ZhimingZhang:18}, an all-fiber mode-locked fs $LG_{0,\pm1}$ vortex laser was proposed, which can generate OAM modes with a temporal duration of $398$ fs. In Ref.~\cite{FengQian2022}, the authors demonstrated a fs vortex laser system based on two-stage optical parametric amplification. In particular, they obtained near-infrared vortex signal pulses with energy, and pulse duration $190$~$\mu$J, and $51$ fs, respectively. Similarly, for the generated near-infrared vortex idler pulses, these values were $158$~$\mu$J, and $48$ fs, respectively. In Ref.~\cite{Qian:20}, Qian et al., for the first time, demonstrated a high-energy, mid-infrared, fs vortex laser based on optical parametric chirped pulse amplification. They were able to create vortex pulses with $9.53$ mJ energy, $119$ fs temporal duration, and $20$ Hz repetition rate. In Ref.~\cite{Zhao:21}, the authors demonstrated the generation of $100$ fs vortices in the $2$~$\mu$m spectral regime by employing a mode-locked solid-state laser and a single-cylindrical-lens converter. This novel method enabled the production of ultrashort vortices with a chirp-free broadband optical spectrum. In Ref.~\cite{HongyuLiu}, the authors demonstrated the generation of high-power, high-order (up to a TC of 30) fs vortex pulses via astigmatic mode conversion from a mode-locked Hermite-Gaussian Yb:KGW laser oscillator. In Ref.~\cite{LinFengRen}, the authors experimentally demonstrated the generation of high-power fs vortex pulses from a Mamyshev oscillator based on few-mode polarization-maintaining Yb-doped fibers. By employing an intracavity transverse spatial mode selection technique, they were successful in creating ultrashort OAM pulses carrying TCs $\pm1$ with a pulse energy in the nanojoule regime, MHz repetition rate and $76$ fs temporal width. In Ref.~\cite{LiuXin:23}, the direct generation of $300$ fs optical vortices with multiple phase singularities from a laser oscillator was reported. In Ref.~\cite{Chen:25}, Chen et al. reported the direct generation of fs LG vortex pulses with tunable TCs from a passive mode-locked Yb:KGW laser oscillator. In Ref.~\cite{Tong:25}, the production of $100-200$ fs vortex pulses with GHz repetition rate, widely tunable wavelength range was demonstrated.

\section{Nonlinear perturbative and non-perturbative optical phenomena driven by vortex beams}\label{Nonlinear}

In the classical realm of linear optics, light behaves as a well-mannered guest within a material. Decades of theoretical and experimental exploration—from the familiar Snell’s law of refraction to Maxwell’s celebrated electromagnetic equations—have shown that this interaction is one of graceful proportionality. When light encounters a material, the material’s response—how it bends, scatters, or absorbs the light—is entirely predictable. In other words, light and matter share a proportional relationship: if a weak light is shone on a material, only a weak light emerges from it. This principle underlies the operation of many everyday optical components, including cameras, eyeglasses lenses, mirrors, and beam splitters. Phenomena such as reflection, refraction, dispersion, interference, and diffraction are all elegantly captured within this linear optics framework.

However, when a material is exposed to intense optical excitation, its response begins to deviate from the predictions of linear theory. The invention of the laser in 1960 marked a revolutionary moment in optics, introducing a light source that was coherent, highly directional, extremely intense, and spectrally pure (i.e., monochromatic). The birth of nonlinear optics can be traced back to a landmark experiment in 1961, when Franken et al. demonstrated that a quartz crystal, illuminated by a 693.4 nm ruby laser, could do more than merely transmit light—it could transform it~\cite{Franken}. The team observed the emission of ultraviolet light at 347 nm, corresponding to the second harmonic of the incident radiation. This discovery—the first observation of a material altering the color of light in such a manner—opened an entirely new frontier in photonics. In essence, nonlinear optics studies phenomena in which intense light modifies the optical properties of a material. Because of its extraordinary intensity, laser light can induce nonlinear optical effects even in weakly nonlinear media. When a laser’s electromagnetic field interacts with matter, the material’s charges respond: electrons are displaced relative to the atomic nuclei, giving rise to a transient, oscillating dipole moment throughout the medium. The induced dipole moment per unit volume, known as the electric polarization ($\bm{P}=N\bm{p}$, where $N$ is the number of elementary dipoles per unit volume and $\bm{p}$ denotes the dipole moment of an individual dipole), plays a central role in governing nonlinear optical phenomena.

As is well known, the relationship between the polarization vector $\bm{P}$ and the electric field vector $\bm{E}$ ceases to be linear when the peak field strength of the driving light reaches values on the order of $10^{7}$–$10^{10}$~V/cm, marking the onset of nonlinear effects. In this regime, $\bm{P}$ no longer scales proportionally with $\bm{E}$, unlike in the linear optics case. However, if the external field strength remains below the level of the Coulomb force binding the electrons within an atom, the field can still be regarded as a perturbation to the atomic system. Under such conditions, the nonlinear polarization can be derived using time-dependent perturbation theory and is generally expressed as:
\begin{eqnarray}
    \bm{P}=\epsilon_{0} \left[\chi^{(1)} \bm{E} + \chi^{(2)} \bm{E}^2 + \chi^{(3)} \bm{E}^3 + \chi^{(4)} \bm{E}^4+\cdots\right],
    \label{eqn44}
\end{eqnarray}
where $\chi^{(1)}$, $\chi^{(2)}$, $\chi^{(3)}$, and $\chi^{(4)}$ denote the linear (first-order), second-order nonlinear, third-order nonlinear, and fourth-order nonlinear optical susceptibility tensors, respectively, and $\epsilon_{0}$ represents the free-space permittivity. In general, the first-order susceptibility is related to the refractive index of the medium through the relation $n = \sqrt{1 + \chi^{(1)}}$. Mathematically, Eq.~(\ref{eqn44}) represents a Taylor series expansion of the polarization $\bm{P}$ around $\bm{E}=0$, which converges under the condition that the interaction energy remains much smaller than the intrinsic energy scale of the atomic system. This description in nonlinear optics is commonly referred to as perturbative nonlinear optics. To gain a clearer understanding of different perturbative nonlinear optical processes, we evaluate the nonlinear polarization—specifically, certain components of it—expressed in Eq.~(\ref{eqn44}) for an external field composed of two superposed fields, i.e., $\bm{E}=\hat{x}E_{1}\cos(\omega_{1}t-k_{1}z+\phi_{1})+\hat{x} E_{2}\cos(\omega_{2}t-k_{2}z+\phi_{2})$, where ($\omega_{1},\omega_{2}$), and ($k_{1},k_{2}$), ($\phi_{1},\phi_{2}$) denote the angular frequencies, wavenumbers, and initial phases of the two fields, respectively. We now compute the first nonlinear contribution to the polarization—namely, the second term in Eq.~(\ref{eqn44}). The general form of the second-order nonlinear polarization can be written as:
\begin{eqnarray}
    \bm{P}^{2}(t)=\epsilon_{0} \chi^{(2)} \bm{E}^{2}(t).
    \label{eqn45}
\end{eqnarray}
Since the electric field is linearly polarized along $\hat{x}$, we may, for simplicity, treat it as a scalar quantity when deriving the new frequency components—bearing in mind, however, that the full tensor form becomes essential when dealing with different polarization states. Accordingly, the scalar electric field can be expressed as $E=E_{1} \cos(\omega_{1}t-k_{1}z+\phi_{1}) + E_{2} \cos(\omega_{2}t-k_{2}z+\phi_{2})$. The nonlinear source polarization is obtained by squaring the electric field:

\begin{eqnarray}
    &&\left[E_{1} \cos(\omega_{1}t-k_{1}z+\phi_{1}) + E_{2} \cos(\omega_{2}t-k_{2}z+\phi_{2})\right]^2=\nonumber \\
    &&\frac{1}{2} \left(E_{1}^2+E_{2}^2\right)+ \frac{1}{2} E_{1}^2 \cos(2\omega_{1}t-2k_{1}z+2\phi_{1})
+\frac{1}{2} E_{2}^2 \cos(2\omega_{2}t-2k_{2}z+2\phi_{2})\nonumber \\
&&+E_{1}E_{2} \cos \left[(\omega_{1}+\omega_{2})t+(k_{1}+k_{2})z + (\phi_{1}+\phi_{2})\right] \nonumber \\
&&+E_{1}E_{2} \cos \left[(\omega_{1}-\omega_{2})t+(k_{1}-k_{2})z + (\phi_{1}-\phi_{2})\right].
    \label{eqn46}
\end{eqnarray}

Therefore, the second-order nonlinear polarization can be written as:

\begin{eqnarray}
    P^{2}(t)&=& \epsilon_{0} \chi^{(2)} \left[\frac{1}{2} \left(E_{1}^2+E_{2}^2\right)+ \frac{1}{2} E_{1}^2 \cos(2\omega_{1}t-2k_{1}z+2\phi_{1}) \right.\nonumber \\
    &+&
    \left.\frac{1}{2} E_{2}^2 \cos(2\omega_{2}t-2k_{2}z+2\phi_{2})\right. \nonumber \\
    &+&
    E_{1}E_{2} \cos \left[(\omega_{1}+\omega_{2})t+(k_{1}+k_{2})z + (\phi_{1}+\phi_{2})\right] \nonumber \\
    &+&
    \left.E_{1}E_{2} \cos \left[(\omega_{1}-\omega_{2})t+(k_{1}-k_{2})z + (\phi_{1}-\phi_{2})\right]\right].
    \label{eqn47}
\end{eqnarray}

We can now interpret each term in Eq.~(\ref{eqn47}): The first term, $\frac{1}{2}(E_{1}^2 + E_{2}^2)$, contains no frequency component and is referred to as the DC term or optical rectification (OR). It merely induces a static field within the nonlinear medium. All other terms, however, represent the generation of new frequency components resulting from the nonlinear interaction between the external field and the medium—a hallmark of nonlinear optical processes. For instance, the second and third terms, $\frac{1}{2}E_{1}^2 \cos(2\omega_{1}t - 2k_{1}z + 2\phi_{1})$ and $\frac{1}{2}E_{2}^2 \cos(2\omega_{2}t - 2k_{2}z + 2\phi_{2})$, respectively, give rise to new frequencies $2\omega_{1}$ and $2\omega_{2}$, which are precisely twice those of the respective fundamental fields—these correspond to second-harmonic generation (SHG). Similarly, the fourth and fifth terms, $E_{1}E_{2}\cos[(\omega_{1} + \omega_{2})t - (k_{1} + k_{2})z + (\phi_{1} + \phi_{2})]$ and $E_{1}E_{2}\cos[(\omega_{1} - \omega_{2})t - (k_{1} - k_{2})z + (\phi_{1} - \phi_{2})]$, respectively, produce new frequencies $\omega_{1} + \omega_{2}$ and $\omega_{1} - \omega_{2}$, corresponding to sum-frequency generation (SFG) and difference-frequency generation (DFG), respectively. It is noteworthy that SHG and OR can be viewed as special cases of SFG and DFG, respectively, when the input field frequencies are identical. Furthermore, the symmetry properties of the material medium play a decisive role in determining which frequency components can actually be generated during the light–matter interaction. In materials possessing inversion symmetry (also known as centrosymmetric media), only odd-order harmonics of the fundamental field can be produced. This arises because spatial inversion symmetry requires all even-order nonlinear susceptibilities to vanish, i.e., $\chi^{(2n)} = 0$ for $n = 1, 2, 3, \ldots$. Hence, to generate even-order nonlinear optical effects, one must break the inversion symmetry of the medium. Finally, it is important to emphasize that in centrosymmetric media, the dominant contribution to the nonlinear polarization originates from the third-order nonlinear susceptibility $\chi^{(3)}$, which gives rise to a variety of optical phenomena such as third-harmonic generation (THG), self-focusing, caused by the Kerr effect in the spatial domain, self-phase modulation (SPM), due to the Kerr effect in the temporal domain, cross-phase modulation (XPM), degenerate four-wave mixing, stimulated Raman scattering, stimulated Brillouin scattering, and two-photon transitions, among others.

The perturbative nonlinear optical processes discussed above all involve the interaction of multiple waves within a nonlinear medium. To fully understand these interactions, two key questions arise: (1) How does the amplitude of each wave grow or decay relative to the others? This describes how energy flows among the interacting waves. (2) What factors govern the phase relationship between the interacting waves inside the nonlinear medium? To address these questions, we derive the inhomogeneous wave equation with the nonlinear polarization as the source term, i.e., $\nabla^{2}\bold{E}-\frac{\epsilon^{(1)}}{c^2}\frac{\partial^2 \bold{E}}{\partial t^2}=\mu_{0}\frac{\partial^2 \bold{P}^{NL}}{\partial t^2}$, using Maxwell’s equations: (1) $\bold{\nabla}\cdot\bold{D}=0$, (2) $\bold{\nabla}\cdot\bold{B}=0$, (3) $\bold{\nabla}\times \bold{E}=-\frac{\partial \bold{B}}{\partial t}$, and (4) $\bold{\nabla}\times \bold{H}=\frac{\partial \bold{D}}{\partial t}$, where $\bold{E}, \bold{D}, \bold{B}$, and $\bold{H}$ denote the electric field, electric displacement, magnetic induction, and magnetic field, respectively. In this derivation, we require that $\bold{\nabla}\cdot\bold{E}=0$ not only because $\bold{\nabla}\cdot\bold{D}=0$ (implying $\bold{\nabla}\cdot\bold{E}=0$ in a linear, isotropic, source-free medium), but also because: (a) we assume $\bold{E}$ to represent a transverse infinite plane wave, and (b) we invoke the slowly varying envelope approximation (SVEA). Moreover, we split the total polarization into its linear and nonlinear parts, $\bold{P}=\bold{P}^{L}+\bold{P}^{NL}$, where $\bold{P}^L$ and $\bold{P}^{NL}$ represent the linear and nonlinear polarization contributions, respectively, and apply the relation $\epsilon^{(1)} = 1 + \chi^{(1)}$. Physically, the inhomogeneous wave equation can be interpreted as follows: the left-hand side describes the linear propagation of the field, while the right-hand side acts as a source term generated by the nonlinear polarization, giving rise to new frequency components—i.e., the essence of nonlinear wave mixing.

\begin{figure}[h!]
\centering 
\includegraphics[width=1\linewidth]{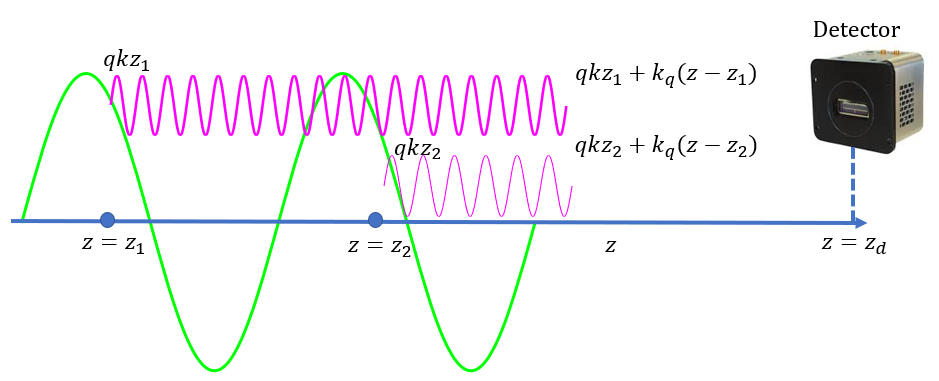}
\caption{Schematic illustration of phase matching for a plane-wave field propagating along the $z$-axis. Atoms located at positions $z=z_{1}$ and $z=z_{2}$ emit radiation that remains in phase as it propagates at the vacuum speed of light. The green line represents the driving plane-wave electric field, while the magenta lines indicate the emitted radiation from different atomic positions.}
\label{Fig28}
\end{figure}

Now, we consider the simplest coupled-wave scenario, namely sum-frequency generation (SFG). We make the following assumptions:
(a) The three interacting waves are monochromatic plane waves with frequencies $\omega_{1}$, $\omega_{2}$, and $\omega_{3}=\omega_{1}+\omega_{2}$, and
(b) The slowly varying envelope approximation (SVEA) holds, i.e., the field amplitude varies slowly over an optical wavelength such that $\left|\frac{\partial^2 E}{\partial z^2}\right|\ll \left|2k \frac{\partial E}{\partial z}\right|$. With these considerations, the complete set of coupled-wave equations for SFG can be written as:
\begin{eqnarray}
    \frac{\partial E_{1}}{\partial z}=- \frac{i\omega_{1}}{n_{1}c} d_{eff} E_{3}E_{2}^{*} \mathrm{e}^{-i\Delta k z} \nonumber \\
    \frac{\partial E_{2}}{\partial z}=- \frac{i\omega_{2}}{n_{2}c} d_{eff} E_{3}E_{1}^{*} \mathrm{e}^{-i\Delta k z} \nonumber \\
    \frac{\partial E_{3}}{\partial z}=- \frac{i\omega_{3}}{n_{3}c} d_{eff} E_{1}E_{2} \mathrm{e}^{i\Delta k z},
    \label{eqn48}
\end{eqnarray}
where $d_{\mathrm{eff}}$ denotes an effective nonlinear coefficient determined by the polarization geometry and the $\chi^{(2)}$ tensor. We further assume that all waves propagate along the $z$-axis with complex amplitudes $E_{1}(z)$, $E_{2}(z)$, and $E_{3}(z)$, and that the refractive indices are given by $n_{1,2,3} = \sqrt{\epsilon^{(1)}(\omega_{1,2,3})}$ at their respective frequencies. The term $\Delta k$, known as the wave-vector mismatch, plays a pivotal role in nonlinear optics—governing the direction of energy flow between waves and determining the conversion efficiency. When $\Delta k \neq 0$, the phases between the waves gradually slip, causing energy to flow back and forth after a propagation distance known as the coherence length, typically written as $L_{\mathrm{coh}}=\frac{\pi}{\Delta k}$. In nonlinear optics, strategies are therefore employed to maximize the coherence length so that a high level of coherence is maintained among the interacting waves. For example, in non-perturbative processes such as high-order harmonic generation (HHG) in atomic gases, efficient generation requires $L_{\mathrm{coh}}>5 L_{\mathrm{abs}}$, where $L_{\mathrm{abs}}$ is the absorption length determined by the medium’s density and photoabsorption cross-section. Maximum conversion efficiency is achieved when the phase mismatch vanishes $\Delta k=0$, i.e. the condition for perfect phase matching. In the plane-wave limit, this condition is naturally satisfied. To illustrate this, we consider the perturbative generation of harmonics—for instance, third-harmonic generation (THG), a $\chi^{(3)}$ process. For efficient detection (see Fig.~\ref{Fig28}), radiation from atoms located at different positions (e.g., $z=z_{1}$ and $z=z_{2}$) must interfere constructively, meaning their emitted phases must match at the detector. Since the driving fields are ideal plane waves, no intrinsic beam phases arise (such as the Gouy phase from focusing or radial phase curvature from a finite beam waist). Thus, harmonics generated at $z=z_{1}$ and $z=z_{2}$ have phases $\phi_{z=z_{1}}=qkz_{1}$, and $\phi_{z=z_{2}}=qkz_{2}$, where $q$ is the harmonic order. After propagating to a detector at $z=z_{d}$, the phase relationship becomes $\phi_{z_{1}}(z=z_{d})=\phi_{z_{2}}(z=z_{d})=qkz$, which is satisfied when $k_{q}=qk$ where $k_{q}$ is the wave vector of the $q$-th harmonic (e.g., $q=2$ for second-harmonic generation). This confirms that perfect phase matching can be intrinsically achieved with plane waves in perturbative nonlinear processes.

All the preceding discussions in this section primarily focused on the plane-wave approximation, where the interacting waves were assumed to propagate along the $z$-axis with flat wavefronts. However, this assumption breaks down in realistic experimental scenarios, where laser beams are tightly focused into nonlinear media. One major motivation for focusing is to achieve higher optical intensity, since nonlinear optical processes are highly sensitive to intensity and a stronger response can be obtained with a higher peak intensity. Although tighter focusing reduces the effective interaction length (i.e., there is a trade-off between achieving high intensity and maintaining a long interaction region), the increase in intensity generally outweighs the loss in interaction length. Importantly, focused beams exhibit spatially varying intensity distributions—localized and non-uniform rather than uniform as in the plane-wave case—so atoms or molecules in the nonlinear medium experience different local intensities, which contribute to phase mismatch. In addition, focused beams possess intrinsic phase contributions, such as the Gouy phase and radial phase variations, which further influence phase matching. Therefore, in nonlinear interactions driven by focused beams, it is essential to properly manage these different sources of phase mismatch to ensure that the overall phase-mismatch factor ($\Delta k$) remains as close to zero as possible (even if not perfectly). 

It is important to note that when focused beams are used to drive nonlinear optical processes, an additional term must be incorporated into the coupled-wave equations to account for diffraction effects. For example, if we consider SFG with focused input fields, the complete set of coupled-wave equations must be modified accordingly, and Eq.~(\ref{eqn48}) would be rewritten as
\begin{eqnarray}
    \frac{\partial E_{1}}{\partial z}+\frac{i}{2k_{1}}\nabla_{T}^{2}E_{1}=- \frac{i\omega_{1}}{n_{1}c} d_{eff} E_{3}E_{2}^{*} \mathrm{e}^{-i\Delta k z} \nonumber \\
    \frac{\partial E_{2}}{\partial z}+\frac{i}{2k_{2}}\nabla_{T}^{2}E_{2}=- \frac{i\omega_{2}}{n_{2}c} d_{eff} E_{3}E_{1}^{*} \mathrm{e}^{-i\Delta k z} \nonumber \\
    \frac{\partial E_{3}}{\partial z}+\frac{i}{2k_{3}}\nabla_{T}^{2}E_{3}=- \frac{i\omega_{3}}{n_{3}c} d_{eff} E_{1}E_{2} \mathrm{e}^{i\Delta k z},
    \label{eqn49}
\end{eqnarray}
where $\frac{i}{2k_{j}}\nabla_{T}^{2}E_{j}$ with $j=1,2,3$ denote the diffracting terms of the focused beams in the SFG process.

In perturbative nonlinear optical processes such as SHG and SFG, various phase-matching techniques-including angle tuning, temperature tuning, and quasi-phase matching-are widely employed to optimize the phase-matching conditions and thereby enhance the nonlinear conversion efficiency.

In Ref.~\cite{Kleinman}, Kleinman et al. theoretically analyzed and experimentally demonstrated SHG using focused Gaussian laser beams (fundamental $TEM_{00}$ mode) in nonlinear crystals such as ammonium dihydrogen phosphate (ADP) and potassium dihydrogen phosphate (KDP), whose dimensions were much larger than the focal region. Their key observations were as follows:
(1) The SHG output exhibited a “half-moon”–like spatial profile with a sharply defined edge. When examining the bright side of this boundary, a series of fringes were observed extending into the illuminated region. Moreover, the fringe pattern was highly dependent on the observation plane: the distribution measured at the crystal surface differed significantly from that observed in the far field.
(2) The phase-matching conditions were strongly influenced by crystal orientation, demonstrating typical angle-tuning-based phase matching in negative uniaxial crystals such as ADP and KDP. In Ref.~\cite{Boyd}, Boyd et al. theoretically investigated the optimization of SHG and parametric generation (PG) driven by a focused Gaussian beam in a uniaxial nonlinear crystal under the small-signal approximation (i.e., assuming negligible pump depletion). Their numerical calculations showed that both the SHG power and the inverse PG threshold depend on the dimensionless focusing parameter $l/b$, where $l$ is the optical path length inside the crystal and $b$ is the confocal parameter determined by the beam minimum waist and wavelength. Importantly, they found that the optimum focusing condition is achieved at $l/b = 2.84$ when double refraction can be ignored.

\subsection{Perturbative nonlinear optical phenomena driven by vortex beams}

The preceding discussion has primarily highlighted the role of Gaussian beams in understanding perturbative nonlinear optical processes such as SHG and PG. Building on this foundation, it is natural to investigate how optical vortices—light beams carrying OAM—affect such interactions. This inquiry raises several key questions:
(1) To what extent is OAM conserved during nonlinear interactions?
(2) How does the TC of a vortex beam influence phase-matching conditions?
(3) Can the unique spatial and phase characteristics of OAM be exploited to improve nonlinear frequency conversion efficiency? To address these questions, we examine perturbative nonlinear processes such as SHG, SFG, and optical parametric generation (OPG) driven by OAM beams in the following subsection. Moreover, a substantial body of literature reports the use of OAM beams to drive additional perturbative nonlinear phenomena—including self-focusing, self-phase modulation, third-harmonic generation, stimulated Raman scattering, and difference-frequency generation, among others.

\subsection{Second-harmonic generation (SHG)}

\begin{figure}[h!]
\centering 
\includegraphics[width=0.9\linewidth]{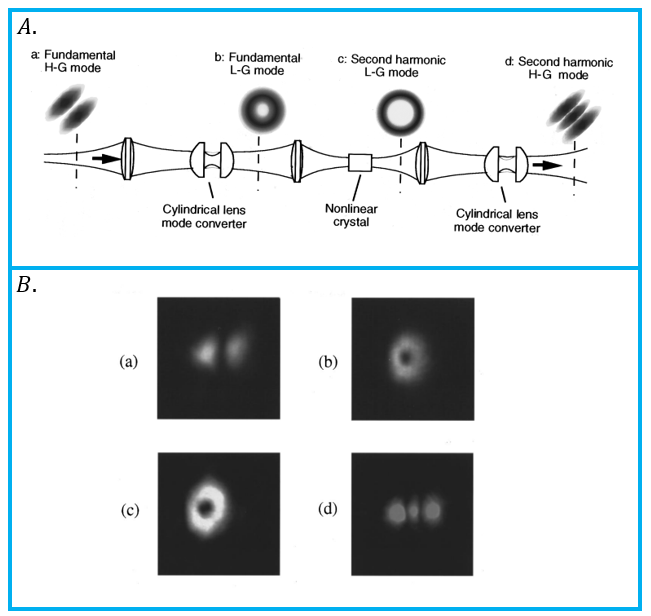}
\caption{Second-harmonic generation (SHG) of OAM-carrying LG modes in a lithium triborate (LBO) crystal.
(A) Experimental setup illustrating the mode transformation during frequency doubling of an LG beam.
(B) (a) Input HG mode, (b) corresponding LG mode after mode conversion, (c) SH LG mode, and (d) SH beam converted back into an HG mode using a mode converter.
Reprinted with permission from Ref.~\cite{Dholkia}.}
\label{Fig29}
\end{figure}

A series of landmark theoretical and experimental studies—namely:
(1) the pioneering SHG experiment in 1961 by Franken et al.~\cite{Franken},
(2) the association between OAM and the helical phase front of light, along with its experimental realization using a mode converter by Allen et al. in 1992~\cite{Allen}, and
(3) the frequency doubling of a laser beam with a phase dislocation on the beam axis and the subsequent formation of two dislocations, demonstrated by Basistiy et al. in 1993~\cite{BASISTIY1}- motivated a collaborative team from the University of St. Andrews (Scotland) and JILA, University of Colorado (USA), to perform the first SHG experiment driven by an OAM beam in 1996~\cite{Dholkia}. Under the assumptions of negligible absorption and negligible depletion of the fundamental OAM beam in the lithium triborate (LBO) crystal, they demonstrated that both the OAM and the frequency (and hence the energy) of the SH signal are doubled with respect to the fundamental beam (see Fig.~\ref{Fig29}). Moreover, they showed that OAM conservation holds regardless of the phase-matching method (angle tuning or temperature tuning) or the nonlinear crystal employed (LBO or KTP). Their analysis further revealed that the beam radius of the SH signal is reduced by a factor of $\sqrt{2}$, implying that the fundamental and SH OAM beams share the same Rayleigh range and thus exhibit identical far-field divergence. The physical interpretation they proposed to explain this OAM up-conversion is as follows: LG beams possess helical wavefronts, meaning that the Poynting vector does not point strictly along the propagation axis. Instead, it contains an azimuthal (spiraling) component, rotating about the beam axis at a rate of $l/k^{(\omega)}r^2$, where $l$ is the topological charge (corresponding to OAM of $l\hbar$ per photon), $k^{(\omega)}$ is the wavenumber, and $r$ is the radial distance from the beam axis. By comparing this rotation rate for the fundamental and SH beams, and using the relation $k^{(2\omega)} = 2k^{(\omega)}$, the OAM up-conversion law naturally follows. The OAM conservation can also be understood within the photon picture: Since SHG is a perturbative nonlinear process, photon interactions must conserve energy. Thus, the creation of one SH photon with energy $2\hbar\omega$ requires the annihilation of two fundamental photons, each with energy $\hbar\omega$. Given that each fundamental photon carries OAM $l\hbar$, destroying two such photons results in the generation of an SH photon carrying OAM $2l\hbar$, consistent with the experimentally observed doubling.

In 1997, Courtial et al. investigated SHG driven by high-order LG modes and confirmed that the OAM conservation law remains valid even when the LG modes possess nonzero radial indices~\cite{Courtial2}. They further demonstrated that the frequency-doubled beam exhibits a Gegenbauer–Gaussian field distribution at the beam waist, which can be expressed as a superposition of LG modes. In Ref.~\cite{Orlov1}, Orlov et al. examined the free-space propagation of SH LG beams and found that a vortical structure consisting of multiple doubly charged vortices forms in the near field of the SH signal. In Ref.~\cite{PETROV1}, Petrov et al. studied SHG driven by optical beams with edge phase dislocations. They reported that no edge dislocation appears in the SH beam when the fundamental Gaussian beam carries an on-axis edge dislocation, whereas a single edge dislocation is generated in the SH output when the fundamental beam exhibits an off-axis edge dislocation. In Ref.~\cite{Oklim}, the authors investigated frequency doubling of vortices with TC $l = 1$ inside a linear astigmatic resonator, demonstrating that the topological instability of the resulting doubly charged harmonic vortices leads to the formation of spatially separated vortex cores capable of rotation. Likewise, in Ref.~\cite{Toda:10}, the dynamics of a paired optical vortex generated via SHG using sub-picosecond laser pulses was examined, revealing that two vortices with a well-defined separation undergo rotation when the nonlinear crystal position is adjusted along the propagation direction. In Ref.~\cite{Beresna}, Beresna et al. demonstrated efficient SHG in atmospheric-pressure air using a tightly focused femtosecond laser beam, achieving two orders-of-magnitude enhancement in normalized conversion efficiency and average SH power. They also showed that a linearly polarized fundamental beam produces a two-lobe SH intensity pattern, whereas a circularly polarized pump yields a ring-shaped SH beam. In Ref.~\cite{Varma2025}, Varma et al. investigated SHG of LG beams interacting with arrays of vertically aligned carbon nanotubes (CNs), revealing that the SH field amplitude can be actively controlled by varying parameters such as TC, radial index, initial beam waist, pump frequency, CN radius, and CN separation. They further demonstrated that LG beams yield significantly higher SH field amplitudes compared to Gaussian beams under similar conditions.
\begin{figure}[h!]
\centering 
\includegraphics[width=1\linewidth]{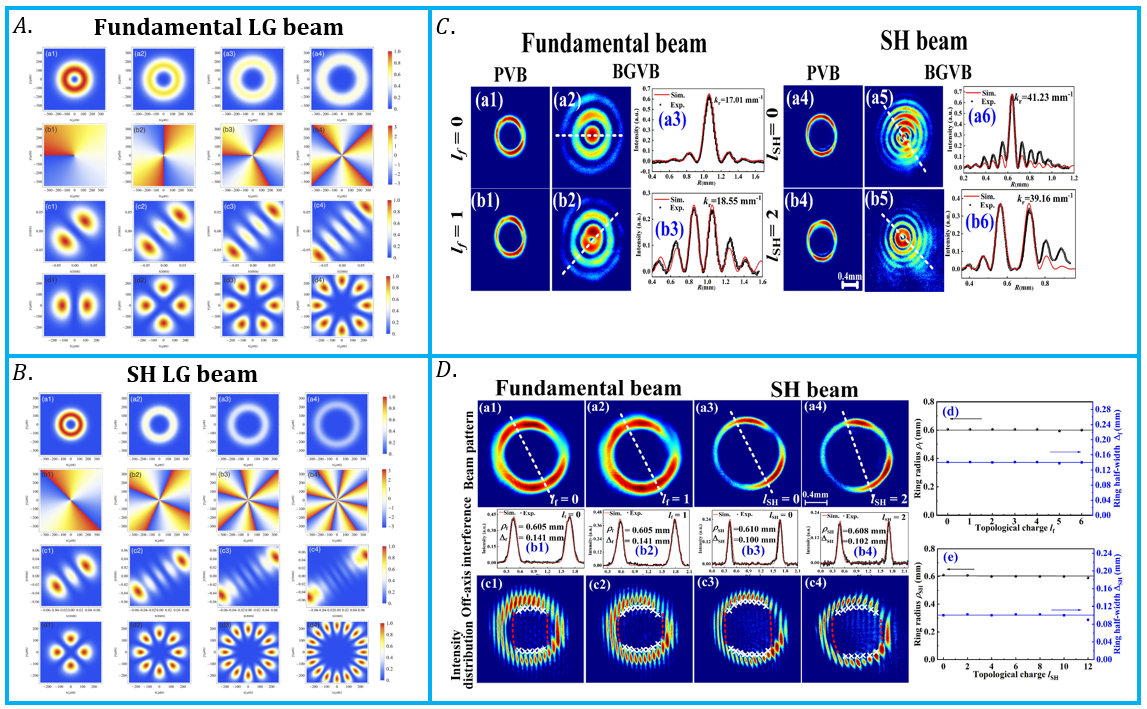}
\caption{Second-harmonic generation of LG and POV beams. (A) Fundamental LG beam carrying TC values $1,2,3,4$: Transverse intensity (first row), phase (second row), TC detection via a tilted convex lens (third row) and mirror-image interference (fourth row). (B) Second-harmonic LG beam carrying TC values$2,4,6,8$: Transverse intensity (first row), phase (second row), TC detection via a tilted convex lens (third row) and mirror-image interference (fourth row). (A)-(B) are taken from Ref.~\cite{BikashSPIE}. (C) Beam patterns and intensity distributions of POV beams and BG beams with different TC values at fundamental/SH wavelengths. (a1, b1,a4, b4) and (a2, b2, a5, b5) are beam patterns of POV and BG beams with different TC values ($l_{f}$ = 0 and 1, $l_{SH}$ = 0 and 2) for the fundamental/SH wave; (a3, b3, a6, b6) are the corresponding intensity distributions; The red solid lines are the simulated results using the standard expression of BG beams. (D) Characteristics of fundamental/SH POVs and fork-shaped interferogram for different TC values.(a1-a4) and (b1-b4) are beam patterns and corresponding intensity distributions for fundamental/SH POV beams, respectively. (c1-c4) are fork-shaped interferogram
between the POV beam and the reference fundamental Gaussian beam. (d) and (e) are variations of beam parameters (ring radius, and ring
half-width) of the fundamental and SH POV beams versus their TC values, respectively. (C)-(D) are taken from Ref.~\cite{Liyueqing}. Reprinted with permission from Ref.~\cite{BikashSPIE,Liyueqing}.}
\label{Fig30}
\end{figure}

In Ref.~\cite{Shiruizhang}, the authors experimentally investigated single-pass SHG driven by LG vortex beams carrying both integer and fractional TC values, and reported that the conversion efficiency decreases with increasing the TC—demonstrating an inverse relationship between the two. They also showed that their approach enables the generation of high-quality short-wavelength vortex beams. In Ref.~\cite{BikashSPIE}, SHG in a potassium titanyl phosphate (KTP) crystal driven by LG beams with integer TC values was theoretically studied, discussing the OAM conservation rule and the divergence characteristics of the fundamental and SH beams (see Fig.~\ref{Fig30}(A)–(B)). In Ref.~\cite{Shan:24}, Shan et al. theoretically examined SHG in two-dimensional (2D) materials driven by LG beams and demonstrated that SHG contributions from electric dipole, electric quadrupole, and magnetic dipole interactions can be distinguished based on beam parameters such as spot size and TC. In Ref.~\cite{Fang:17}, the authors experimentally studied SHG of a high-order LG mode in a periodically poled KTP (PPKTP) crystal under quasi-phase-matching and confirmed that the frequency, OAM, and radial index of the SH beam are exactly twice those of the fundamental mode. This is particularly attractive since it enables simultaneous generation of short-wavelength LG beams with both higher TC values and higher radial index. In Ref.~\cite{Bovino:11}, Bovino et al. investigated SHG of femtosecond fractional TC vortices in a noncollinear configuration and found that the resulting SH field always carries zero OAM. In Ref.~\cite{ApurvChaitanya:15}, the authors studied SHG of ultrafast optical vortices in a bismuth triborate (BIBO) crystal, showing that the SHG efficiency decreases with increasing TC of the fundamental beam—consistent with the growth of beam size with $l$. They also demonstrated that optical vortices exhibit a larger angular acceptance bandwidth than Gaussian beams, while the acceptance remains independent of the vortex TC.

In Ref.~\cite{Dimitrov}, the authors experimentally demonstrated a distinctive approach for generating BG beams using the SHG process. They showed that:
(1) The Fourier transform of the SH signal of ring-shaped, azimuthally modulated necklace beams—created by the interference of two vortices with equal but opposite topological charges—leads to the formation of BG beams; (2) The SHG process effectively removes the azimuthal phase dislocations contained in the necklace beams; and (3) For an input beam with a TC of $l = 16$, the divergence of the resulting central peak was measured to be approximately $180$~$\mu$rad. In Ref.~\cite{JARUTIS3}, Jarutis et al. investigated SHG of high-order Bessel beams (generated from LG vortices using an axicon) and highlighted the crucial role of the zeroth-order coherent background (i.e., the zeroth-order Bessel component) in the breakup of a $2l$-charged vortex into $2l$ singly charged vortices, where $l$ is the TC of the fundamental beam. In Ref.~\cite{Belyi}, Belyi et al. theoretically and experimentally demonstrated that higher-order Bessel beams can efficiently participate in SHG while exhibiting addition or subtraction of wavefront dislocations depending on the sign of the interacting topological charges. Furthermore, in Ref.~\cite{Liyueqing}, SHG of POV beams was studied in a single-pass configuration under the small-signal approximation. In contrast to LG-driven SHG under identical conditions, the conversion efficiency of POV-driven SHG was found to be independent of the TC. The authors also showed that: (1) The SH efficiency can be controlled through POV beam parameters—specifically, the ring radius and ring width, which exhibit an inverse dependence; and
(2) In the far field, SH POV beams transform into their Fourier counterparts, i.e., BG vortex beams (see Fig.~\ref{Fig30}(C)–(D)).

\subsection{Sum-frequency generation (SFG)}
In 1997, the research group led by A. Stabinis (Laser Research Centre, Vilnius University) experimentally investigated collinear type-II phase-matched SFG of optical vortices in a KDP crystal and demonstrated that the SFG process enables the generation of output vortices with different topological charges~\cite{BERZANSKIS1}. In 1998, the same group theoretically and experimentally examined sum-frequency mixing of optical vortices in type-I critically phase-matched KDP crystals and analyzed the influence of walk-off on vortex interactions~\cite{BERZANSKIS2}. They showed that the vortex dynamics in the mixing process depend strongly on walk-off, which can break the interaction symmetry, leading to a range of vortex interaction effects, including:
(1) the breakup of a higher-order vortex into multiple singly charged vortices, (2) particle-like behaviors such as vortex attraction and repulsion, and (3) the appearance of vortex–antivortex pairs when pump depletion is considered.
\begin{figure}[h!]
\centering 
\includegraphics[width=1\linewidth]{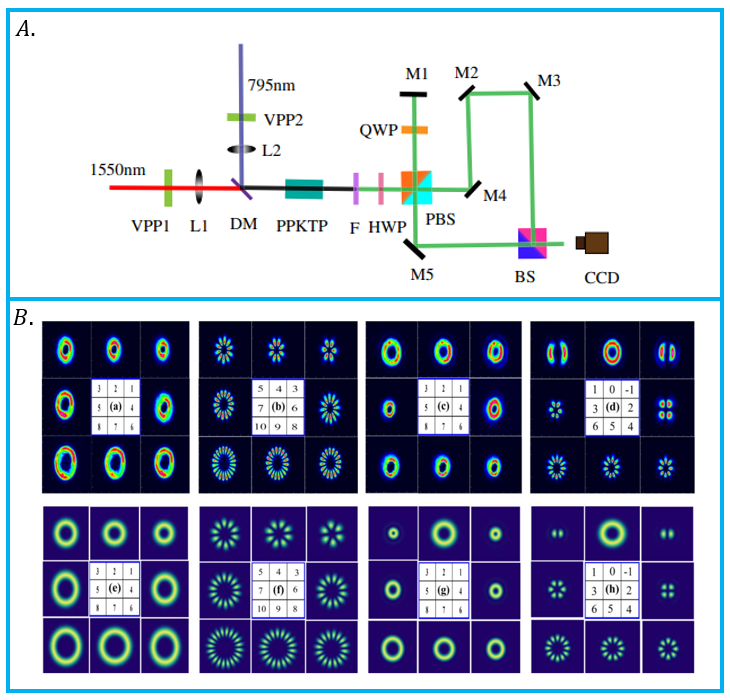}
\caption{Sum-frequency generation of vortex beams.
(A) Experimental setup. VPP1 and VPP2: vortex phase plates; L1 and L2: lenses; BS (PBS): beam splitter (polarizing beam splitter); M1–M4: mirrors; F: filter; PPKTP: periodically poled KTP crystal; DM: dichroic mirror; QWP (HWP): quarter-wave plate (half-wave plate); CCD: charge-coupled device camera.
(B) Experimental results (a–d) and corresponding simulations (e–h) when both pump beams carry OAM. The 795 nm pump beam carries OAM $+2$ in (a) and (b), and OAM $-2$ in (c) and (d). Panels (a) and (c) show the spatial intensity distributions of the SFG output, while (b) and (d) show the corresponding interference patterns.
Reprinted with permission from Ref.~\cite{YanLi:15}.}
\label{Fig31}
\end{figure}

Furthermore, in Ref.~\cite{YanLi:15}, Li et al. theoretically and experimentally investigated SFG in quasi-phase-matched PPKTP crystals (see Fig.~\ref{Fig31}(A) for the experimental setup) under two excitation conditions:
(I) Only one of the two fundamental beams carries OAM (an LG mode at 1550 nm), while the other is a Gaussian beam at 795 nm;
(II) Both fundamental beams carry OAM (two LG modes). Their key results were: (1) In the first configuration, the SFG output at 525.5 nm remains in an LG mode carrying the same OAM as the 1550 nm LG beam, consistent with OAM conservation. Additionally, the SFG beam size increases monotonically with the OAM of the 1550 nm beam.
(2) In the second configuration, the SFG output carries OAM equal to the sum of the OAM values of the two fundamental beams, irrespective of the sign of their individual OAMs—except when the two fundamental beams possess equal-magnitude but opposite-sign OAM, in which case the output behavior changes (see Fig.~\ref{Fig31}(B)).

In Ref.~\cite{LeeAndrew2016}, Lee et al. demonstrated sum-frequency mixing of vortex beams inside an intra-cavity self-Raman laser operating in a linear configuration. A type-I, temperature-tuned, phase-matched LBO crystal was used to investigate SFG between the fundamental and Stokes fields, both carrying vortex characteristics. They observed fluctuations in both the intensity and TC of the SFG field, accompanied by similar variations in the topological charges of the near-infrared fundamental and Stokes fields. These effects were attributed to the competition between two intra-cavity nonlinear mechanisms: stimulated Raman scattering and SFG.

\subsection{Parametric down conversion (PDC) with vortex beams}

In $\chi^{(2)}$-governed nonlinear optical processes such as SHG and SFG—where two low-frequency photons combine to produce a higher-frequency photon—we discussed how the phase relationship among the three interacting waves determines the direction of energy transfer. Efficient conversion from low- to high-frequency waves occurs when the phase-mismatch factor approaches zero. Depending on the relative phases of the interacting waves, it is also possible to reverse the direction of energy flow, whereby a high-frequency photon generates two lower-frequency photons inside a nonlinear crystal. This process is known as parametric downconversion (PDC). The term “parametric” emphasizes that the internal energy states of the crystal remain unchanged: the crystal provides its nonlinear response but neither absorbs nor emits net energy. In other words, energy is conserved entirely within the optical fields (i.e., among the interacting photons). The term “downconversion” refers to the fact that the energies of the output photons—commonly called the signal and idler photons—are lower than that of the input pump photon. For PDC to occur, both energy, $\hbar \omega_{p}=\hbar \omega_{s}+\hbar \omega_{i}$, and momentum, $\hbar \bm{k}_{p}=\hbar \bm{k}_{s}+\hbar \bm{k}_{i}$, must be conserved, where $(\omega_{p}, \bm{k}_{p})$, $(\omega_{s}, \bm{k}_{s})$, and $(\omega_{i}, \bm{k}_{i})$ denote the angular frequencies and wave vectors of the pump, signal, and idler photons, respectively. Unlike SHG and SFG, the outcomes of PDC are less restricted: the signal and idler frequencies $\omega_{s}$ and $\omega_{i}$ may vary over a range of values as long as their sum equals $\omega_{p}$. A similar flexibility exists for their momenta $(\bm{k}_{s}, \bm{k}_{i})$ and for their OAM values $(l_{s}, l_{i})$, since OAM is directly connected to the azimuthal component of $\bm{k}$, as discussed in earlier sections.

\begin{figure}[h!]
\centering 
\includegraphics[width=1\linewidth]{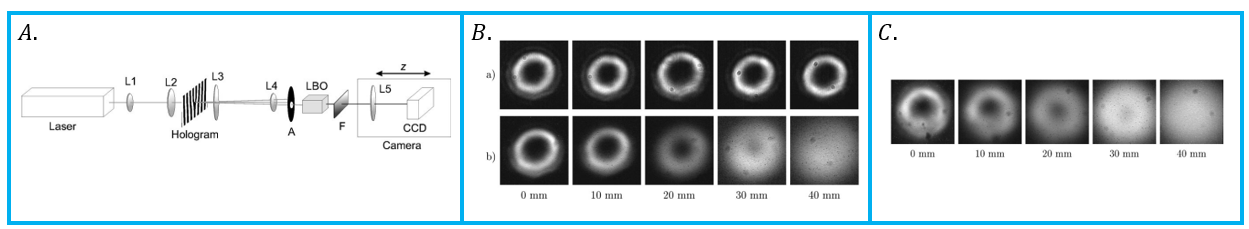}
\caption{Parametric downconversion of optical vortex beams.
(A) Experimental setup: L1–L5 are lenses, A is an aperture, and F is an optical filter. (B) (a) Intensity profile of the green pump beam carrying topological charge $l=2$, and (b) intensity profiles of the downconverted infrared beam recorded at various propagation distances from the back face of the nonlinear crystal. (C) Intensity profiles of the downconverted beams generated by a pump beam carrying topological charge $l=1$, shown for multiple propagation distances.
Reprinted with permission from Ref.~\cite{Jarlt}.}
\label{Fig32}
\end{figure}

In 1999, Arlt et al. investigated spontaneous parametric downconversion (SPDC) of LG vortex beams in an LBO crystal configured for type-I temperature-tuned phase matching (see Fig.~\ref{Fig32}(A) for the experimental setup)~\cite{Jarlt}. Their key observations were: (1) Beam structure evolution: In the degenerate SPDC regime—where the signal and idler photons share equal energy—the collimated LG pump beam maintains its spatial structure after exiting the crystal. However, the downconverted beam undergoes noticeable structural evolution: very close to the crystal exit, the intensity profile resembles that of the LG pump (with a clear central null), but farther away it gradually degrades toward a Gaussian-like profile exhibiting non-zero central intensity (see Fig.~\ref{Fig32}(B)). The same behavior was observed for non-degenerate SPDC (see Fig.~\ref{Fig32}(C)). (2) Absence of classical-field OAM conservation: The OAM was not conserved as a directly observable quantity in the classical fields. This result stands in clear contrast to SHG and SFG processes, where OAM is typically conserved within the interacting waves. (3) Lack of spatial coherence transfer: The spatial coherence of the LG pump beam is not preserved in the individual downconverted beams (signal or idler). Each downconverted beam exhibits low spatial coherence and can be represented as an incoherent superposition of multiple spatial modes. Consequently, the OAM of the pump is not transferred to either the signal or idler beam individually.

However, in 2002, Caetano et al. demonstrated that OAM is conserved in the stimulated parametric downconversion process~\cite{Caetano}. In this configuration, along with the pump laser (either a Gaussian or LG beam), a second auxiliary laser (also either Gaussian or LG) is aligned to overlap spatially with one of the downconverted modes (here, the signal beam), thereby inducing stimulated emission in the idler beam. To suppress spontaneous emission, the auxiliary beam power is made significantly higher than that of the pump. As a consequence, the idler beam exhibits completely modified intensity and spectral properties. In particular, they found that due to OAM conservation, $l_{p}=l_{s}+l_{i}$, where $l_{p}$, $l_{s}$, and $l_{i}$ denote the OAM values of the pump, auxiliary (signal), and idler beams, respectively, the idler output forms a doughnut-shaped intensity profile and propagates as an LG mode with a well-defined nonzero OAM. Notably, for the idler to exhibit a vortex structure (i.e., $l_{i} \neq 0$), at least one of the input beams must carry OAM—either $(l_{p} = 0,, l_{s} \neq 0)$ or $(l_{p} \neq 0,, l_{s} = 0)$. Furthermore, Ref.~\cite{Oliveira} established additional selection rules beyond total angular-momentum conservation in stimulated PDC. They showed that the radial mode properties of the idler beam depend sensitively on the TC values of the pump and auxiliary beams. Importantly, they demonstrated that an idler beam with both nonzero net OAM and nonzero radial order can be generated even when both input beams carry zero radial order.

\begin{figure}[h!]
\centering 
\includegraphics[width=0.5\linewidth]{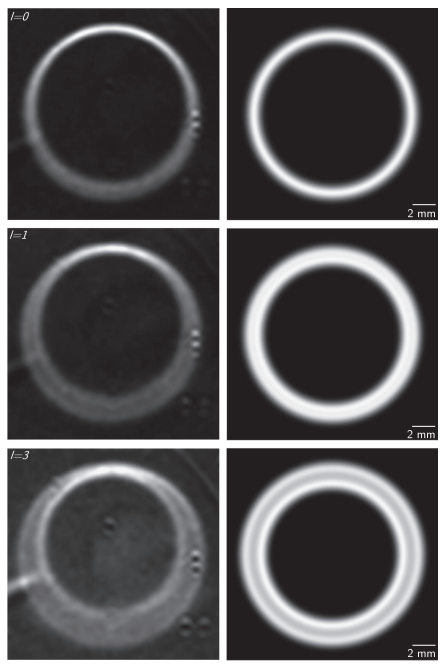}
\caption{Spontaneous parametric down conversion of Gaussian and higher-order optical vortex beams. First column: experimental results, second column: simulation results. (First row) $l=0$, (Second row) $l=1$, and (Third row) $l=3$. Reprinted with permission from Ref.~\cite{PRABHAKAR201464}.}
\label{Fig33}
\end{figure}
  
In Ref.~\cite{PRABHAKAR201464}, the authors investigated the spatial distribution of degenerate signal–idler photon pairs generated via SPDC when pumped either by a Gaussian beam or by higher-order optical vortices. They observed that, when driven by vortex beams, the SPDC output exhibits two concentric bright emission rings, with non-zero intensity between them. Moreover, the full width at half maximum (FWHM) of the rings increases monotonically with the topological charge of the pump vortex beam (see Fig.~\ref{Fig33}). In Ref.~\cite{Jabir2016}, perfect vortex beams were used as the pump in the SPDC process. The main findings were:
(1) The angular spectrum of the SPDC photons is independent of the OAM of the pump photons;
(2) The SPDC photons exhibit an asymmetric angular spectrum when spatial walk-off in the nonlinear crystal is unavoidable. However, the degree of asymmetry can be controlled by adjusting the vortex radius of the pump beam—an inverse relationship was demonstrated between these two quantities.

\subsection{Non-Perturbative nonlinear optical phenomena driven by vortex beams}

In the previous subsection, we discussed perturbative nonlinear optical processes, where the electric field of light is treated as a small perturbation to the native state of a system, and the induced polarization is expressed as a power-series expansion in the electric field. This description holds when the external field strength is much weaker than the internal atomic fields that bind electrons. In this regime, nonlinear interactions can be intuitively understood in the photon picture—where discrete numbers of photons are absorbed or emitted—and the efficiency of an $n$-th order nonlinear process scales as the $n$-th power of the driving field intensity (e.g., SHG intensity scales as the square of the fundamental intensity). However, when the driving field becomes extremely intense—as achieved with modern ultrafast, high-power lasers—the electric field approaches the magnitude of the atomic binding field. In this limit, the perturbative expansion fails to converge, even when higher-order terms are included. The system can no longer be viewed as undergoing small, probabilistic transitions between unperturbed states; instead, the strong laser field significantly alters the electronic potential landscape itself. In other words, perturbation theory breaks down, and a fundamentally different treatment becomes necessary. This is where the non-perturbative regime emerges. Here, the oscillating electric field directly governs the electron motion, and the light–matter interaction cannot be described solely by photon counting. A distinguishing characteristic of non-perturbative processes lies in their intensity scaling: the yield of an $n$-th order non-perturbative process typically scales as the $p$-th power of the driving intensity, where $p < n$. Prominent examples of non-perturbative strong-field phenomena include high-order harmonic generation (HHG)~\cite{McPherson:87,Ferray1988}, above-threshold ionization (ATI)~\cite{Agostini}, non-sequential double ionization (NSDI)~\cite{Fittinghoff}, and laser-induced electron diffraction (LIED)~\cite{ZUO1996313}. A rapidly developing frontier in strong-field physics involves driving HHG with structured light, particularly vortex beams carrying OAM. In this review, we survey advances in OAM-driven HHG across both gaseous and solid-state media. To set the stage, we first provide a brief overview of the fundamental mechanisms of HHG before exploring how these dynamics are modified under vortex-beam excitation.

\begin{table}
\tbl{History of attosecond pulse generation from gaseous media.}
{\begin{tabular}{|c|c|c|} 
\hline
Year & Pulse Duration (as) & Reference   \\
\hline
2001   & 250 & P. M. Paul et al. \cite{Paul}\\
2001   & 650 & M. Hentschel et al. \cite{hentschel_attosecond_2001} \\
2004   & 250 & R. Kienberger et al. \cite{kienberger_atomic_2004}    \\
2006   & 130 & G. Sansone et al. \cite{Sansone}\\
2008   & 80  & E. Goulielmakis et al. \cite{Goulielmakis}\\
2012   & 67 & K. Zhao et al. \cite{KunZhao:12}\\
2017   & 53 & J. Li et al. \cite{li_53}\\
2017   & 43 & T. Gaumnitz et al. \cite{Gaumnitz:17}\\
2024   & 51 & X. Wang et al. \cite{Xiaowei_2024}\\
2025   & 25 & J. Gao et al. \cite{Jingsong_25atto}\\
\hline
\end{tabular}}
\label{TB1}
\end{table}

Manipulating exposure time has long been the primary technique for capturing motion in both static and dynamic scenes. This idea can be traced back to 1872, when Eadweard Muybridge answered a well-debated question of his era—“Does a horse in full gallop ever have all four hooves off the ground at once?”—by reducing the exposure time to less than one-tenth of a second, revealing that the answer is indeed yes. Since then, significant effort has been devoted to achieving even shorter exposure times. This naturally raises a fundamental question: Can exposure times be reduced enough to resolve electron motion inside atoms, molecules, and solids? From the Bohr model of hydrogen, we know that an electron on the first Bohr orbit takes approximately 152 attoseconds ($1$ as $=10^{-18}$ s) to complete one classical revolution around the nucleus. Thus, resolving electron dynamics requires exposure times on the attosecond scale. However, directly generating attosecond-duration laser pulses from a laser oscillator is currently impractical, even using state-of-the-art gain media such as titanium-doped sapphire. A typical Ti:sapphire laser operates near a central wavelength of $\sim 800$ nm, corresponding to an optical period of about 2.67 femtoseconds ($1$ fs $=10^{-15}$ s). Since a pulse cannot practically be shorter than a single optical cycle, conventional lasers are fundamentally limited to durations above the femtosecond scale. Therefore, new light sources are required to access attosecond temporal resolutions. High-order harmonic generation (HHG) provides exactly such a route: a tabletop source of coherent extreme-ultraviolet (XUV) radiation—with temporal durations reaching the attosecond regime, and in some cases extending into the soft X-ray region—enabling direct measurements of ultrafast electron dynamics~\cite{Antoine,Paul}. In recognition of the significance of these ultrafast light sources, the Nobel Prize in Physics 2023 was awarded to Pierre Agostini, Ferenc Krausz and Anne L’Huillier “for experimental methods that generate attosecond pulses of light for the study of electron dynamics in matter.” Their pioneering work underpins the very concept of HHG-based attosecond sources and highlights the capability of HHG to provide the temporal resolution needed to track electrons in action. In Table~\ref{TB1}, we show the evolution of attosecond pulses generation from gaseous media. The table summarizes the rapid progress in attosecond pulse generation over the past two decades, highlighting a steady reduction in achievable pulse durations. Early demonstrations in 2001 reported pulses in the few-hundred–attosecond range, marking the first experimental access to electron dynamics on their natural timescale. Subsequent advances in laser technology, phase control, and gating techniques led to a continuous shortening of pulse durations, reaching sub-100 as by 2008 and approaching the few-tens–of-attoseconds regime in the 2010s. The most recent results illustrate both sustained refinement and renewed breakthroughs, culminating in the generation of pulses as short as 25 as in 2025. Overall, the evolution captured in the table reflects the maturation of attosecond science from proof-of-principle experiments to highly controlled sources capable of probing and controlling ultrafast electronic dynamics with unprecedented temporal resolution.

HHG is an extremely nonlinear and non-perturbative optical process that arises when an intense laser field $\left(\gtrsim10^{13}–10^{14},\mathrm{W/cm^2}\right)$ is tightly focused into a target medium and generates radiation at new frequencies—typically integer multiples of the driving field frequency. The target medium may consist of atoms, molecules, or condensed phase such as solids and liquids. Notably, in condensed systems the threshold intensity for HHG can be 1–2 orders of magnitude lower than in gaseous media due to the stronger electronic confinement and higher density. HHG from atomic gases was first observed independently by McPherson et al. and Ferray et al. in the late 1980s~\cite{McPherson:87,Ferray1988}. In the weak-field regime, the generation of low-order harmonics can be described perturbatively: an atom simultaneously absorbs $n$ photons of energy $\hbar\omega$ and emits a photon of energy $n\hbar\omega$ due to energy conservation. Because the probability of absorbing $n$ photons rapidly decreases with increasing $n$, the resulting harmonic spectrum exhibits an exponential decay with harmonic order. However, this perturbative picture breaks down in the strong-field regime. Experimentally, HHG spectra reveal a distinct structure: the harmonic intensity initially stays nearly constant across a broad range of harmonic orders—forming the plateau—before dropping sharply at a characteristic cutoff beyond which no emission occurs~\cite{Jeffrey}. The emergence of this plateau is a hallmark of non-perturbative light–matter interaction. Moreover, harmonics within the plateau display similar intensity-scaling behavior, unlike perturbative harmonics where the scaling follows the exact harmonic order. It is also important to note that typical HHG spectra in atomic gases contain only odd-order harmonics, a consequence of the spatial inversion symmetry of atoms and the temporal symmetry of a linearly polarized driving field.

Immediately following the first experimental evidence of HHG in atomic gases, researchers worldwide sought to uncover the fundamental mechanisms responsible for generating high-order harmonics. In 1993, Paul B. Corkum introduced a semi-classical model—famously known as the three-step model—to theoretically interpret the characteristic HHG spectra~\cite{PBCorkum}. In this model, the laser field is treated classically, while the target atoms are treated quantum mechanically. Despite being a single-atom description, it remarkably explains the origin of both the plateau and the cut-off observed in experimentally measured HHG spectra. The three-step model describes the generation of coherent ultrashort XUV or soft X-ray radiation through the following sequence:
(1) Tunnel ionization of the electron wave packet from the parent atom (a quantum process),
(2) Acceleration in the continuum solely under the influence of the strong driving laser field (well described by Newtonian mechanics), and
(3) Recombination of the electron with its parent ion-core, releasing the excess kinetic energy gained while in the continuum as a high-harmonic photon (again a quantum process). Laser-induced ionization lies at the heart of strong-field physics and depends sensitively on both the intensity and frequency of the laser field. At moderate intensities, atoms ionize via multiphoton absorption. However, at sufficiently high intensities, the strong optical field distorts the Coulomb potential to form a transient tunneling barrier within a fraction of an optical cycle, enabling the electron to tunnel out on a timescale shorter than the laser period. This regime corresponds to adiabatic tunneling, where the laser frequency is much smaller than the electron tunneling frequency, allowing sufficient time for the electron to escape through the barrier. Keldysh was the first to classify the multiphoton and tunneling regimes of strong-field ionization via the Keldysh parameter $\gamma$~\cite{Keldysh}, defined as $\gamma=\sqrt{I_{p}/2U_{p}}=\omega \frac{\sqrt{2I_{p}}}{E_{0}}$ (expressed in atomic units), where $I_{p}$ is the ionization potential of the target atom, $U_{p} = \frac{E_{0}^2}{4\omega^2}$ is the ponderomotive energy (i.e., the cycle-averaged kinetic energy of a free electron in the laser field), $\omega$ is the laser frequency, and $E_{0}$ is the peak field amplitude.

In the context of HHG, $\gamma \gg 1$ ($\gamma \ll 1$) corresponds to a regime where multiphoton ionization (tunneling ionization) dominates the generation process. The condition $\gamma \ll 1$ can be achieved either by increasing the laser intensity or by reducing the laser frequency—both conditions favoring tunnel ionization. By applying classical Newtonian mechanics to describe the electron’s motion in the continuum (for example, extracting its trajectory and return time information), one arrives at a universal cut-off law for gas-phase harmonics, $\hbar \omega_{\mathrm{cut-off}}=I_{p}+3.17 U_{p}$, which is rigorously valid for linearly polarized driving fields. Notably, two distinct electron trajectories—commonly labeled short and long, depending on how long the electron resides in the continuum—can yield the same return energy for a given harmonic order. Because these trajectories have different ionization and recombination times, the resulting harmonic emission is intrinsically chirped. Since $I_{p}$ is fixed for a given atomic species, the HHG cut-off can be extended by either increasing the laser intensity or decreasing the laser frequency. However, neither parameter can be varied indefinitely: (1) Excessively high laser intensities (e.g., $\gtrsim10^{16}~\mathrm{W/cm^2}$) can lead to full ionization of the target medium, generating a large density of free electrons that degrade phase matching and introduce incoherence in the emitted radiation. (2) Increasing the laser wavelength causes strong quantum diffusion of the electron wave packet, reducing the recombination probability and limiting the harmonic yield. Thus, a careful balance between intensity and wavelength must be maintained to maximize HHG efficiency. Beyond this trade-off, numerous additional strategies have been developed to extend the HHG cut-off (noting that a broader phase-locked plateau directly enables the generation of shorter attosecond pulses).

In 1994, Maciej Lewenstein and collaborators introduced a quantum formulation of the semi-classical three-step model: the strong-field approximation (SFA)~\cite{Lewenstein}. Within SFA, one solves the time-dependent Schrödinger equation under the single-active-electron (SAE) approximation, using a wavefunction ansatz that includes both the ground state and continuum states. The laser-atom interaction is then encoded in the computation of the time-dependent dipole moment, subject to the following key assumptions: (1) Only the ground state is considered bound, and intermediate bound-state excitations are neglected (valid when $\hbar\omega \ll I_{p}$). (2) Ground-state depletion is ignored as long as the laser intensity remains well below the saturation intensity.
(3) The influence of the parent ion on the continuum electron is neglected when $U_{p} \gg I_{p}$. Under these conditions, the dipole moment consists of three contributions—two transition dipole terms associated with ionization and recombination, and a rapidly varying phase term accounting for the semi-classical action accumulated during continuum propagation. These components correspond directly to the three steps of the classical picture of HHG. Due to the presence of rapidly oscillating phase factors, the stationary-phase (saddle-point) approximation is applied to obtain analytical expressions for the dipole moment. Enforcing stationarity in momentum, ionization time, and recombination time yields the familiar quantum analogs of the recollision conditions: conservation of energy at ionization and recombination, and the requirement that the electron return to the ion-position during recombination. A refined version of SFA, known as extended SFA (SFA+), was later proposed by Pérez-Hernández et al.~\cite{Perez-Hernandez:09}. This approach incorporates (1) bound-state excitations and (2) laser dressing of the ground state in the dipole calculation. For high-order harmonics, SFA+ has been shown to closely reproduce full time-dependent Schrödinger equation (TDSE) results. In addition to SFA-based techniques, several related models—including the quantum-orbit model~\cite{Milosevic} and the quantitative re-scattering model~\cite{ALE,Abro}—have been further developed to accurately describe single-atom responses in strong fields.

All the discussions above have focused exclusively on the microscopic aspect of HHG. However, HHG is inherently a collective nonlinear phenomenon, and thus a complete description requires going beyond single-atom response calculations. In realistic experiments, a gas jet or gas cell containing roughly $10^{12}$ atoms is placed in the laser interaction region, and the detected HHG signal arises from the coherent superposition of radiation emitted by all atoms within that volume. Because the atoms occupy different spatial positions, they experience different local intensities—and therefore acquire different phases—from the driving laser field. As a result, the harmonic fields emitted from different regions of the medium generally do not remain perfectly phase-aligned, leading to phase mismatch, which can drastically reduce the overall conversion efficiency. Therefore, it is always necessary to carefully tune various beam and medium parameters to counteract phase-mismatch contributions arising from multiple sources, such as beam focusing (i.e., the Gouy phase), curved wavefronts of the driving beam (i.e., the radial phase), dispersion induced by neutral atoms, plasma dispersion due to free electrons generated during HHG, and the intrinsic dipole phase (which depends on the strong-field quantum path and the local laser intensity). Previous studies have shown that positioning the HHG medium after the laser focus suppresses contributions from long electron trajectories, resulting in short-trajectory dominance across all detection angles. In contrast, placing the target at the focal plane yields significant contributions from both long and short trajectories. When the medium is positioned before focus, long (short) trajectories dominate for low (high) divergence angles. Additionally, it has been demonstrated that when HHG is driven by a fundamental Gaussian beam, the harmonic divergence decreases with increasing harmonic order, reflecting the tighter recollision geometry of higher-energy electrons.

These discussions naturally raise several key questions about HHG driven by structured light: (1) What changes occur in the HHG process when spatially structured vortex beams are used instead of fundamental Gaussian beams? (2) How can a rigorous theoretical framework be developed for ultrashort vortex–matter interactions? (3) Is OAM conserved among the interacting fields, as typically observed in SHG and SFG?
(4) How does the divergence of the generated vortex harmonics scale with harmonic order? To address these questions, in the next subsection we provide an extensive review of HHG in atomic gases driven by vortex beams.

 \subsection{Gas-phase HHG driven by vortex beams}

The history of spatially structured light–driven HHG in atomic gases traces back to 2012, when Zürch et al. conducted the first experiment using a linearly polarized vortex beam carrying an OAM of $l = 1$~\cite{Zürch2012}. However, the results were somewhat controversial. Although the generated harmonics exhibited clear vortex features in both the near- and far-field regions, the OAM conservation law ($l_{q} = q \times l$, where $l_{q}$, $l$, and $q$ denote the OAM per photon of the $q^{\text{th}}$ harmonic, that of the driving field, and the harmonic order, respectively) appeared to be violated at the far field. Specifically, it was observed that the OAMs of all high-order harmonics were equal to that of the driving field ($l_{q} = l = 1$). The authors attributed the decay of highly charged vortices (formed near the generation plane) into singly charged vortices (observed in the far field) to strong background beam modulation and perturbations inherent to the HHG process.

\begin{figure}[h!]
\centering 
\includegraphics[width=1\linewidth]{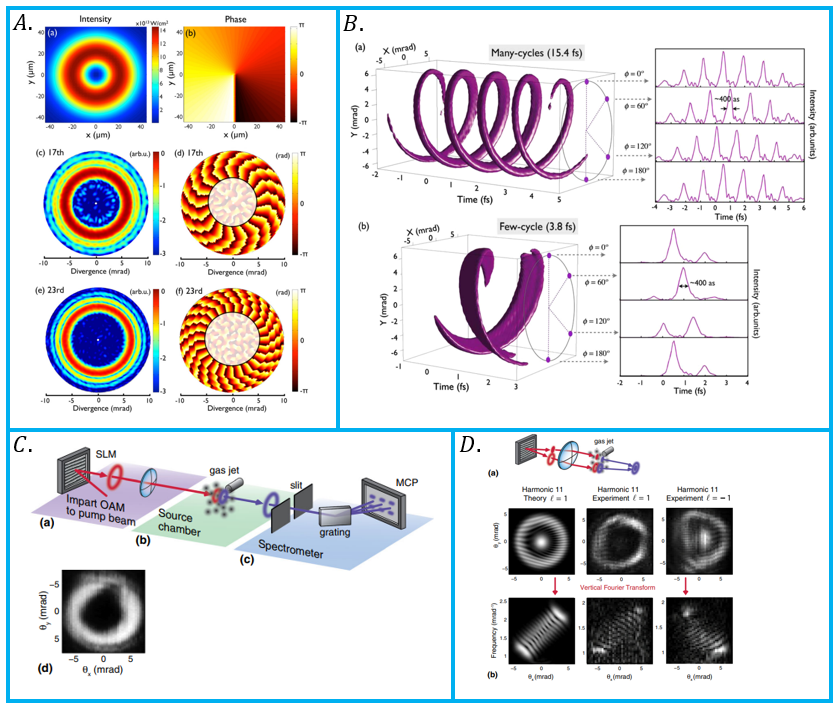}
\caption{Theoretical description of harmonic vortices and their experimental realization.
(A) Transverse (a) intensity and (b) phase profiles of the $LG_{1,0}$ mode of the fundamental beam at the focus position. (c,e) Intensity and (d,f) phase angular profiles for the 17th–23rd harmonics.
(B) Temporal evolution of the high-harmonic signal driven by (a) a multi-cycle (15.4 fs) and (b) a few-cycle (3.8 fs) pulse. The emitted XUV radiation carries OAM in the form of a helical attosecond structure. Panels (A)–(B) are adapted from Ref.~\cite{Carlos1}.
(C) OAM is imparted to the driving beam using a spatial light modulator. (a) High-order harmonics are generated in argon (b) and measured with a spectrometer (c). (d) Intensity profile of the 13th harmonic, representative of all harmonics exhibiting the same characteristic ring-shaped structure.
(D) Generating harmonics from two distinct beams at focus produces an OAM beam and a Gaussian reference beam (a). (b) Their far-field interference results in an n-fork interference pattern, as shown in the upper left image, where $n$ corresponds to the OAM per photon of the nth harmonic. The vertical Fourier transform of this pattern reveals a frequency gradient across the beam with $n + 1$ discrete steps, as shown in the lower left image. (C)-(D) are taken from Ref.~\cite{Gariepy}. Reprinted with permission from Ref.~\cite{Carlos1,Gariepy}.}
\label{Fig34}
\end{figure}

These findings sparked a series of theoretical and experimental investigations into HHG driven by vortex beams. In Ref.~\cite{Carlos1}, García et al. developed a theoretical framework known as the quantum SFA model to simulate HHG in gas jets and track the propagation of the generated radiation from each emission point to the detector. Their calculations showed that highly charged vortices indeed survive propagation and that the OAM of the $q^{\text{th}}$ harmonic scales as $l_{q} = q \times l$. Furthermore, they demonstrated that different harmonic orders exhibit nearly identical divergences (see Fig.~\ref{Fig34}(A)), in stark contrast to HHG driven by Gaussian beams, where divergence typically decreases with increasing harmonic order. By coherently combining multiple high harmonics, they predicted the formation of an attosecond light spring (see Fig.~\ref{Fig34}(B))—a helical train of attosecond pulses delayed along the azimuthal coordinate. In this picture, a fixed azimuthal angle corresponds to a sequence of attosecond bursts, while spatial integration over the entire light spring results in a femtosecond-scale envelope. The OAM within such an attosecond light spring varies continuously because each harmonic’s OAM increases linearly with its order.

In 2014, Gariepy et al.~\cite{Gariepy} provided the first clear experimental confirmation of OAM conservation in HHG driven by linearly polarized vortex beams (see Fig.~\ref{Fig34}(C)). To characterize the OAM of the harmonic vortices, they employed a self-referenced interferometric method in which the harmonic vortex was interfered with a tilted Gaussian beam of the same wavelength, producing fork-like dislocations in the resulting interferogram. By counting the number of fringes above and below the dislocation line, they determined the OAM of each harmonic order (see Fig.~\ref{Fig34}(D)). However, this technique becomes increasingly challenging for very high-order harmonics, since the fringe spacing narrows drastically with decreasing wavelength, making the patterns difficult to resolve experimentally.

Controlling the OAM of vortex harmonics is as crucial as their generation. In Ref.~\cite{Kong2017}, Kong et al. proposed a novel scheme to produce harmonics with tunable orbital angular momenta. In their approach, a strong linearly polarized infrared Gaussian beam drives the HHG process in a noble gas medium, while a weak linearly polarized vortex beam—of the same frequency as the Gaussian driver—acts as a control field. They derived a selection rule for the OAM of the generated harmonics, given by $l_{q}=ml_{1}+nl_{2}$, where $l_{q}$, $l_{1}$, and $l_{2}$ denote the OAM of the $q^{\text{th}}$-order harmonic (XUV light), the OAM of the Gaussian driving beam (which is zero), and the OAM of the weak vortex beam, respectively, while $m$ and $n$ represent the numbers of photons absorbed from the Gaussian and vortex beams. This relation clearly shows that the OAM of the generated harmonics can be precisely tuned by adjusting the OAM of the weak vortex beam and by controlling the relative number of photons absorbed from each of the two driving fields.

Furthermore, using this scheme, the authors demonstrated the generation of XUV vortices featuring the smallest possible central dark core. A natural question then arises: what happens if two vortex beams drive the harmonic generation process? Does the previously discussed OAM selection rule still hold under such conditions? To address these questions, Rego et al. examined the case where a superposition of two linearly polarized vortex beams drives HHG~\cite{Rego1}. They derived a new OAM selection rule expressed as $l_{q}=(q-n)l_{1}+nl_{2}+m(l_{2}-l_{1})$, where $(q - n)$ and $n$ represent the number of photons absorbed from the vortex beams carrying OAMs $l_{1}$ and $l_{2}$, respectively, and $m$ is an integer determined by a strong-field quantum-path parameter and the beam intensity. This generalized rule naturally reduces to the single-beam case when the two driving beams possess identical OAMs ($l_{1} = l_{2}$), recovering the standard OAM selection law for vortex-beam–driven HHG.

In addition, they discussed how the intrinsic dipole phase of the harmonics plays a pivotal role in imparting a strong non-perturbative twist to the XUV vortices. In Ref.~\cite{Carlos2}, the authors introduced a new theoretical framework, known as the thin-slab model (TSM), to study intense ultrashort vortex light–matter interactions and to elucidate how the generated harmonics inherit the vortex characteristics of the driving field. In this model, the conventional HHG target is replaced by an infinitesimally thin two-dimensional slab, oriented perpendicularly to the propagation direction of the driving field. It is assumed that harmonic generation occurs precisely at the slab’s position—thus, the thin slab acts as a planar source of harmonics.

In the vicinity of the slab, the amplitudes and phases of the generated harmonics (referred to as near-field harmonics) are typically expressed as the driving field amplitude raised to the power of $p$—a scaling factor that remains nearly constant across plateau harmonics—and as the product of the harmonic order with the driving field’s phase plus the intrinsic dipole phase, respectively. Subsequently, the Fraunhofer diffraction formalism is employed to compute the complex field amplitudes of the far-field harmonics.

Notably, because of the slab’s infinitesimal thickness, longitudinal phase-matching is neglected, whereas transverse and azimuthal phase-matching are included to account for the structured transverse intensity of the driving vortex beam. The strength of this model lies in its ability to cleanly separate the long and short quantum-path contributions to HHG, thereby offering a clearer understanding of the complex spatial features observed in the far-field intensity distributions of different harmonic orders.

It is also important to emphasize that the results obtained from the TSM have been successfully validated against more sophisticated approaches, such as the quantum SFA model, which includes a discretized treatment of the HHG target. Using the TSM, the authors demonstrated several key features:
(1) The harmonic emission can be manipulated by adjusting the relative position between the laser focus and the target. Specifically, when the target is placed after the laser focus, short quantum-path contributions are emitted at smaller divergence angles compared to the long-path ones. Conversely, when the target is positioned before the laser focus, the long-path contributions are emitted at smaller angles than the short ones.
(2) Plateau harmonics exhibit nearly identical divergence angles, independent of their order.

In Ref.~\cite{ChengJin}, Jin et al. developed a new theoretical framework to study HHG driven by vortex beams in atomic targets. They solved the full three-dimensional Maxwell’s wave equations along with the Huygens integral to describe the propagation of harmonic fields in both the gas target and vacuum. A detailed phase-matching analysis for HHG driven by linearly polarized Laguerre–Gaussian beams revealed that when the gas target is placed after the laser focus, both short and long electron trajectories contribute to harmonic emission, leading to harmonics with similar divergence and a single-ring far-field intensity pattern. Importantly, they found that the diameter of this far-field ring does not scale with the harmonic order, but rather with the OAM of the driving field as $\sqrt{l}$.

In a complementary study, Sanson et al.~\cite{Sanson:20} examined how aberrations in the driving vortex beam influence the intensity and phase of the generated vortex harmonics. Later, Romain et al.~\cite{Romain} established a connection between the short and long quantum paths of HHG and the radial index of the emitted XUV vortices. They showed that the modal composition of the XUV radiation can be tuned by controlling the relative weight of different quantum-path contributions during the generation process.

In Ref.~\cite{Turpin2017}, Turpin et al. drove HHG using conical refraction (CR) beams carrying fractional OAM and demonstrated that the resulting harmonics in the XUV and soft X-ray spectral regions preserve the transverse structure of the driving CR beam, confirming the conservation of fractional OAM during HHG.

\begin{figure}[h!]
\centering 
\includegraphics[width=1\linewidth]{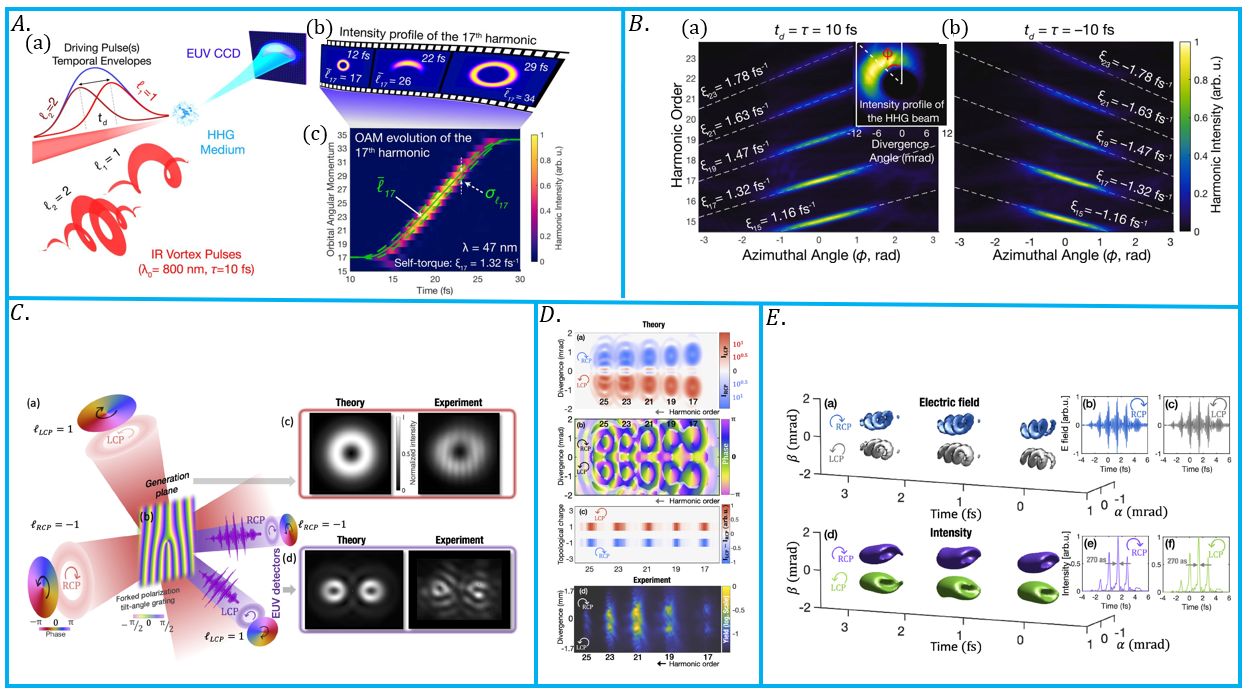}
\caption{Self-torque of light and generation of twisted attosecond pulses.
(A) Two time-delayed, collinear IR pulses with identical wavelengths but different OAM values are focused into an argon gas target (HHG medium) to generate harmonic beams exhibiting self-torque, as shown in (a). (b) Predicted evolution of the intensity profile of the 17th harmonic at three distinct instants during the emission process. (c) Temporal evolution of the OAM of the 17th harmonic for two driving pulses, each with a duration of 10 fs and a relative delay of 10 fs.
(B) Simulated spatial HHG spectra along the azimuthal coordinate when the time delay between the driving pulses is (a) 10 fs and (b) −10 fs. The self-torque of light introduces an azimuthal frequency chirp that varies for each harmonic, as indicated by the gray dashed lines. Panels (A)–(B) are adapted from Ref.~\cite{Regoself}.
(C) Polarization tilt–angle fork-grating scheme for generating attosecond vortex pulse trains.
(D) Spectrally and spatially resolved (a) intensity and (b) phase profiles for the RCP and LCP harmonic components. (c) Topological charge content of the harmonic beams obtained via azimuthal Fourier transform of the theoretical simulation results. (d) Spectrally and spatially resolved experimental characterization of the high-harmonic output, showing two well-defined harmonic beams with substantial spectral overlap.
(E) Far-field spatiotemporal structure of the electric field from numerical simulations, showing the generation of two separated attosecond vortex pulse trains with opposite handedness in both SAM and OAM in (a). (b) and (c) show the real parts of the RCP and LCP attosecond pulse trains at the spatial position of maximum intensity. (d) Far-field spatiotemporal intensity distribution from numerical simulations. (e) and (f) Intensity of the RCP and LCP attosecond pulse trains at the position of maximum intensity. (C)-(E) are taken from Ref.~\cite{Twistedatto}. Reprinted with permission from Ref.~\cite{Regoself,Twistedatto}.}
\label{Fig35}
\end{figure}

Other types of vortex beams have also been employed to investigate HHG in atomic media, including Bessel–Gauss beams~\cite{Han:23}, perfect optical vortex beams~\cite{BikashPRR}, ring Pearcey–Gaussian–vortex beams~\cite{DanLi}, elegant Laguerre–Gaussian beams~\cite{Granados1}, and double-ring vortex or optical ring-lattice beams~\cite{BikashOL}. For example, in Ref.~\cite{BikashPRR}, perfect vortex beams were proposed to overcome limitations associated with conventional vortex drivers. In traditional vortex beams, the size of the central dark core and the peak intensity scale monotonically with the OAM, resulting in reduced peak intensity and enlarged ring size at high $l$ values. When the intensity falls below the HHG threshold, harmonic generation becomes inefficient. Although tighter focusing could mitigate this issue, it often induces strong spin–orbit (SAM–OAM) coupling, leading to unwanted spatiotemporal vortex (STOV) structures that carry OAM in the transverse direction. In contrast, perfect vortex beams maintain OAM-independent intensity and ring size distributions, making them ideal for driving HHG even at high OAM values. Using the TSM, Das et al. demonstrated that harmonics generated with perfect vortex beams possess nearly identical divergence and strictly follow the OAM up-conversion rule~\cite{BikashPRR}. Moreover, by driving the HHG process with two time-delayed vortex beams differing by one unit of OAM, it becomes possible to generate self-torqued light, wherein the OAM of the emitted XUV beam evolves dynamically over sub-femtosecond timescales (see Fig.~\ref{Fig35}(A)–(B))~\cite{Regoself}.

All the preceding discussions focused primarily on driving HHG with linearly polarized vortex beams and generating linearly polarized vortex harmonics. A natural question then arises: what happens if we replace the linearly polarized driver with a circularly polarized vortex beam (either left- or right-handed)? Can a single circularly polarized vortex beam drive the HHG process? Will the generated vortex harmonics also exhibit circular polarization? And if so, what are the corresponding selection rules for the orbital (OAM) and spin (SAM) angular momenta?
These questions can be understood intuitively from a photon-picture perspective. Each photon carries a SAM of either $+1$ or $-1$, depending on its polarization. In a circularly polarized vortex beam, all photons share the same SAM. Consequently, if an electron absorbs multiple photons from such a beam, their SAMs add coherently. However, since the emitted high-harmonic photons can only possess SAM values of $\pm1$, it becomes impossible to generate harmonics using a single circularly polarized vortex beam. The same reasoning applies to single circularly polarized Gaussian beams, which also fail to drive HHG effectively.
Nevertheless, when two counter-rotating circularly polarized vortex beams are used as drivers, circularly polarized harmonics can indeed be produced. In Ref.~\cite{Kong2}, Kong et al. explored how the interplay between SAM and OAM plays a critical role in shaping the macroscopic wavefront of the emitted high-order harmonics. They proposed a bi-color driving scheme to generate spectrally separated XUV beams carrying OAM.
Building on this, Paufler et al.~\cite{Paufler_2019} simulated HHG driven by counter-rotating circularly polarized vortex beams in a bi-color ($\omega$–$2\omega$) configuration and derived the corresponding selection rules for energy, OAM, and SAM as $\omega_{q}=m\omega+n2\omega$, $l_{q}=ml_{1}+nl_{2}$, where $m$ and $n$ denote the number of photons absorbed from the beams carrying OAMs $l_{1}$ and $l_{2}$, respectively. They showed that SAM conservation suppresses harmonic orders that are integer multiples of three and demonstrated that the OAM of the generated harmonics can be precisely controlled by the OAMs of the driving beams—allowing spatial separation of left- and right-circularly polarized harmonics. In Ref.~\cite{Twistedatto}, Alba de las Heras et al. reported the first experimental demonstration of twisted attosecond pulse trains (distinct from attosecond light springs) using two circularly polarized vortex beams with opposite helicities in both SAM and OAM within a non-collinear geometry. They showed that simultaneous conservation of linear momentum, SAM, and OAM leads to the spatial separation of left- and right-circularly polarized harmonics: all left-circular harmonics carry an OAM of $+1$, while all right-circular harmonics carry an OAM of $-1$—a necessary condition for generating twisted attosecond pulses with well-defined OAM values (see Fig.~\ref{Fig35}(C)–(E)). More recently, in Ref.~\cite{delasHeras2024}, the same authors theoretically demonstrated that by employing two counter-rotating circularly polarized vortex beams carrying identical OAM values, the polarization state of XUV beams with self-torque can be continuously tuned—from linear to circular—offering a new route for dynamic control over the polarization and angular momentum structure of attosecond light.

Up to this point, our discussions have primarily centered on vortex beams with spatially homogeneous polarization profiles, i.e., scalar vortex beams. A natural next question arises: What happens if we drive the HHG process using vortex beams with spatially inhomogeneous polarization distributions—namely, vector vortex beams (VVBs)?
In simple terms, VVBs can be regarded as the superposition of two counter-rotating circularly polarized fields with distinct TC values. This combination endows them with spatially varying polarization distributions and a net TC, distinguishing them fundamentally from scalar vortices. One may then ask: can this additional degree of freedom—the spatially inhomogeneous polarization—be exploited to enhance the harmonic conversion efficiency? And how does the spatial structure of the harmonics driven by such beams appear?
To address these questions, de las Heras et al.~\cite{delasHerasVVB} investigated HHG driven by VVBs and reported several key findings:
(1) The generated harmonics exhibit vectorial vortex characteristics, preserving the complex polarization topology of the driving field.
(2) The TC, defined via the geometric Pancharatnam phase, scales linearly with the harmonic order.
(3) The conversion efficiency of the HHG process increases with the TC of the driving VVB, enabling the efficient generation of high-TC VVBs in the XUV spectral regime.
(4) By coherently synthesizing multiple high-order harmonics, it becomes possible to generate helical attosecond structures with spatially varying polarization.
Complementarily, in Ref.~\cite{Pandey2}, Pandey et al. conducted the first experimental demonstration of high-harmonic generation driven by VVBs and found that:
(1) The HHG process allows tunable control over the TC, polarization state, and OAM helicity of the generated XUV VVBs.
(2) The TC of the $q^{\text{th}}$-order harmonic scales linearly with the Pancharatnam TC of the driving VVB—not with the individual TCs of its left- and right-circularly polarized components—highlighting the geometric nature of OAM transfer in such structured light–matter interactions.

Recently, HHG in atomic gases driven by spatiotemporal optical vortices (STOVs) has attracted considerable research interest from the strong-field physics community. In contrast to spatial vortices, which carry longitudinal OAM, STOVs are characterized by transverse OAM and an intrinsic spatiotemporal coupling (STC). In Ref.~\cite{MartínHernández}, the authors theoretically and experimentally investigated on STOVs driven HHG and observed that: (1) The non-perturbative interaction of intense, infrared STOVs with an atomic target is capable of producing XUV light beams showing intrinsic STC at nanometric spatial and attosecond temporal scales. (2) The conservation of transverse OAM is followed. (3) Since the generation of XUV STOVs via HHG is highly sensitive to the experimental conditions, it was particularly shown that the inhomogeneity parameter- which typically describes non-elliptical STOVs- greatly influences the production of XUV lights both in the spatiotemporal and spatiospectral regimes. (4) A connection between the TCs carried by XUV STOVs and their spectral bandwidth is established, which facilitates the experimental characterization of XUV STOVs. Furthermore, in Ref.~\cite{Yiqifang}, the authors theoretically examined how the interplay between the microscopic and macroscopic characteristics of STOVs in HHG gives rise to intriguing phenomena, including spatial–spectral tilt and fine interference structures in the spatially resolved harmonic spectra. Moreover, by implementing a bi-color scheme with two STOV pulses possessing opposite helicity in both SAM and OAM,  they showed that the TCs and spectral features of the generated high harmonics can be readily controlled.  

\subsection{Solid-state HHG driven by vortex beams}

The remarkable coherence properties of high-order harmonic generation (HHG) have long been harnessed to propel the frontiers of ultrafast science, enabling the imaging of molecular orbitals~\cite{Itatani2004} and the exploration of multielectron dynamics in molecules~\cite{Shiner2011}. One of the most celebrated applications of HHG is the generation of attosecond light pulses, which constitute the cornerstone of attosecond science. These pulses are now routinely employed in pump–probe experiments, providing unprecedented temporal resolution to investigate ultrafast processes in dilute gases and complex biomolecular systems. Despite its success in atomic gases, HHG still faces several intrinsic limitations:
(1) Low conversion efficiency, primarily due to the low atomic density in the interaction region and phase mismatch among the generated harmonics, which arises from multiple factors such as beam focusing (Gouy phase), the curved wavefront of the driving beam, dispersion from neutral atoms and free electrons, and the intrinsic dipole phase.
(2) High laser intensity requirements to reach photon energies in the extreme ultraviolet (XUV) regime.
(3) Strong XUV absorption in air, necessitating the use of high-vacuum beamlines, which makes the experimental setup bulky and complex. This naturally leads to a fundamental question: Can the well-established HHG formalism in gases be extended to probe solid-state materials? The answer is an emphatic yes. A major breakthrough came in 2011, when Ghimire et al. demonstrated the generation of high harmonics (up to the 25$^{\text{th}}$ order) by focusing an intense mid-infrared laser beam (wavelength $3.25~\mu$m) onto the exit face of a ZnO crystal~\cite{Ghimire2011}. Remarkably, the generated harmonics extended well beyond the minimum bandgap energy of ZnO ($\sim3.34$ eV). Moreover, the dependence of the harmonic yield on the driving-field intensity exhibited non-perturbative scaling laws, and the harmonic cutoff was found to scale linearly with the laser’s electric-field strength—distinctly different from the quadratic dependence characteristic of gas-phase HHG. The observed anisotropy of the harmonic yield with crystal rotation, in contrast to the isotropic response in gases, further indicated a fundamentally different generation mechanism. The emergence of HHG in solids offers several compelling advantages:
(1) Lower intensity requirements—typically one to two orders of magnitude below those needed for gas-phase HHG, limited mainly by the damage threshold of the material, which depends on its properties and the laser parameters (wavelength, pulse duration, and repetition rate).
(2) Enhanced conversion efficiency, owing to the much higher atomic density of solids compared with gases.
(3) Simplified experimental setup, as mid-infrared driving lasers generate harmonics spanning the visible to vacuum-ultraviolet (VUV) range, eliminating the need for vacuum chambers, differential pumping stages, or gas cells.
(4) Intrinsic spectroscopic capabilities, since solid-state HHG can serve as a powerful probe for reconstructing electronic band structures and studying ultrafast carrier dynamics in materials. Collectively, these features make solid-state HHG a compact, efficient, and versatile tabletop source of coherent radiation, bridging attosecond science with condensed-matter physics.

The ``three-step model'' has proven remarkably successful in describing HHG in gases. By simplifying  the complex laser-matter interaction into three fundamental steps—tunnel ionization, laser-driven acceleration, and radiative recombination—it effectively captures a wide range of experimentally observed phenomena and continues to yield reliable predictions. However, elucidating the generation mechanisms of HHG in solids is considerably more complex than in atomic systems. This complexity arises from the defining characteristics of solids, including their diverse crystallographic symmetries, anisotropic optical responses, and the coexistence of intricate many-body interactions among charge carriers, defects, and lattice vibrations. The emission of high harmonics in solids is governed by a complex interplay of multiple competing mechanisms. Significant research efforts are therefore dedicated to identifying the dominant mechanism based on material properties and driving field parameters.

The electronic properties of solid-state materials are fundamentally defined by their band structure, consisting of a fully occupied valence band and an unoccupied conduction band, separated by a finite energy gap. When exposed to a strong laser field, electrons can be promoted from the valence to the conduction band through nonlinear optical processes such as multiphoton or tunnel ionization. Once excited, the laser field drives a complex, highly nonlinear electron dynamics within the solid, which can be separated into two main contributions:
(1) Intraband transitions, corresponding to the acceleration of electrons (or holes) within a single band, and
(2) Interband transitions, corresponding to quantum-mechanical recombination between electrons in the conduction band and holes in the valence band.
The coherent superposition of these two processes gives rise to the emission of high-harmonic radiation~\cite{Vampa_2017}, and both mechanisms are intrinsically coupled. A simple yet powerful model to visualize HHG in solids considers a material with a single valence and a single conduction band—a simplification that captures the essential physics of the laser–solid interaction. In the first step, an electron is promoted from the valence to the conduction band via strong-field-induced tunneling. Since the tunneling rate depends exponentially on the minimum band gap, this process predominantly occurs near the $\Gamma$ point, where the gap is smallest. The promotion of the electron leaves behind a hole in the valence band. In the second step, the laser field drives oscillations of both the electron and the hole within their respective bands. These oscillations generate a nonlinear current—the intraband current—that contributes to high-harmonic emission. In the final step, the electron in the conduction band can recombine with the hole in the valence band, emitting a high-harmonic photon whose energy is determined by the instantaneous band gap at the time of recombination.
Although this picture bears a conceptual resemblance to the three-step model in gas-phase HHG, several key differences arise, namely (a) Electron dynamics and band dispersion: In gases, once ionized, the electron moves freely in the continuum under the sole influence of the laser field, corresponding to a parabolic energy–momentum dispersion. In solids, however, the electron experiences the periodic potential of the crystal lattice and interacts with other electrons, phonons, and defects, resulting in a non-parabolic band dispersion. This nonlinearity induces a complex relationship between the electron’s group velocity and the driving field, generating high-frequency components—often termed velocity harmonics—that form part of the intraband contribution. Additionally, nonlinearity in the tunneling process itself gives rise to harmonic emission analogous to Brunel harmonics in gases and (b) Hole dynamics:
In atomic HHG, the residual ion (hole) is effectively stationary and represented by a flat band in momentum space. In solids, by contrast, the hole also participates in the dynamics, moving in momentum space in the same direction as the electron due to their opposite charge and effective mass.

Beyond these conceptual distinctions, several other key differences exist between HHG in solids and in gases, namely (1) multiple plateaus: In solids, multiple harmonic plateaus can appear when electrons in higher conduction bands recombine with holes in the valence band, significantly extending the harmonic cutoff~\cite{Yuchuan}, (2) reflection geometry: HHG in solids can occur in reflection geometries~\cite{Ghimire2011,Kolesik:25}, which are not accessible in gas-phase experiments, (3) cutoff scaling: In gases, the harmonic cutoff scales quadratically with the driving wavelength, while in solids, a clear and universal scaling law between cutoff energy and wavelength has not yet been established, (4) ellipticity dependence: In gas-phase HHG, the harmonic yield drops rapidly with increasing ellipticity of the driving field due to reduced recombination probability. In solids, however, this suppression is much weaker, and in some cases, the yield can even increase with ellipticity~\cite{Ghimire2011,Naotaka,Sekiguchi} and (5) polarization control: Owing to the intrinsic symmetry of solids, it is possible to generate circularly polarized harmonics using a single circularly polarized field. For example, Saito et al. demonstrated circularly polarized harmonic generation in GaSe crystals driven by circularly polarized mid-infrared pulses~\cite{NariyukiSaito:17}. Interestingly, every $3n$ harmonic (with $n = 1, 2, 3, ...$) was suppressed due to the threefold rotational symmetry of GaSe—a behavior impossible to realize in gas-phase HHG. Overall, these unique features make solid-state HHG not merely a condensed-matter analogue of gas-phase HHG, but a rich and fundamentally distinct phenomenon governed by the interplay of band structure, symmetry, and many-body effects.

A wide range of theoretical frameworks—including the time-dependent Schrödinger equation (TDSE)~\cite{Xinqiangwang,Mengxi}, density matrix approaches~\cite{Vampa2}, the semiconductor Bloch equations (SBEs)~\cite{Golde}, and time-dependent density functional theory (TDDFT)~\cite{Tancogne,Tancogne2}—have been extensively employed to describe the interaction of intense laser fields with solids. In parallel, HHG has been the focus of vigorous experimental and theoretical investigation across a broad spectrum of materials~\cite{Mrudul,ChangLee2024,You2017,yuki}. Despite this extensive body of research, the overwhelming majority of studies have relied exclusively on Gaussian driving beams. 

In Table~\ref{TB2}, we summarize the evolution of attosecond pulse generation from solid-state materials. As is evident, attosecond pulse generation in solids is less mature than in the gas phase, which is reflected in the comparatively longer pulse durations reported to date. Early demonstrations, such as attosecond emission from SiO$_2$ nanofilms in 2016, achieved pulse durations on the order of several hundred attoseconds, establishing the feasibility of solid-state high-harmonic generation as a compact ultrafast light source. More recent results in wide-bandgap materials such as ZnO and MgO continue to show pulse durations in the several-hundred to near–thousand–attosecond range, underscoring both the challenges and the distinct physical mechanisms inherent to solids, including band-structure effects, dephasing, and many-body interactions. Overall, the table highlights that while solid-state HHG has made important strides, further advances in material engineering, phase control, and dispersion management are required before solid-based attosecond sources can approach the temporal resolution routinely achieved in gas-phase systems.

\begin{table}[ht!]
\tbl{History of attosecond pulse generation from solid-state media.}
{\begin{tabular}{|c|c|c|c|} 
\hline
Year & Material &Pulse Duration (as) & Reference   \\
\hline
2016   &SiO$_2$ nanofilm& 472 & M. Garg et al. \cite{Garg_Nature_2016}\\
2025   &ZnO& 950 & A. Nayak et al. \cite{AttoSolid} \\
2025   & MgO& 700 & Z. Chen et al. \cite{ZhaopinAtto}    \\
\hline
\end{tabular}}
\label{TB2}
\end{table}

While the interaction of vortex beams with crystalline media is well established in the weak-field regime (as discussed in the context of SHG, SFG, and parametric down-conversion), the interaction of ultrashort vortex beams with solids in the strong-field HHG regime remains largely unexplored. This gap naturally raises several key questions:
(1) Does the characteristic intensity-null structure of vortex beams survive the HHG process in solids and imprint onto the emitted harmonics?
(2) Is the OAM conserved during the process?
(3) How does the OAM of the driving field influence the harmonic yield?
(4) Do the higher-order harmonics exhibit similar divergence trends as those observed in gas-phase HHG?
(5) Does the OAM of the driving field affect the intraband and interband contributions differently?

To address these questions, Gauthier et al. experimentally investigated OAM-driven HHG in solids, using a ZnO crystal as the target medium~\cite{Gauthier:19}. Their key findings can be summarized as follows:
(1) The ring thickness of the harmonic vortices decreases with increasing harmonic order, attributed to the nonlinear response of the crystal to the annular intensity profile of the driving vortex beam.
(2) The HHG process in solids amplifies mode imperfections of the driving vortex beam, thereby reducing the mode purity of the emitted harmonics.
(3) The central dark core of the harmonic vortices expands with harmonic order.
(4) The OAM conservation law—a linear scaling of harmonic OAM with harmonic order—is strictly obeyed, mirroring the behavior observed in atomic gases. Motivated by these observations, Granados et al.~\cite{Granados_2025} developed a theoretical framework unifying the SBEs for solids with the TSM established for gases. This hybrid SBE+TSM approach successfully reproduced the experimentally observed spatial features of vortex harmonics reported by Gauthier et al.~\cite{Gauthier:19}. Building on this foundation, the same group later proposed a scheme for generating twisted attosecond pulses—essentially, attosecond light springs—from solid-state media~\cite{Granados_US}. More recently, Jianing et al.~\cite{Jianing} theoretically investigated HHG in solids driven by spatiotemporal optical vortex beams using the semiconductor quantum-orbit model, and demonstrated the simultaneous conservation of spin angular momentum (SAM) and transverse OAM during the harmonic generation process. Together, these pioneering studies mark the emergence of a new research frontier—vortex-beam-driven HHG in solids—that bridges structured light physics and ultrafast condensed-matter dynamics. In Ref.~\cite{KoheiNagai}, the role of spin-orbit interaction in the production of high harmonics with spatial structures from a solid was thoroughly studied and the conservation of total angular momentum was established.  
\begin{figure}[h!]
\centering 
\includegraphics[width=0.6\linewidth]{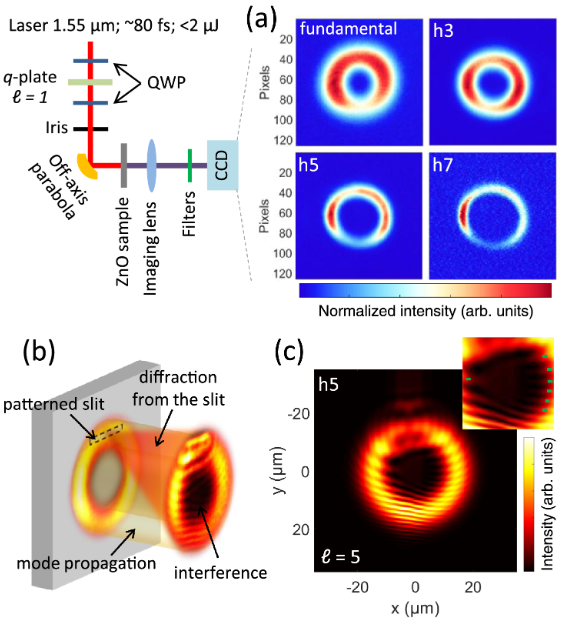}
\caption{HHG from a ZnO semiconductor using OAM beams. (a) Sketch of the experimental set-up (left) and transverse intensity distributions of the fundamental beam, third, fifth, and seventh harmonics, recorded at the CCD plane (right). (b) Topological charge of the modes is defined using self-referenced interferences. (c) Simulation result with the experimental parameters of the self-referenced interferogram for the fifth harmonic. Reprinted with permission from Ref.~\cite{Gauthier:19}.}
\label{Fig36}
\end{figure}

\section{Conclusions and Outlook}

In this review, we have summarized the fundamental aspects of optical vortices, emphasizing how the orbital angular momentum (OAM) of light constitutes an additional degree of freedom that can be effectively used to control light–matter interactions in both linear and nonlinear regimes. We revisited the main classes of vortex beams, including Laguerre–Gaussian (LG), Bessel–Gauss (BG), Perfect Optical Vortex (POV), and Lorentz–Gauss (LGa) beams, and discussed their generation, detection, and propagation characteristics in different media such as free space, gradient-index, and chiral media. The analysis demonstrates that the spatial structure and TC of these beams strongly influence their propagation behavior and focusing properties, providing unique advantages for a wide range of optical applications.

A particular focus was placed on nonlinear optical processes, where the use of vortex beams has led to new insights into light–matter coupling at high intensities. In the perturbative regime, vortex beams have been employed to study harmonic generation, four-wave mixing, and self-focusing, where the conservation and transfer of OAM play a key role in determining the spatial and polarization characteristics of the generated fields. In the non-perturbative regime, the driving of high-order harmonic generation (HHG) with vortex beams has attracted growing attention, both in gases and in solids. In gaseous media, it has been shown that the OAM of the fundamental field can be coherently transferred to the emitted harmonics according to well-defined selection rules, producing high-frequency radiation with helical phase fronts. These XUV or soft-x-ray vortex harmonics exhibit a characteristic ring-shaped intensity profile and carry quantized OAM values that scale with the harmonic order. Such beams provide new opportunities for studying chiral photoionization, orbital current generation, and angular-momentum–resolved ultrafast spectroscopy.

In condensed-matter systems, HHG driven by structured light reveals additional effects that arise from the interplay between the beam topology and the crystal symmetry. The combination of the OAM of the driving field with the lattice periodicity allows for the emergence of OAM-dependent selection rules and polarization control of the emitted harmonics. This connection between structured driving fields and solid-state band structures offers a promising approach to explore angular-momentum transport, topological transitions, and light-induced symmetry breaking in solids. Moreover, the ability to generate high-order harmonics at relatively low intensities suggests a feasible path toward compact sources of coherent vortex radiation extending into the extreme-ultraviolet spectral region.

Future work will likely concentrate on several directions. First, extending the control of OAM in strong-field interactions to the attosecond domain would allow for the generation of helical attosecond pulses and the study of ultrafast dynamics in chiral and magnetic materials. Second, the development of adaptive and integrated devices—such as metasurfaces, spatial light modulators, and photonic waveguides—may enable precise tailoring of the amplitude, phase, and polarization of vortex beams in compact configurations. Finally, theoretical and numerical studies, including fully quantum and semiclassical approaches, will be essential to describe the coupling of OAM with charge, spin, and lattice degrees of freedom in complex materials.

In conclusion, optical vortices represent a mature and versatile tool in modern optics. Their ability to carry and transfer OAM has not only deepened our understanding of light–matter interaction but has also opened new perspectives for strong field and ultrafast physics. The continued development of vortex-beam-driven nonlinear optics, especially in the context of HHG in gases and solids, will further extend the reach of structured light into new spectral, temporal, and material regimes.

\section*{Disclosure statement}

No potential conflict of interest was reported by the author(s).

\section*{Funding}

 We acknowledge support by the National Key Research and Development Program of China (Grant No.~2023YFA1407100), Guangdong Province Science and Technology Major Project (Future functional materials under extreme conditions - 2021B0301030005), the Guangdong Natural Science Foundation (General Program project No. 2023A1515010871), the National Natural Science Foundation of China (Grant No. 12574092) and the Guangdong Provincial Quantum Science Strategic Initiative GDZX2504001.



\end{document}